%% file: desy-06-164.tex
\documentclass[12pt]{article}
\usepackage{bigstrut}
\usepackage{lineno} 
\usepackage{epsfig}
\usepackage{amsmath}
\usepackage{hhline}
\usepackage{amssymb}
\usepackage{times}
\usepackage{cite}

\newlength{\dinwidth}
\newlength{\dinmargin}
\newcommand{\fig}{Figure}

\newcommand{\Mycaption}[1]{\caption{\em #1}}
\newcommand{\FIXME}[1]{}

\newcommand{\eV}{\mbox{\rm{e}\hspace{-0.08em}\rm{V}}}
\newcommand{\MeV}{\rm M\eV}
\newcommand{\GeV}{\rm G\eV}
\newcommand{\pb}{\rm pb}
\newcommand{\cm}{\rm cm}
\newcommand{\grad}{\ensuremath{^\circ\!}}     

\newcommand{\pom}{\ensuremath{I\!\!P}}

\newcommand{\pt}{\ensuremath{p_t}}
\newcommand{\kt}{\ensuremath{k_t}}

\newcommand{\ccb}{${c\bar{c}}$}
\newcommand{\dzero}{\ensuremath{D^{0}}}
\newcommand{\dstar}{\ensuremath{D^*}}

\newcommand{\dstarpm}{\ensuremath{D^{*\pm}}}

\newcommand{\ptds}{\ensuremath{p_t(\dstar )}}
\newcommand{\etads}{\ensuremath{\eta(\dstar )}}

\newcommand{\xpom}{\ensuremath{x_{\pom}}}
\newcommand{\my}{\ensuremath{M_Y}}
\newcommand{\mx}{\ensuremath{M_X}}
\newcommand{\zpom}{\ensuremath{z_{\pom}}}
\newcommand{\zpomobs}{\ensuremath{z_{\pom}^{obs}}}
 
\def\hera{{\sc HERA}}

\def\pythia{{\sc pythia}}
\def\heracles{{\sc heracles}}
\def\rapgap{{\sc rapgap}}
\def\hvqdis{{\sc hvqdis}}
\def\fmnr{{\sc fmnr}}

\def\gsim{\,\lower.25ex\hbox{$\scriptstyle\sim$}\kern-1.30ex%
\raise 0.55ex\hbox{$\scriptstyle >$}\,}
\def\lsim{\,\lower.25ex\hbox{$\scriptstyle\sim$}\kern-1.30ex%
\raise 0.55ex\hbox{$\scriptstyle <$}\,}
\newcommand {\lapprox} {\raisebox{-0.7ex}{$\stackrel {\textstyle<}{\sim}$}}

\begin{document}  
\begin{titlepage}

\begin{flushleft}

DESY 06-164 \hfill ISSN 0418-9833 \\
October 2006
\end{flushleft}

\vspace{2cm}
\noindent

\begin{center}
\begin{Large}

  \textbf{\boldmath Diffractive Open Charm Production in \\ 
          Deep-Inelastic Scattering and Photoproduction \\
          at HERA \\
  }

\vspace{2cm}

H1 Collaboration

\end{Large}
\end{center}

\vspace{2cm}

\begin{abstract}
\noindent
Measurements are presented of diffractive open charm production at \hera\@. 
The event topology is given by $e p \rightarrow e X Y$ where the system $X$ 
contains at least one charmed hadron and is well separated by a large rapidity 
gap from a leading low-mass proton remnant system $Y$\@. 
Two analysis techniques are used for the cross section measurements. In the 
first, the charm quark is tagged by the reconstruction of a $\dstarpm$\!(2010) 
meson. This technique is used in deep-inelastic scattering (DIS) and 
photoproduction ($\gamma p$). In the second, a method based 
on the displacement of tracks from the primary vertex is used to measure the 
open charm contribution to the inclusive diffractive cross section in DIS.
The measurements are compared with next-to-leading order QCD predictions based 
on diffractive parton density functions previously obtained from a QCD analysis 
of the inclusive diffractive cross section at H1. 
A good agreement is observed in the full kinematic regime, which supports the 
validity of QCD factorization for open charm production in diffractive DIS 
and $\gamma p$.
\end{abstract}

\vspace{1.5cm}

\begin{center}
Submitted to Eur. Phys. J. {\bf C}
\end{center}

\end{titlepage}

%
%
%
\begin{flushleft}
  \input{h1auts}
\end{flushleft}

\newpage

\section{Introduction}
\label{sec:introduction}

%
%
Diffractive processes in positron-proton ($ep$) collisions are those 
where the hadronic final state is separated by a large gap in rapidity, 
without hadrons, into two systems $X$ and $Y$, where the system $Y$ 
may consist only of a proton or low mass system. The system $X$ is 
known as the photon dissociative system. The diffractive event signature 
is understood to arise from a color singlet exchange between the two 
systems $X$ and $Y$\@. 

In quantum chromodynamics (QCD), the theory of strong interactions, the 
hard scattering collinear factorization theorem~\cite{collins_1998} 
predicts that the cross section for diffractive deep-inelastic $ep$ 
scattering (DIS) factorizes into a set of universal diffractive parton 
density functions (DPDFs) of the proton and process-dependent hard 
scattering coefficients. Next-to-leading order (NLO) DPDFs have been 
determined by QCD fits to the measured cross sections of inclusive 
diffractive scattering at HERA~\cite{h1_diff_incl_1997,h1_diff_incl_2005} 
within the factorizable Pomeron 
model~\cite{resolved_pomeron} 
and using the DGLAP evolution 
equations~\cite{dglap}\@. 
The DPDFs  have been found to be dominated by gluons, which carry 
$\approx\!70$~\% of the momentum of the diffractive exchange. 

If QCD factorization is fulfilled, NLO QCD calculations based on the 
diffractive parton density functions of~\cite{h1_diff_incl_1997,h1_diff_incl_2005}, 
should be able to predict the production rates of more exclusive 
diffractive processes like dijet and open charm production in shape 
and normalization. For diffractive dijet production this has been 
tested in photoproduction ($\gamma p$) and in 
DIS~\cite{h1_diff_dijet}\@. 
In the regime of DIS the predictions of QCD have been found to be in 
good agreement with the experimental results. 

\begin{figure}[htbp]
  \begin{center}
    \includegraphics[width=0.43\textwidth]{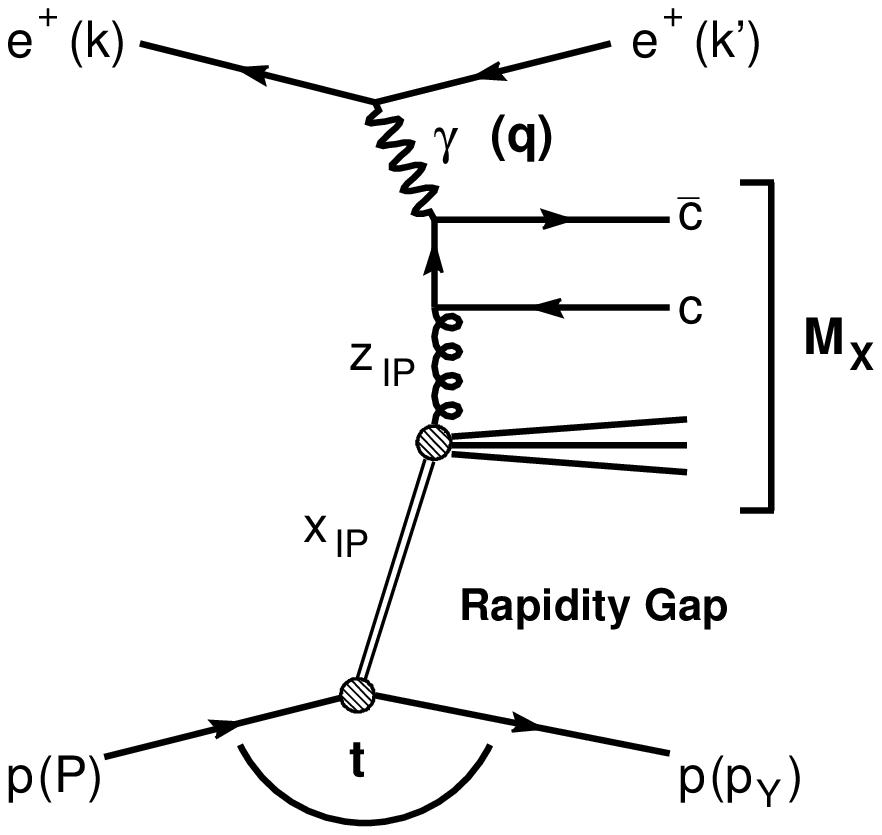}
    \includegraphics[width=0.43\textwidth]{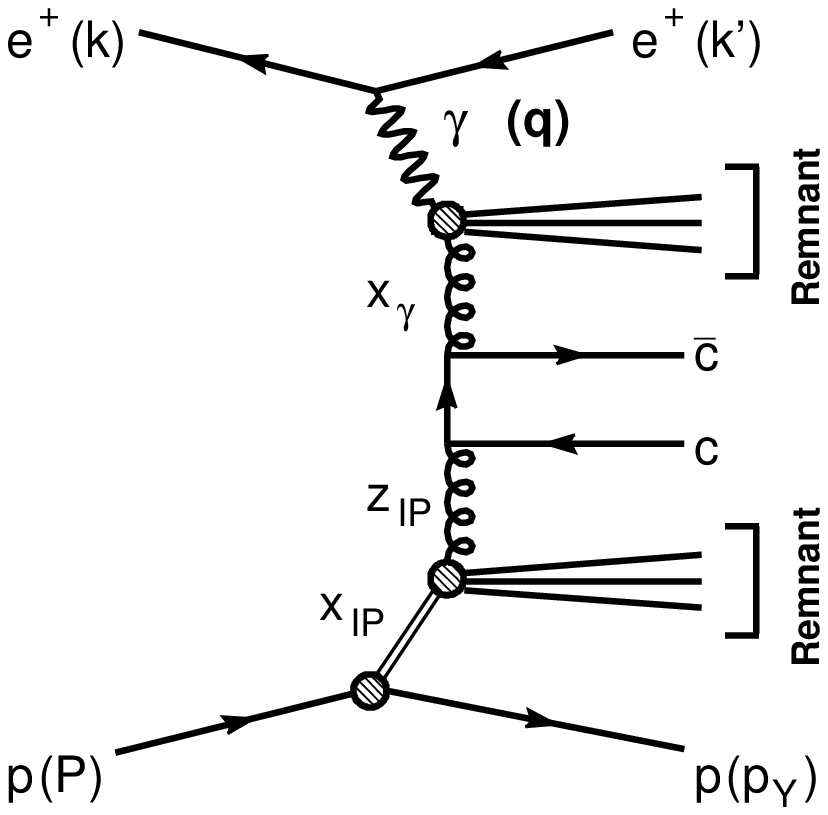}
    \setlength{\unitlength}{\textwidth}
    
    \begin{picture}(0,0)
      \begin{Large}
        \put(-0.47,0.025){\bfseries a)}
        \put( 0.04,0.025){\bfseries b)}
      \end{Large}
    \end{picture}
    \Mycaption{
      The main processes of diffractive open charm production at \hera \, 
      in the collinear factorization approach. Figure a) shows the direct 
      process where the photon enters the hard scatter itself. Figure b) 
      shows the resolved photon process where only a reduced fraction 
      $x_{\gamma}$ of the photon's momentum takes part in the hard scatter.
      }
    \label{fig:feyn}
  \end{center}
\end{figure}

In the collinear factorization approach diffractive open charm production 
at \hera \, is expected mainly to proceed via  boson gluon 
fusion (BGF) as depicted in \fig~\ref{fig:feyn}a. Thus it is directly 
sensitive to the gluon content of the diffractive exchange, which is only 
determined indirectly and for low momentum fractions $\zpom$ of the gluon 
in inclusive diffractive scattering via scaling violations~\cite{h1_diff_incl_2005}\@. 
In the BGF process a charm quark anti-quark pair (\ccb) is produced of 
which one quark couples to the photon with virtuality $Q^{2}$ and the other 
to a gluon that emerges from the diffractive exchange.

In \fig~\ref{fig:feyn}a the ``direct photon'' process is shown, where the 
photon itself enters the hard scatter, which is expected to be dominant for 
\ccb-production in DIS and photoproduction. In photoproduction, however, 
the quasi real photon may also evolve into a hadronic structure, as indicated 
in \fig~\ref{fig:feyn}b, before it enters the hard scatter. In this case only 
a fraction $x_{\gamma}<1$ of the photon's momentum takes part in the 
scattering process; the rest forming a remnant. In these ``resolved photon'' 
processes initial state interactions may take place between the photon- and 
the proton-remnant systems, destroying the rapidity gap signature and thus 
the diffractive nature of the process. A breakdown of QCD factorization has 
been observed for diffractive dijet production in $p \bar{p}$ collisions at 
the Tevatron~\cite{cdf_diff_dijets_2000}, where the prediction overestimates 
the observed rate by approximately one order of magnitude. 
Diffractive open charm production is especially suitable for testing a 
potential suppression of the direct photon component of the production mechanism 
in photoproduction.

In an alternative theoretical approach DPDFs are not introduced and diffractive scattering 
is explicitly modeled by the perturbative exchange of a colorless gluon state 
(two gluons or a gluon ladder). Formulated in the proton rest frame the 
``two-gluon'' state of the proton can couple directly to the \ccb \, pair 
($\gamma^{*}p\rightarrow c \bar{c} p$) or to a $c \bar{c} g$ color dipole 
fluctuation of the photon ($\gamma^{*}p\rightarrow c \bar{c} g p$)~\cite{bjklw}\@. 
The gluon density of the proton is usually determined from fits to the inclusive 
DIS cross section in the $k_{t}$-factorization~\cite{jung_dis03} scheme. 
A model combining the perturbative two-gluon approach with the collinear 
factorization scheme, which has also been used to fit the HERA diffractive 
DIS cross sections, is given in~\cite{mrw}\@. 

Two methods to identify charm production are presented in this paper. In the 
first method the charm quark is tagged by the reconstruction of $\dstar$ mesons. 
The measurement is performed in DIS and, due to the high selectivity of the
 trigger, extended to $\gamma p$, where it represents the first cross section
measurement of diffractive open charm production at \hera\@. 
In DIS it supercedes a 
former analysis of H1~\cite{h1_diff_dstar_2001} with increased statistics and 
with reduced systematic uncertainties. 
A similar measurement in DIS was performed by the 
ZEUS collaboration~\cite{zeus_diff_dstar_2004}. 
The results are presented in the form 
of integrated and differential $\dstar$ cross sections and in DIS are extrapolated 
into the unmeasured phase space of the $\dstar$ meson using NLO QCD calculations 
in order to determine the open charm contribution to the diffractive cross 
section. 
The second method, which was used to measure the total 
inclusive charm and beauty cross sections in 
DIS~\cite{h1_incl_charm_highq2_2004,h1_incl_charm_lowq2_2005}, 
is used here for the first time in diffractive DIS.
In this method, referred to in the following as the `displaced track analysis', 
the charm quark is identified by the reconstruction of tracks, which are displaced 
from the interaction vertex, that arise due to long lived charmed hadrons. This 
method is used in a kinematic region with high acceptance for the decay products 
of charmed hadrons within the silicon vertex detector of H1, which is used in the 
reconstruction of these tracks. With this method it is thus possible to measure 
the total open charm contribution to the diffractive cross section with small 
extrapolations from QCD calculations. 

%
%
In section~\ref{sec:kinematics} the kinematic variables used throughout the 
paper are introduced. A short discussion of the H1 detector and the event 
selection are given in sections~\ref{sec:detector} and~\ref{sec:selection}, 
followed by a description of the event simulation in section~\ref{sec:simulation} 
and the cross section determination with the two independent methods in 
section~\ref{sec:charm_selection}\@. The NLO QCD calculations and the 
comparison of the measured cross sections with the NLO QCD calculations 
are discussed in sections~\ref{sec:nlo_calculations} and~\ref{sec:results}\@.

\section{Kinematics of Diffractive \pmb{$ep$} Scattering}
\label{sec:kinematics}

%
%
Due to the diffractive nature of the process the 
photon (with four-momentum $q$) and the proton (with four-momentum $P$) 
dissociate into two distinct hadronic systems $X$ and $Y$ (with four-momenta 
$p_{{\rm X}}$ and $p_{{\rm Y}}$, respectively), which are separated by a 
large gap in rapidity between the final state hadrons.
%
%
The kinematics of the inclusive $ep$ scattering are fully determined by the 
negative squared four momentum transfer of the exchanged photon $Q^{2}$, 
the squared center of mass energy of the $ep$ scattering process $s$ and 
the inelasticity $y$\@. In addition, the following variables are defined 
to characterize the diffractive nature of the process

\begin{equation}
  \label{equ:define_diff_kine}
  \my^{2} = p_{Y}^{2} \ ;                           \qquad
  t = (P-p_Y)^2 \       ;                           \qquad     
  \xpom = \frac{q \cdot (P - p_{Y})}{q \cdot P} \ ; \qquad     
  \beta = \frac{Q^{2}}{ 2q \cdot (P - p_{Y})},
\end{equation}
where $\my$ and $t$ denote the invariant mass of the system $Y$ and the 
squared four-momentum transferred at the proton vertex, respectively. 
The variable $\xpom$ can be interpreted as the longitudinal momentum 
fraction of the diffractive exchange with respect to the proton. The 
variable $\beta$ (which is only defined in DIS) corresponds to the 
Bjorken $x$ variable from inclusive scattering taken with respect to the 
diffractive exchange. The quantities $t$ and $\my$ are constrained to be 
small by the experimental selection and are integrated over implicitly.
For the $\dstar$ analyses the observable $\zpomobs$ is introduced as

\begin{equation}
 \label{equ:define_zpom}
 \zpomobs = \frac{Q^2 + \hat{s}^{{\rm obs}}}{\xpom \cdot y \cdot s}, 
\end{equation}
where $\hat{s}^{{\rm obs}}$ is a hadron level estimate of the invariant 
mass of the \ccb \, pair emerging from the hard scattering process. It is 
reconstructed from the scattered positron and the kinematics of the 
reconstructed $\dstar$ meson including an approximate correction of the 
momentum of the $\dstar$ meson to the momentum of the charm 
quark~\cite{h1_diff_dstar_2001}\@. In direct BGF processes $\zpomobs$ is a 
direct estimator for the longitudinal momentum fraction $\zpom$ of the 
gluon that enters the scattering process with respect to the momentum of 
the diffractive exchange. In resolved processes $\zpom$ cannot be 
disentangled by the reconstruction method from the momentum fraction 
$x_{\gamma}$ that enters the hard scattering process from the photon 
side.

\section{The H1 Detector}
\label{sec:detector}

A detailed description of the H1 detector can be found in~\cite{h1_detector_1}\@. 
Only the components most relevant for this analysis are briefly discussed 
here. The coordinate system is centered at the nominal $ep$ interaction 
point with the $z$-axis pointing along the beam direction of the outgoing 
proton, also referred to as the `forward' direction in the following. 
Charged particles emerging from the interaction region are measured by the 
Central Tracking Detector (CTD), which covers a range of $-1.74 < \eta < 1.74$ 
in pseudorapidity~\footnote{The pseudorapidity $\eta$ of an object detected 
with polar angle $\theta$ is defined as $\eta = - \ln \ \tan (\theta / 2)$.}. 
The CTD comprises two large cylindrical Central Jet drift Chambers (CJCs) 
and two $z$ chambers situated concentrically around the beam-line within 
a solenoidal magnetic field of $1.15~{\rm T}$\@. It also provides triggering 
information based on track segments measured in the $r$-$\phi$ plane of the 
CJCs, and on the $z$ position of the event vertex obtained from the double 
layers of two multi-wire proportional chambers (MWPCs). 
The CTD tracks are linked to hits in the central silicon tracker (CST)~\cite{H1_cst} 
to provide precise spatial track reconstruction. The CST consists of two layers 
of double-sided silicon strip detectors surrounding the beam pipe, with a coverage 
of $-1.3 < \eta < 1.3$ in pseudorapidity for tracks passing through both layers.

The tracking detectors are surrounded by a Liquid Argon calorimeter (LAr) in 
the forward and central region ($-1.5 < \eta < 3.4$) and by a lead-scintillating 
fiber calorimeter (SpaCal) with electromagnetic and hadronic sections in the 
backward region~\cite{h1_spacal} ($-4 < \eta < -1.4$)\@. These calorimeters 
provide energy and angular reconstruction for final state particles from the 
hadronic system. DIS events are identified by the energy deposits of the 
scattered positron in the SpaCal calorimeter. Photoproduction events are 
selected with a crystal \v{C}erenkov calorimeter located close to the beam 
pipe at $z = -33.4~{\rm m}$ in the positron direction (electron tagger), which 
measures the energy deposits of positrons scattered by angles of less than 
$5~{\rm mrad}$\@. Another \v{C}erenkov calorimeter located at $z=-103~{\rm m}$ 
is used to determine the $ep$ luminosity by detecting the radiated photon 
emitted in the Bethe-Heitler process ($ep \to ep\gamma$)\@.

For the rapidity gap selection a set of detectors close to the beam pipe in 
the forward direction is used. The Forward Muon Detector (FMD) is located at 
$z=6.5~{\rm m}$ and covers a pseudorapidity range of $1.9< \eta <3.7$\@. It 
may also detect particles produced at larger $\eta$ due to secondary scattering 
within the beam pipe. A PLUG hadronic sampling calorimeter allows energy 
measurements in the range of $3.5<\eta <5.5$\@. Finally, particles in the region 
of $6.0 \ \lapprox \ \eta \ \lapprox \ 7.5$ can be detected by the Proton Remnant 
Tagger (PRT), a set of scintillation counters surrounding the beam pipe at 
$z=26~{\rm m}$\@.

\section{Event Selection}
\label{sec:selection}

The data presented in this analysis were collected over the years 1999 and 2000 
and correspond to an integrated luminosity of \mbox{$\mathcal{L} = 47.0 \, \pb^{-1}$}\@
for the $\dstar$ analyses and \mbox{$48.3 \, \pb^{-1}$}\@  for the displaced 
track analysis. At this time \hera \! was operated with positrons of energy 
$27.6~\GeV$ and protons of energy $920~\GeV$\@ so that the center of mass 
energy of the $ep$ collision is $\sqrt{s}=318~\GeV$\@.

DIS events are triggered by an electromagnetic energy cluster in the SpaCal 
calorimeter. In the $\dstar$ analyses the trigger further requires a charged 
track signal in the CTD and a reconstructed event vertex, while a looser track 
requirement of hits in the MWPCs is used in the trigger for the displaced track 
analysis. In the offline analyses the scattered positron is selected as an 
electromagnetic SpaCal cluster with energy $E'_{e}>8~\GeV$\@. 
Photoproduction events are suppressed by requiring $\sum_{i} (E_{i} - p_{z,i}) > 35~\GeV$. 
Here, $E_i$ and $p_{z,i}$ denote the energy and longitudinal momentum
components of a particle and the sum is over all final state particles
including the scattered positron and the hadronic final state
(HFS). The HFS particles are reconstructed using a combination of
tracks and calorimeter deposits in an energy flow algorithm that
avoids double counting.
The $z$ position of the interaction vertex is required to lie within $\pm 35~\cm$ 
($\pm 20~\cm$) of the center of the CTD for the $\dstar$ (displaced track) analyses, 
where the reduced range in the displaced track analysis is chosen in order to match 
the smaller acceptance of the CST.
The kinematic variables of the DIS scattering process $Q^{2}$ and $y$ are 
reconstructed using a method which uses the angle of the hadronic final 
state in addition to the energy and the polar angle of the scattered 
positron~\cite{h1_diff_incl_2005}\@. The accepted kinematic range in DIS is 
restricted to $2 < Q^{2} < 100~\GeV^{2}$ \, and $0.05 < y < 0.7$\@ for the 
$\dstar$ analysis and to $15 < Q^{2} <100~\GeV^{2}$ \, and $0.07 < y < 0.7$\@ 
for the displaced track analysis, where the reduced kinematic range 
in the displaced track analysis is chosen such that the direction of the 
quark struck by the photon mostly lies within the angular acceptance of the 
CST and that the HFS has 
a significant transverse momentum.

Photoproduction events are selected by a trigger that requires a scattered positron 
to be measured in the electron tagger, a charged track signal in the CTD and a 
reconstructed event vertex. The events have passed an additional online software 
filter that selects events with candidates for charmed hadron decays by calculating 
the invariant mass of track combinations. The inelasticity $y$ is reconstructed 
from the energy of the scattered positron and is restricted to the range 
$0.3 < y < 0.65$\@. The photon virtuality is experimentally restricted to 
$Q^{2} < 0.01~\GeV^{2}$\@.

In all analyses presented in this paper diffractive events are selected by the 
absence of hadronic activity above noise thresholds in the most forward part of 
the LAr calorimeter ($\eta > 3.2$) and in the forward detectors. This selection 
ensures that the gap between the systems $X$ and $Y$ spans more than $4$ units 
between $\eta = 3.2$  and $7.5$ in pseudorapidity. As $\my$ is not directly 
measurable by this method the data are corrected to a visible range of 
$\my < 1.6~\GeV$ \, and $|t| < 1~\GeV^{2}$, consistent with former 
measurements~\cite{h1_diff_dijet,h1_diff_dstar_2001,h1_diff_incl_2005}, with 
the help of Monte Carlo simulations. The variable $\xpom$ is calculated from

\begin{equation}
\label{equ:calc_xpom_php2}
  \xpom = \frac {Q^2+\mx^{2} }{s \cdot y} ; 
  \qquad
  \mx^{2} = \sum_{i} \left( E_{i}^{2} - p_{i,x}^{2} -p_{i,y}^{2} -p_{i,z}^{2} \right),
\end{equation}
where the sum for the calculation of $\mx$ runs over all HFS objects in the 
system $X$\@. Each of the presented analyses is restricted to $\xpom < 0.04$, 
which suppresses contributions from non-diffractive scattering and secondary 
Reggeon exchanges.  The displaced track analysis is further restricted to 
$\mx > 6 \ {\GeV}$.

\section{Event Simulation and Acceptance Correction}
\label{sec:simulation}

The data are corrected for trigger efficiencies, detector acceptances,
efficiencies, and migration effects due to the finite resolution of the H1 
detector using a Monte Carlo simulation.
All the generated events are passed through a detailed simulation of the 
detector response based on the GEANT simulation program~\cite{geant_manual} 
and reconstructed using the same reconstruction software as used for the 
data. For the event simulation residual noise contributions in the LAr 
calorimeter and the forward detectors are taken into account.

Events are generated using the \rapgap \ event generator~\cite{rapgap}, which 
simulates the process $e^+ p \rightarrow e^+ Xp$ with $\xpom < 0.15$, assuming 
proton vertex factorization. Both Pomeron and 
Reggeon sub-leading exchanges are included.
The $t$ dependence is of the form ${\rm d} \sigma / {\rm d} t \propto e^{B_{EL} t}$ 
with a slope parameter $B_{EL} = 6 \ {\GeV^{-2}}$\@. For the simulation 
of diffractive events containing charm quarks  \rapgap\ implements the BGF 
process in leading order (LO) of pQCD. For the $\dstar$ analyses LO DPDFs 
are taken from a former analysis of H1~\cite{h1_diff_incl_1997}\@. For the 
displaced track analysis the DPDFs are taken from~\cite{h1_diff_incl_2005}\@. 
To simulate higher order effects of QCD, parton showers are included in the 
calculations. Fragmentation is performed according to the Lund string 
model~\cite{frag_lund_bowler}\@. In DIS \rapgap \ is interfaced to the QED 
simulation program \heracles~\cite{heracles} to evaluate the radiative effects 
of QED. 
For diffractive photoproduction the contributing diagrams of  charm 
excitation and other resolved photon processes are included in the event 
generation, using the LO parton 
distribution functions for the resolved photon obtained 
in~\cite{res_phot_grv}\@.
In the excitation processes the charm quark is treated as a massless parton
in the resolved photon, 
whereas in all other processes the charm mass is taken into
account in the calculations. The resolved processes are found to contribute less then 
$10\%$ of the charm signal and to be mainly concentrated at large values of 
$\xpom$ and small values of $\ptds$ and $\zpomobs$\@. 

Due to the limited detector acceptance in the forward region of H1 it is not 
possible to efficiently detect a break-up of the proton into a low mass resonant 
state $Y$\@. To keep the uncertainties arising from such proton dissociation 
processes small 
the measurement is integrated over the region $\my < 1.6 \ {\GeV}$ and 
$|t| < 1 \ {\GeV^2}$. Diffractive proton dissociative events in the region 
$\my < 5~\GeV$ are simulated using \rapgap \ with a cross section dependence 
of the form $e^{B_{PD}t}$ with $B_{PD} = 1.6~\GeV^{-2}$ and an approximate 
$\my$ dependence of the form ${\rm d} \sigma / {\rm d} \my^2 \propto 1 / \my^2$~\cite{diffvm_1}\@.
The correction factor $\delta^{p {\rm dis}}$ for migrations across the 
measurement boundary is 
evaluated in the simulation for each kinematic bin. In the simulations the 
ratio of proton elastic to proton dissociative interactions is taken to be 
$1:1$, which is in accordance with the inclusive measurements 
of~\cite{h1_diff_incl_2005,h1_diff_elasp_2005}\@. The value of
$\delta^{p {\rm dis}}$ is found to be in the range $0.88-0.97$\@.

Non-diffractive events with $\my > 5~\GeV$ or $\xpom>0.15$ are simulated 
by \rapgap \ in DIS and by the event generator \pythia~\cite{pythia} in 
photoproduction. The non-diffractive background contribution in the final 
event selections is estimated to be less than $3\%$ for all data samples.

\section{Open Charm Selection}
\label{sec:charm_selection}

Charm quarks are selected by two independent methods. In the first method 
they are selected by the full reconstruction of $\dstar$ mesons. This 
provides a clear signature, which enables the tagging of charm quarks in 
DIS and photoproduction.
In the second method the more general character of the long lifetime of 
charmed hadrons is used, by reconstructing the displacement of tracks from 
the primary vertex in the CST of H1, similarly to inclusive charm production 
measurements in~\cite{h1_incl_charm_highq2_2004,h1_incl_charm_lowq2_2005}\@. 
This provides the advantage of a high acceptance for charm quarks and small 
correction factors for extrapolations to the full phase space. It is therefore 
especially suited for a measurement of the total diffractive charm cross 
section.

\subsection{Diffractive \pmb{$\dstar$} Analyses}
\label{sec:dstar_method}

\begin{figure}[htbp]
  \begin{center}
    \includegraphics[width=0.48\textwidth]{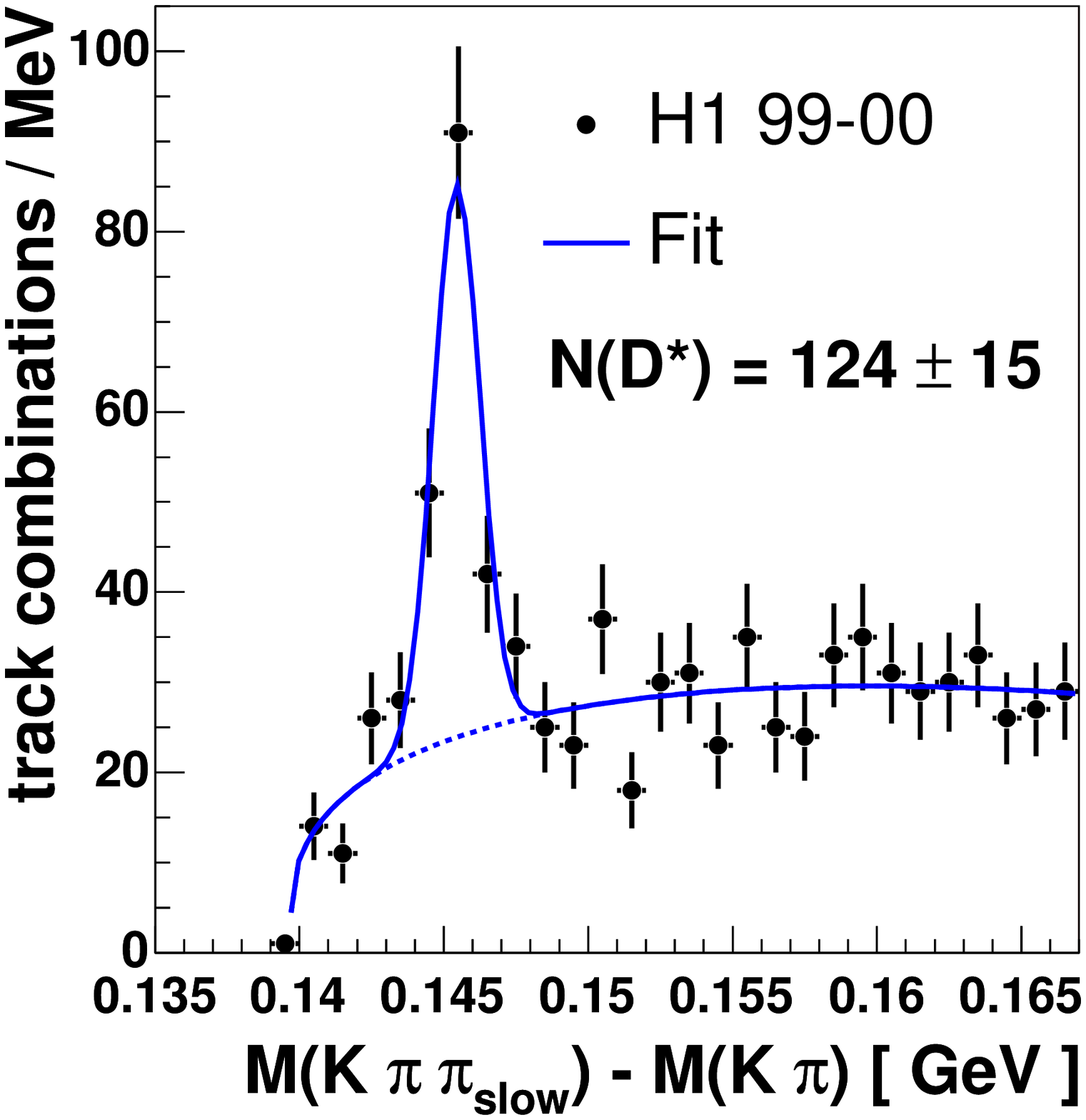}
    \includegraphics[width=0.48\textwidth]{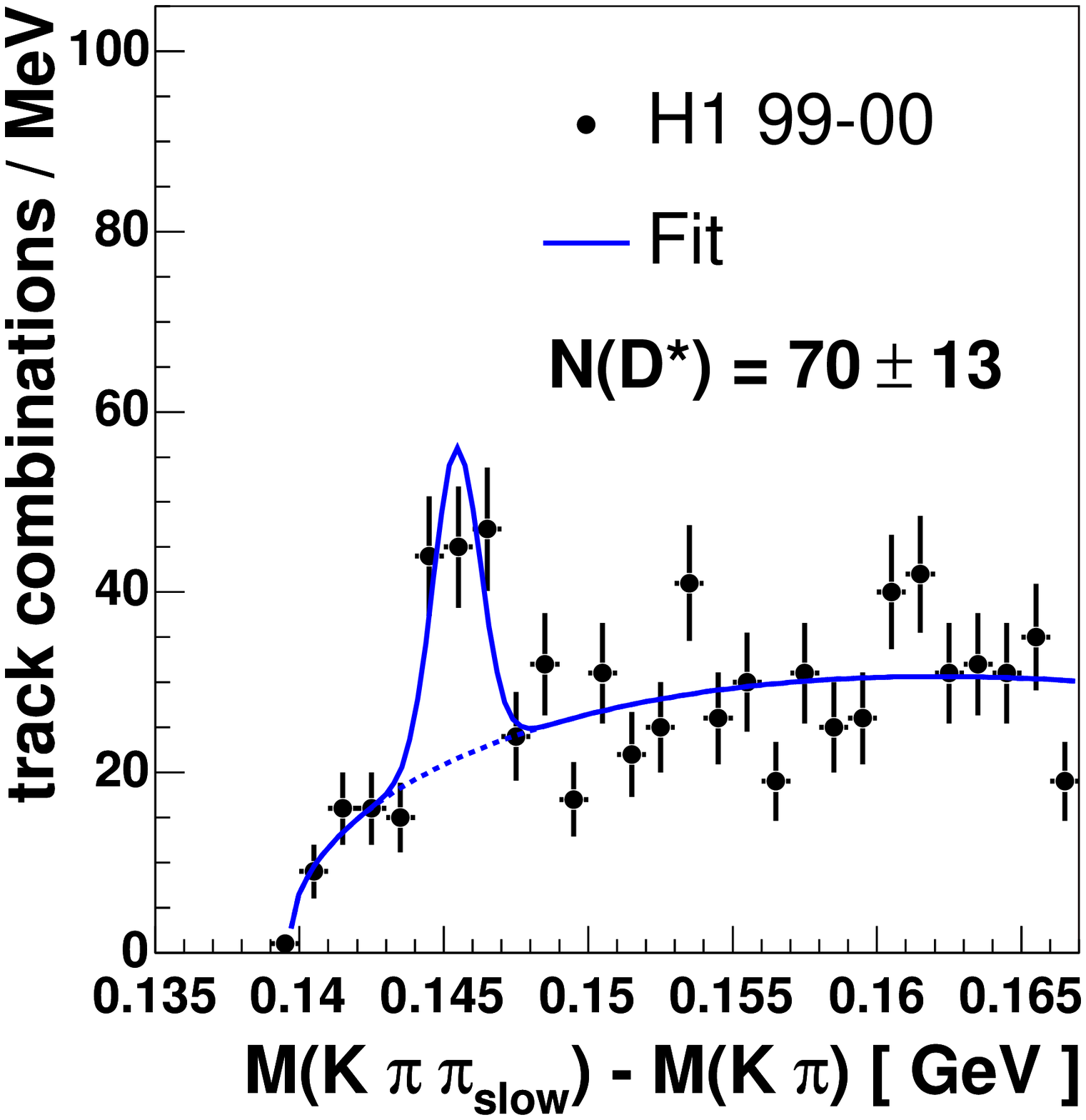}
    \setlength{\unitlength}{\textwidth}
    
    \begin{picture}(0,0)
      \begin{Large}
        \put(-0.37, 0.41){\bfseries DIS}
        \put( 0.12, 0.41){\bfseries $\pmb{\gamma p}$}

        \put(-0.47, 0.05){\bfseries a)}
        \put( 0.01, 0.05){\bfseries b)}
      \end{Large}
    \end{picture}
    
    \Mycaption{
      The $\Delta M$ distribution for each track combination that passes the selections 
      described in sections~\ref{sec:selection} and~\ref{sec:dstar_method} for (a) DIS  
      and (b) photoproduction. The parameterization used to obtain the number of 
      reconstructed $\dstar$ mesons shown in the plot is described in the text.
      }
    \label{fig:diff_yield}
  \end{center}
\end{figure}

In the $\dstar$ analyses $\dstarpm$ mesons are fully reconstructed using 
the decay channel

\begin {equation}
\label{equ:dstar_decay}
 D^{*+} \rightarrow D^0 \pi^+_{slow} \rightarrow (K^- \pi^+) \pi^+_{slow  } \
\;  \rm{(+ C.C.)},
\end {equation}
which has a branching ratio of $2.57\%$~\cite{pdg_2004}\@. The decay 
products are detected in the CTD. To ensure good detection efficiency 
and to reduce combinatorial background, the tracks are required to lie 
within an angular range of $20 \grad < \theta <160 \grad$ and to have 
a transverse momentum $\pt$ relative to the beam axis of at least 
$120~\MeV$ for the $\pi_{slow}$, $300~\MeV$ for the other pion and 
$500~\MeV$ for the $K$ candidate. The invariant mass of the $K \pi$ 
combination has to be consistent with the $\dzero$ mass within 
$\pm 80~\MeV$\@. The transverse momentum and pseudorapidity of the 
reconstructed $\dstar$ meson candidate are restricted to $\ptds > 2~\GeV$ \, 
and $|\etads| < 1.5$\@. The distributions of the mass difference 
$\Delta M = M(K^{\mp}\pi^{\pm}\pi^{\pm}_{slow}) - M(K^{\mp}\pi^{\pm})$ 
for all track combinations which fulfill the above requirements for all 
selected events in DIS and photoproduction are shown in \fig~\ref{fig:diff_yield}\@. 
The number of $D^{*\pm}$ mesons is determined by fitting these 
distributions with a Gaussian function for the signal plus a background 
parameterization given by $N(\Delta M - m_{\pi})^{u_{e}} (1- u_{s} \, (\Delta M)^2)$, 
where $m_{\pi}$ denotes the mass of the charged pion and $N$, $u_{e}$ 
and $u_{s}$ are free parameters. The position and the width of the 
Gaussian function are fixed to values taken from higher statistics 
samples where no diffractive cuts were applied. The resulting numbers 
of identified $\dstarpm$ mesons in DIS and photoproduction are 
summarized in table~\ref{tab:int_cross_section}\@.

Differential cross sections are obtained from the fitted number of 
$\dstar$ mesons in each measurement bin. A correction is applied for 
mass reflections originating from decays of the $\dzero$ meson other 
than that given in equation~\ref{equ:dstar_decay}, which has been estimated 
to be $3.5\%$ of the $\dstar$ signal~\cite{h1_mass_reflections_2000}\@. 
A correction factor of $\simeq 0.95$ for the effects of initial and final 
state QED radiation is applied to the DIS cross sections. The cross sections
are bin center corrected using \rapgap \ to determine the point in the 
bin at which the bin-averaged cross section equals the differential 
cross section.

\subsection{Displaced Track Analysis}
\label{sec:lifetime_method}

The production of open charm in diffraction is also investigated using 
a largely independent method, which has been used in~\cite{h1_incl_charm_highq2_2004} 
and~\cite{h1_incl_charm_lowq2_2005} to measure the total inclusive charm 
and beauty cross sections in DIS. This method distinguishes events 
containing heavy quarks from those containing only light quarks by 
reconstructing the displacement of tracks from the primary vertex in the 
transverse plane (impact parameter), caused by the long lifetimes of the 
charm and beauty flavored hadrons, using the precise spatial information 
from the CST of H1. Due to the low beauty fraction in the diffractive data 
sample, it is not possible to make a measurement of the beauty cross 
section and only a measurement of the charm cross section is presented 
in this paper.

As in~\cite{h1_incl_charm_highq2_2004,h1_incl_charm_lowq2_2005} the primary 
event vertex in the $r$-$\phi$ plane is reconstructed from all tracks 
(with or without hits in the CST) using the information on the position 
and transverse extent of the beam interaction region. For the analysis,
tracks are selected if they have a transverse momentum of more than 
$0.5~\GeV$ and at least two associated hits in the CST. The impact parameter 
of a track is defined as the distance of closest approach (DCA) of the track 
to the primary vertex point in the transverse plane. 

In order to determine a signed impact parameter ($\delta$) for a
track, the azimuthal angle of the struck quark $\phi_{\rm quark}$ must
be determined for each event. To do this, jets with a minimum $\pt$ of
$2.5 \ {\GeV}$, in the angular range $15^\circ < \theta < 155^{\rm
o}$, are reconstructed using the invariant $\kt$ 
algorithm~\cite{kt_algorithm}
in the laboratory frame using all reconstructed HFS particles. The
angle $\phi_{\rm quark}$ is defined as the $\phi$ of the jet with the
highest transverse momentum or, if there is no jet reconstructed in
the event, as $180^\circ-\phi_{\rm elec}$, where $\phi_{\rm elec}$ is
the azimuthal angle of the scattered positron in degrees. Monte Carlo 
simulations indicate that $\approx 78\%$ of all charm events have at 
least one reconstructed jet in the kinematic region described above. 
The direction defined by the primary vertex and $\phi_{\rm quark}$ in 
the transverse plane is called the 'quark axis'. 
If the angle between the quark axis and
the line joining the primary vertex to the point of DCA 
of the track is less than
$90^\circ$, $\delta$ is defined as positive, and is defined as
negative otherwise. Tracks with azimuthal angle outside $\pm 90^\circ$ 
of $\phi_{\rm quark}$ are rejected.
The estimated error on $\delta$ is denoted as $\sigma(\delta)$\@. 

\begin{figure}[htbp]
  \begin{center}
    \includegraphics[width=.48\textwidth]{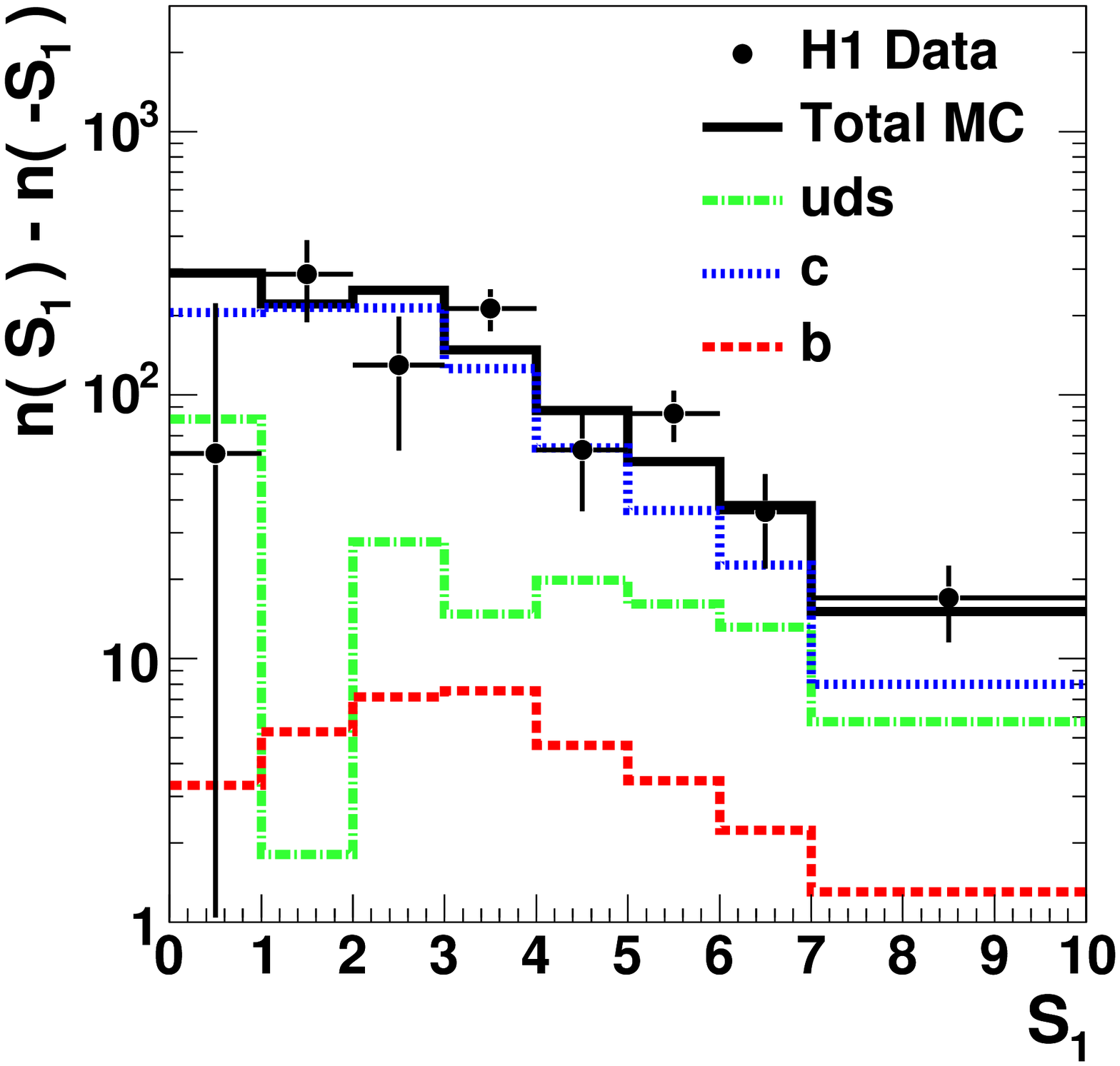}
    \includegraphics[width=.48\textwidth]{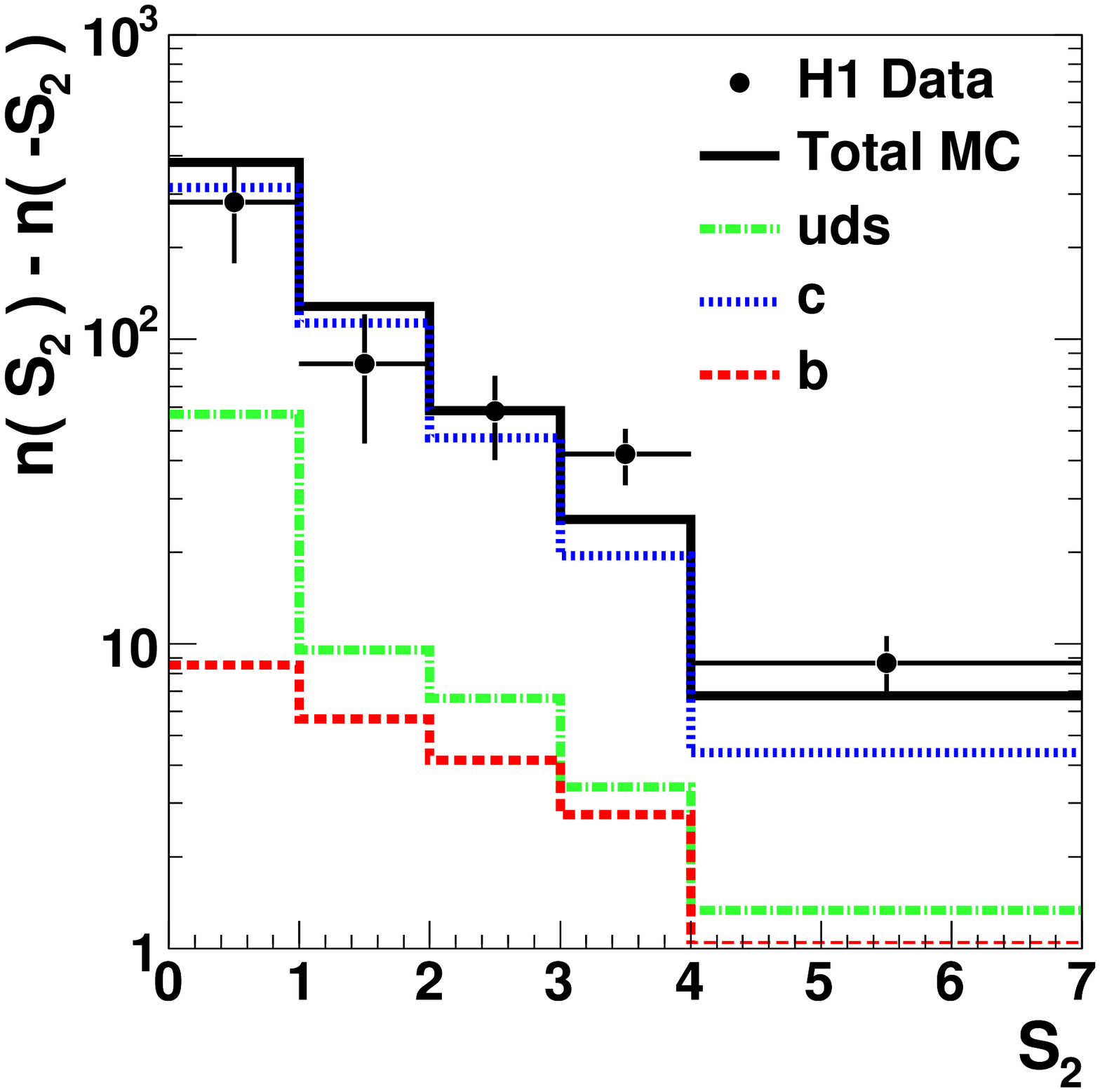}
    \setlength{\unitlength}{\textwidth}
    
    \begin{picture}(0,0)
      \begin{Large}
        \put(-0.49,0.11){\bfseries a)}
        \put( 0.00,0.11){\bfseries b)}
      \end{Large}
    \end{picture}

    \Mycaption{
      The subtracted significance distributions of (a) $S_1$ and (b) $S_2$\@. Included in the figure 
      is the result from the fit of the Monte Carlo distributions of the various quark flavors to the 
      data.
      }    
    \label{fig:negsub}
  \end{center}
\end{figure}

To distinguish between the charm and light quark flavors a similar 
method to that in~\cite{h1_incl_charm_lowq2_2005} is used. The quantity 
$S_1$ ($S_2$) is defined as the significance ($\delta/\sigma(\delta)$) 
of the track with the highest (second highest) absolute significance 
that is associated to the quark axis. In the present analysis $S_3$, 
which is the significance of the track with the third highest absolute 
significance, is not used due to lower statistics than 
in~\cite{h1_incl_charm_lowq2_2005}\@. Events where $S_1$ and $S_2$ have 
opposite signs are excluded from the $S_2$ distribution, but contribute 
to the $S_1$ distribution. The distributions of $S_1$ and $S_2$ are 
shown in~\fig~\ref{fig:significance} for the kinematic region given in 
section~\ref{sec:selection}\@. A reasonable description of the data by 
the simulation is observed. 
The light quark significance distributions are approximately symmetric 
around zero, whereas the charm distributions have an excess in the 
positive bins compared with the negative. It is thus possible to 
substantially reduce the uncertainty due to the resolution of $\delta$ 
and the light quark normalization, by subtracting the contents of the 
negative bins in the significance distributions from the contents of the 
corresponding positive bins. The subtracted distributions are shown 
in~\fig~\ref{fig:negsub}\@. 

The fractions of charm and light quark 
flavors in the data are extracted in three $\mx^{2}$ intervals using 
a least squares simultaneous fit to the subtracted $S_1$ and $S_2$ 
distributions (as shown in~\fig~\ref{fig:negsub}) and the total number 
of reconstructed diffractive events before any track selection. The 
significance distributions of the charm, beauty and light flavors, as 
predicted by the Monte Carlo simulation for the luminosity of the data, 
are used as templates. In each interval the charm and light quark 
flavor contributions from the Monte Carlo simulation are scaled by 
factors $P_c$ and $P_l$, respectively, to give the best fit to the 
observed subtracted $S_1$, $S_2$ distributions and the total number 
of events. Since the same event may enter the $S_1$ and the $S_2$ 
distributions, it was checked using a high statistics Monte Carlo 
simulation that this has a negligible effect on the results of the 
fits with the statistics of the present data. Only the statistical 
errors of the data and the Monte Carlo simulation are taken into 
account in the fit. The beauty scale factor is fixed to $P_b=1$,
and varied in the evaluation of the systematic uncertainties (see 
section~\ref{sub:systematics}). The results of the fit to the 
complete data sample are shown in~\fig~\ref{fig:negsub}\@. The fit 
gives a good description of all significance distributions, with a 
$\chi^2/ n.d.f$ of $18.1/12$\@. 
Values of $P_c=0.77 \pm 0.09$ and $P_l=0.97 \pm 0.03$ are obtained. 
It can be seen that the resulting distributions are dominated by 
charm quark events, the light quarks contributing only a small 
fraction, mainly due to strange hadrons, for all values of the 
significance.  The beauty contribution forms a small fraction overall, 
but increases with increasing $\mx^{2}$\@. Acceptable $\chi^2$ values 
are also found for the fits to the samples in the separate $\mx^{2}$ 
intervals.  

The results of the fit in each $\mx^{2}$ interval are converted to a 
measurement of the diffractive differential cross section using:

\begin{equation}
\frac{{\rm d}^{3}\sigma^{c\bar{c}}_{D}}{{\rm d} 
Q^{2}\, {\rm d} y\,{\rm d} M_{X}^2}  = \frac{P_{c} N^{\rm MC gen}_{c}    
\delta^{\rm rad}_c \delta^{p {\rm dis}}\delta^{\rm BCC}_c} {\mathcal{L} \cdot {\rm BV}},
\end{equation}
where $N^{\rm MC gen}_{c}$ is the number of generated charm events 
expected from the Monte Carlo simulation in each bin with volume 
{\rm BV} corresponding to the luminosity of the data $\mathcal{L}$\@. 
A bin center correction $\delta^{\rm BCC}_c$ in the range $0.89-1.21$ 
is calculated using the NLO QCD expectation to correct the bin averaged 
cross section to the cross section at a specified point in $Q^{2}$, $y$ 
and $\mx^{2}$\@. A correction factor of $\delta^{\rm rad}_c \simeq 0.93$ 
for initial and final state QED radiation is applied.
The correction factor for proton dissociation 
$\delta^{p {\rm dis}}$ is described in section~\ref{sec:simulation}.

Measurements of the ratio of the diffractive charm cross section to 
the total diffractive cross section are made where the total diffractive 
cross section is determined using

\begin{equation}
\frac{{\rm d}^{3}\sigma_{D}}{{\rm d} 
Q^{2}\, {\rm d} y\,{\rm d} M_{X}^2}  = \frac{N^{\rm rec} N^{\rm MC gen}    
\delta^{\rm rad} \delta^{p {\rm dis}} \delta^{\rm BCC}} {N^{\rm MC rec} \mathcal{L} \cdot {\rm BV}},
\end{equation}
where $N^{\rm rec}$ is the number of reconstructed data events in the 
bin after the event selection described in section~\ref{sec:selection}, 
$N^{\rm MC rec}$ ($N^{\rm MC gen}$) is the number of reconstructed 
(generated) Monte Carlo events in the bin; $\delta^{\rm rad}$ and 
$\delta^{\rm BCC}$ are the radiative correction and bin center correction 
for inclusive diffraction, respectively. The ratio is then given by:

\begin{equation}
f^{c\bar{c}}_D = \frac{{\rm d}^{3}\sigma^{c\bar{c}}_{D}}{{\rm d} 
Q^{2}\, {\rm d} y\,{\rm d} M_{X}^2}    / \frac{{\rm d}^{3}\sigma_{D}}{{\rm d} 
Q^{2}\, {\rm d} y\,{\rm d} M_{X}^2}.
\end{equation}

\subsection{Systematic Uncertainties}
\label{sub:systematics}

\begin{figure}[htbp]
  \begin{center}
    \includegraphics[width=.48\textwidth]{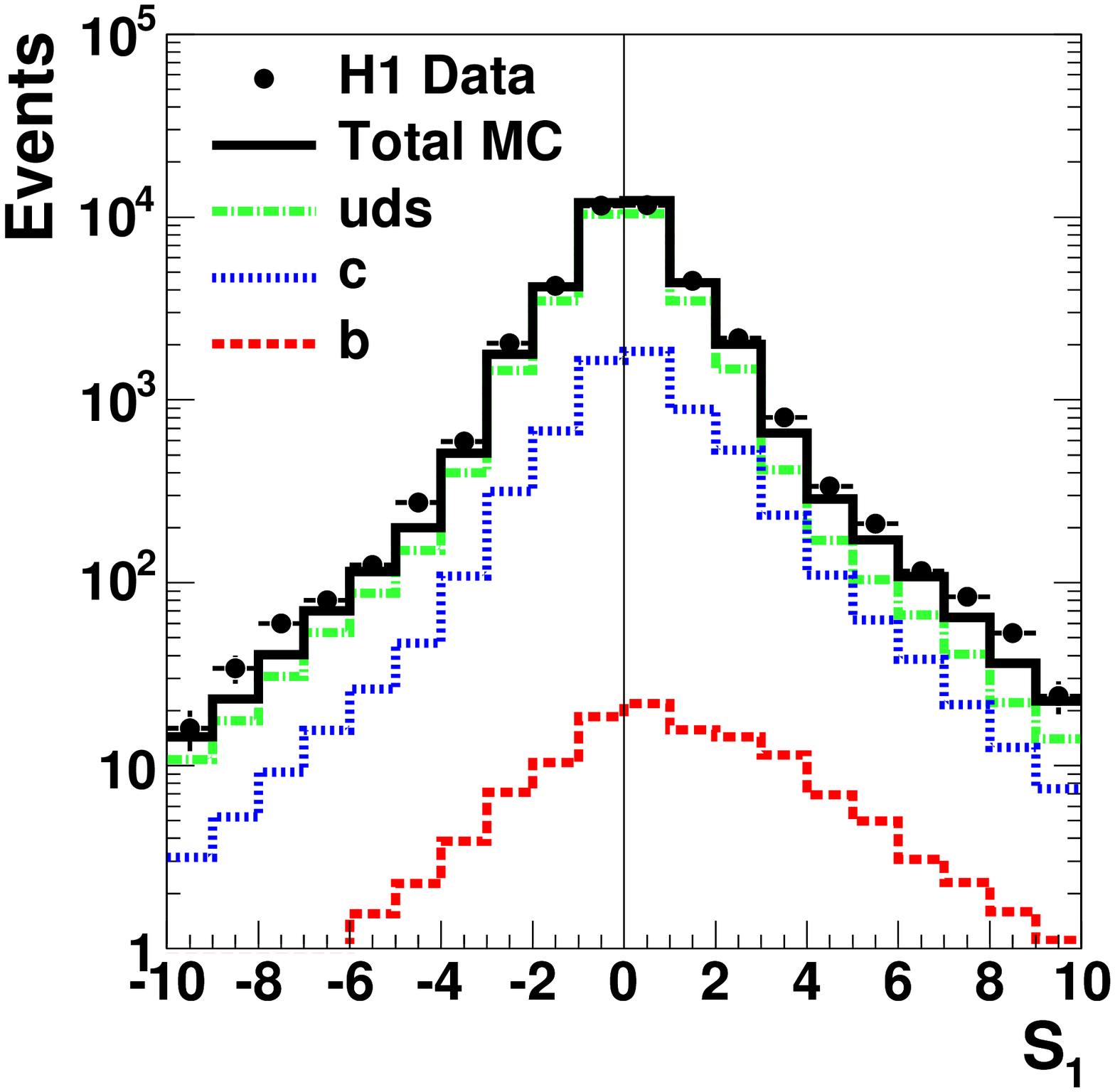}
    \includegraphics[width=.48\textwidth]{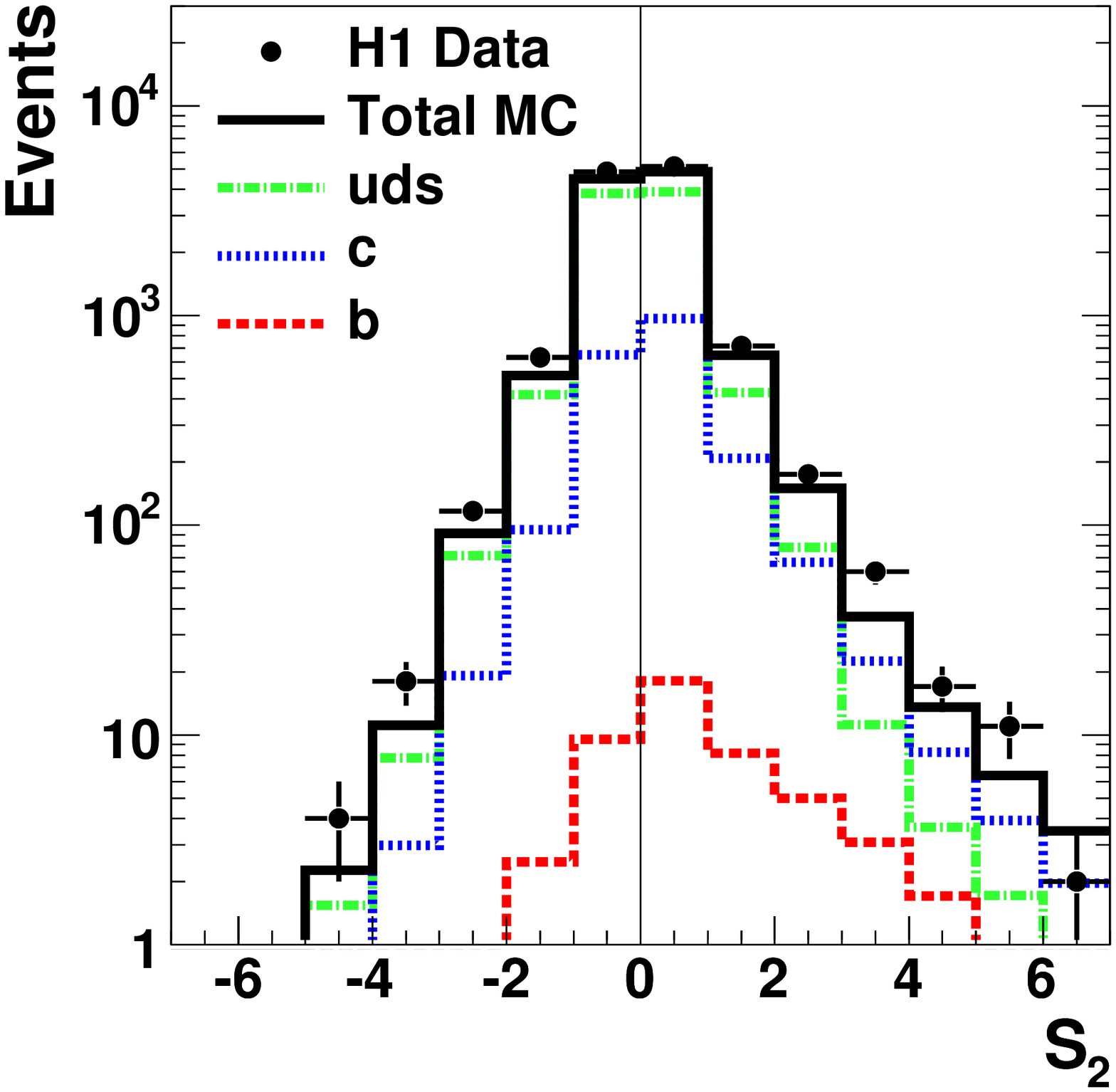}
    \setlength{\unitlength}{\textwidth}
    
    \begin{picture}(0,0)
      \begin{Large}
        \put(-0.49,0.11){\bfseries a)}
        \put( 0.00,0.11){\bfseries b)}
      \end{Large}
    \end{picture}

    \Mycaption{
      The significance $\delta /\sigma(\delta)$ distribution (a) of the highest absolute significance 
      track ($S_1$) and (b) of the track with the second highest absolute significance ($S_2$)\@.
      Included in the figure is the expectation from the Monte Carlo simulation program \rapgap\@ for 
      light, charm  and beauty quarks. The contributions from the various quark flavors are shown after 
      applying the scale factors obtained from the fit to the subtracted significance distributions 
      of the data shown in \fig~\ref{fig:negsub}\@.
      }    
    \label{fig:significance}
  \end{center}
\end{figure}

The following sources of systematic uncertainty for the two different 
analysis methods are taken into account; the estimated values are given 
in table~\ref{tab:sys_uncertainties}:

\begin {itemize}
\item The simulated trigger efficiencies for the $\dstar$ analyses are 
compared with the efficiencies determined from data using monitor trigger 
samples. Within the statistics of these data samples the simulated trigger 
efficiencies are found to agree with the data, with a remaining uncertainty 
in the range $\pm (3 - 5)\%$ depending on the analysis. 
For the displaced track analysis an uncertainty of $1\%$ is assigned 
as determined from the data.

\item For the DIS measurements the reconstructed polar angle and the energy 
of the scattered positron are varied within the estimated uncertainties 
of $\pm 1~{\rm mrad}$ for the angular measurement and $\pm 1\%$ for the 
energy scale of the SpaCal, leading to an uncertainty of $\pm 2 \%$ on 
the cross section measurements. In photoproduction a variation within the 
estimated uncertainty of $\pm 1.5\%$ on the energy scale of the crystal 
\v{C}erenkov calorimeter of the electron tagger results in an uncertainty 
of $\pm 2 \%$\@.

\item The uncertainty of the track reconstruction efficiency and uncertainties 
related to the signal extraction for the $\dstar$ analyses have been determined 
by analyzing inclusive $\dstar$ samples as in~\cite{h1_incl_dstar_php_2005} 
and are estimated to be $\pm 6\%$ for the reconstruction efficiency of the three 
daughter tracks of the $\dstar$ meson in the CTD and $\pm 6\%$ for the signal 
extraction. The uncertainty on the correction for mass reflections is estimated 
to be $\pm 1.5\%$~\cite{h1_mass_reflections_2000}\@.

For the displaced track analysis a track efficiency uncertainty of $\pm 2\%$ 
due to the CTD and of $\pm 1\%$ due to the CST is estimated, resulting in an 
uncertainty of $\pm 2\%$ on the cross sections. An uncertainty in the resolution 
of $\delta$ of the tracks is estimated by varying the resolution by an amount 
that encompasses the differences between the data and the simulation (see 
\fig~\ref{fig:significance}). This is achieved by applying an additional 
Gaussian smearing in the Monte Carlo simulation of $\pm 200~\mu{\rm m}$ to 
$5\%$ of randomly selected tracks and $\pm 25\mu{\rm m}$ to the rest, resulting 
in an error of $2\%$ on the cross sections.

\item The effect of a $\pm 4\%$ uncertainty in the energy scale of the hadronic 
final state leads to a change of the cross section in the range $\pm (1 - 3)\%$ 
depending on the analysis.

\item The uncertainty in the acceptance and migration corrections due to 
uncertainties in the physics models for diffractive charm production is 
estimated by varying the shape of various kinematic distributions 
in the Monte Carlo simulation within 
limits set by the present measurements. Reweighting the shapes of the $\xpom$, 
$\beta$ and $Q^{2}$ distributions by $(\frac{1}{\xpom})^{\pm 0.25}$, $\beta^{\pm 0.3}$ 
and $(1+\log_{10}(Q^{2}/{\GeV^2}))^{\pm1}$ in DIS results in an uncertainty 
of $\pm 5\%$ on the total cross section for the $\dstar$ analysis and 
$\pm (12 - 18)\%$ for the displaced track analysis. A variation of the $\xpom$ 
and $y$ distributions by $(\frac{1}{\xpom})^{\pm 0.25}$ and $(\frac{1}{y})^{\pm 0.2}$ 
in photoproduction results in an uncertainty of $\pm 1\%$ on the total 
cross section. The uncertainty on the fraction of the Reggeon contribution 
is estimated by varying its normalization in the simulation by $\pm 100\%$, 
which leads to an uncertainty of $\pm 1\%$ ($\pm 4\%$) for the $\dstar$ analyses 
in DIS (photoproduction) and  $\pm (1 - 9)\%$ for the 
measurement bins of the displaced track analysis. A variation of the $t$ 
distribution by $e^{\pm 2 \, t}$ for proton elastic scattering and the 
$M_Y$ and the $t$ distribution by $(\frac{1}{\my^{2}})^{\pm 0.3}$ and 
$e^{\pm 1 \, t}$ for proton dissociative scattering as well as the ratio 
of proton elastic to proton dissociative scattering between $1:2$ and $2:1$ 
results in a systematic uncertainty on the cross sections 
in the range $\pm (4 - 5)\%$\@.

\item The uncertainties for residual noise in the FMD and the PLUG calorimeter 
in the simulation are estimated using a set of randomly triggered events during 
the data taking period and result in a combined uncertainty of $\pm 1.5\%$\@. 
The tagging efficiency of the PRT for proton dissociative systems with 
$\my > 1.6~\GeV$ or $|t|>1~\GeV^{2}$ in the simulation is adjusted with the 
help of an independent non-diffractive data sample with activity in the forward 
part of the LAr calorimeter and the FMD, where such events are enriched. The 
effect of the remaining uncertainty on this efficiency on the cross section 
measurements is estimated by varying the simulation within the statistical 
accuracy of the measured efficiency and is estimated to lie between $\pm (7 - 9)\%$\@. 
The uncertainty on the tagging efficiency of the FMD is estimated to be 
$\pm 10\%$~\cite{h1_diff_incl_2005}\@. The effect on the cross sections is 
$\pm 1\%$\@. The residual influence of non-diffractive background from events 
without a rapidity gap is estimated by assigning a $\pm 100\%$ uncertainty 
to the corresponding event samples in the \rapgap \ simulation. This leads 
to an uncertainty on the cross sections in the range $\pm (1 - 3)\%$\@.

\item The uncertainty of the charm fragmentation scheme is estimated by 
changing the pa\-ra\-me\-tri\-za\-tion of the longitudinal fragmentation 
function from the Lund-Bowler model~\cite{frag_lund_bowler} to Peterson 
functions with $\epsilon=0.078$ ($\epsilon=0.058$)~\cite{frag_peterson} 
in the simulation of the events for the  $\dstar$ (displaced 
track) analyses, which results in an uncertainty on the cross section of 
$\pm 1\%$ ($\pm 4 \%$) for the $\dstar$ analyses in DIS (photoproduction) 
and of $\pm 7\%$ in the displaced track analysis.

\item For the displaced track analysis the uncertainties on the lifetimes 
of the various $D$ mesons, decay branching fractions and mean charge 
multiplicities are estimated by varying the input values of the Monte Carlo 
simulation by the errors on the world average measurements. The values and 
the uncertainties for the lifetimes of the $D$ mesons are taken 
from~\cite{pdg_2004} and those from the branching fractions of charm quarks 
to hadrons from~\cite{charm_fractions}\@. They are consistent with measurements 
in DIS at HERA~\cite{h1_incl_dmesons_2005}\@. The values and the uncertainties 
for the mean charged track multiplicities for charm quarks are taken 
from~\cite{track_multiplicities}\@. A combination of all these uncertainties 
results in an error of $3\%$ on the cross sections. For the $\dstar$ analyses 
the uncertainty of $\pm 2.5\%$ on the branching fraction for the decay channel
in equation~\ref{equ:dstar_decay} is taken from~\cite{pdg_2004}\@.

\item The uncertainty on the asymmetry of the $\delta$ distribution for the 
light quarks in the displaced track analysis is estimated by repeating the 
fits with the subtracted light quark significance distributions (shown in 
\fig~\ref{fig:negsub}) changed by $\pm 50\%$\@. The light quark asymmetry is 
checked to be within this uncertainty by comparing the asymmetry of the Monte 
Carlo events to that of the data, in the region of $0.1<|\delta|<0.5~{\rm cm}$, 
where the light quark asymmetry is enhanced. This results in an uncertainty 
on the cross section of $\pm 4\%$ at high $\mx$  and of $\pm 16\%$ at low 
$\mx$\@.

The uncertainty on the beauty contribution for the displaced track analysis 
is estimated by repeating the fits with the subtracted beauty quark significance 
distributions (shown in \fig~\ref{fig:negsub}) changed by $^{+400}_{-100}\%$, 
which results in an negligible error on the cross section at low $\mx$ 
increasing to $^{-14}_{+ \ 5}\%$ and $^{-40}_{+13}\%$ in the middle and high $\mx$ 
bins, respectively.

\item An uncertainty on the quark axis in the displaced track analysis is 
estimated by shifting it by $\pm 2\grad$ ($\pm 5\grad$) for events with 
(without) a reconstructed jet. These shifts have been estimated 
in~\cite{h1_incl_charm_lowq2_2005} by comparing 
the difference between $\phi_{\rm quark}$ and the track azimuthal angle in 
data and Monte Carlo simulation. The resulting error on the cross sections is 
$\pm 3\%$. 

\item The uncertainty in the calculation of QED radiative effects is found
to be $\pm 2\%$ in DIS.

\item The uncertainty in the bin center correction for the displaced track 
analysis is estimated by varying the shape 
of the $Q^2$, $\beta$ and $\xpom$ distributions
of the NLO QCD expectation. 
This leads to a $\pm (8 - 10)\%$ uncertainty 
on the cross sections.

\item The uncertainty of the luminosity determination is estimated to be 
$\pm 1.5\%$\@.
\end {itemize}

The total systematic uncertainty for each data point has been obtained by 
adding all individual contributions in quadrature. For the $\dstar$ analyses 
it ranges between $15\%$ and $30\%$ for the differential cross sections and 
amounts to $\pm 15\%$ for the integrated cross section in both kinematic 
regimes. For the displaced track analysis it ranges between $26 \%$ and 
$47 \%$ for the three points of the inclusive charm cross section measurement.

\section{QCD Calculations}
\label{sec:nlo_calculations}

%
%
\subsection{NLO Calculations in Collinear Factorization}
\label{sec:nlo_collinear}

The measured charm cross sections are compared with NLO QCD calculations 
based on two alternative sets of diffractive parton density functions 
from H1~\cite{h1_diff_incl_2005}\@ which both provide a good description 
of the inclusive diffractive DIS data. As default the standard 
parameterization H1 2006 DPDF Fit A is chosen. The alternative set 
of DPDFs (H1 2006 DPDF Fit B) is obtained from a slightly different 
parameterization of the gluon density at the starting scale of the fit 
procedure. It leads to a steeper fall-off of the gluon density at higher 
values of $\zpom$\@.
In the fit to the inclusive diffractive DIS data~\cite{h1_diff_incl_2005}\@ 
charm quarks are treated as massive, appearing via BGF-type
processes up to order 
$\alpha_s^2$~\cite{h1_nlo_fit}\@. The quark mass is 
set to $m_{c} = 1.4~\GeV$ and the scale for heavy flavor production to 
$\mu_{r} = \mu_{f} = 2m_{c}$\@.

In order to be able to compare the measured $\dstar$ cross section to 
the results based on the NLO QCD fits diffractive versions of the programs 
\hvqdis~\cite{hvqdis_incl,hvqdis_diff} in DIS and \fmnr~\cite{fmnr_incl_1,fmnr_incl_2} 
in photoproduction are used. The renormalization and the factorization 
scales are set to $\mu_{r} = \mu_{f} = \sqrt{Q^{2} + 4m_{c}^{2}}$ in DIS 
and to $\mu_{r} = \mu_{f} = \sqrt{p_{t}^{2} + 4m_{c}^{2}}$ in photoproduction, 
respectively. For both calculations the charm mass is chosen to be 
$m_{c} = 1.5~\GeV$\@. 
The calculations result in predictions for the production of charm quarks. 
To obtain predictions for a measurement of $\dstar$ meson production 
hadronization corrections evaluated using the LUND hadronization model as 
implemented in \rapgap \, are applied. For the longitudinal fragmentation 
Peterson functions are used with $\epsilon = 0.035$ as suggested for NLO 
predictions by~\cite{frag_nason_oleari}\@. For the calculation of these 
corrections parton showers are included to simulate the higher order 
effects of QCD in the event generation of the LO Monte Carlo program. 
To estimate the uncertainty of the NLO calculations the renormalization 
and the factorization scales are simultaneously varied by factors of $1/2$ 
and $2$, the charm mass is varied by $\pm 0.2~\GeV$ and the Peterson 
fragmentation parameter $\epsilon$ is varied by $\pm 0.025$\@. The 
uncertainties originating from all these variations are added in quadrature. 
They result in a combined uncertainty on the theoretical integrated 
$\dstar$ meson cross 
section of $\approx 25\%$ in DIS and $\approx 22\%$ in photoproduction.

\subsection{Two Gluon Exchange Models}
\label{sec:nlo_2gluon}

The measured $\dstar$ cross sections are compared with QCD calculations 
based on the perturbative two gluon approach of `BJKLW'~\cite{bjklw} 
using the $\kt$ unintegrated gluon density J2003 set2 evolved by the 
CCFM~\cite{ccfm} evolution 
equations obtained from fits~\cite{jung_dis03}\@ to the inclusive DIS 
cross section. These calculations are applicable only in the region 
of small $\xpom$ ($\xpom<0.01$), where contributions from secondary 
Reggeon exchanges can be neglected. To ensure that the perturbative 
calculations are applicable a cut on the transverse momentum of the 
gluon of $\pt^{g}>2.0\, {\GeV}$ for the process $\gamma^{*} p \to 
c \bar{c} g p$ is applied.

\newpage

\subsection{The MRW Model}
\label{sec:nlo_mrw}

The measurements of the diffractive charm cross section in DIS are also compared with 
the approach of `MRW'~\cite{mrw} which can be considered to be a hybrid
of the two approaches described in sections~\ref{sec:nlo_collinear}
and~\ref{sec:nlo_2gluon}. 
The parameters of the input DPDFs were determined from a fit to the H1 
inclusive diffractive data~\cite{mrw}.  
At low $\beta$, charm is produced via a `resolved Pomeron' mechanism by
BGF-type processes calculated up to order $\alpha_s^2$, as in the approach of 
section~\ref{sec:nlo_collinear}  At high $\beta$, the perturbative 
two-gluon state participates directly in the hard interaction
via `photon--Pomeron' fusion.  This `direct Pomeron' contribution is similar to the
$\gamma^*p\to c\bar{c}p$ contribution of the BJKLW model and depends on the 
square of the gluon distribution of the proton.

\section{Results}
\label{sec:results}

In DIS the integrated cross section of diffractive $\dstarpm$ production 
in the kinematic range of $2 < Q^{2} < 100~\GeV^{2}$, $0.05 < y < 0.7$, 
$\xpom < 0.04$, $\my < 1.6~\GeV$, $|t| < 1~\GeV^{2}$, $\ptds > 2~\GeV$ \, 
and $|\etads| < 1.5$ is measured to be

\begin {equation}
 \sigma (ep \rightarrow eD^{*\pm}X'Y)_{\rm{DIS}} = 234 \pm 29 ({\rm stat.}) \pm 34 ({\rm syst.})\ \pb,
\end {equation}
which is in good agreement with the measurement in the same kinematic 
range in the previous analysis by H1~\cite{h1_diff_dstar_2001}.

In photoproduction the integrated $ep$ cross section of diffractive 
$\dstarpm$ production in the kinematic range of $Q^{2} < 0.01~\GeV^{2}$, 
$0.3 < y < 0.65$, $\xpom < 0.04$, $\my < 1.6~\GeV$, $|t| < 1~\GeV^{2}$, 
$\ptds > 2~\GeV$ \, and $|\etads| < 1.5$ is measured to be

\begin{equation}
 \sigma (ep \rightarrow eD^{*\pm}X'Y)_{{\gamma p}} = 265 \pm 50 ({\rm stat.}) \pm 41 ({\rm syst.})\ \pb.
\end{equation}
A comparison of the measured integrated cross sections 
in DIS and photoproduction with the predictions 
of the NLO calculations for the two sets of H1 2006 DPDFs (Fit A and Fit B)~\cite{h1_diff_incl_2005}\@ 
is given in table~\ref{tab:int_cross_section}\@. A 
good agreement between the data cross sections and the NLO QCD calculations
is observed.

\begin{figure}[htbp]
  \begin{center}
    \includegraphics[width=.48\textwidth]{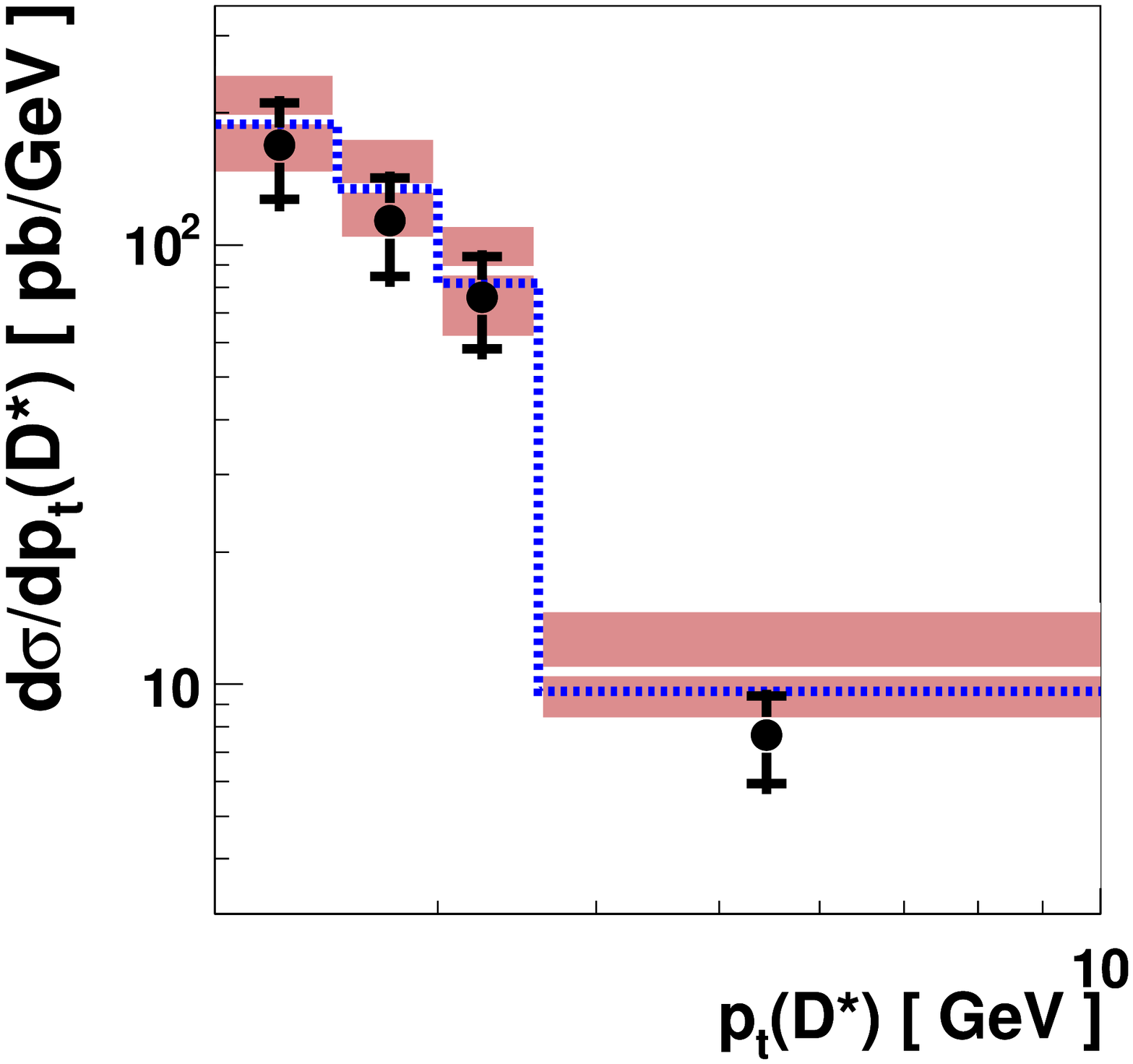}
    \includegraphics[width=.48\textwidth]{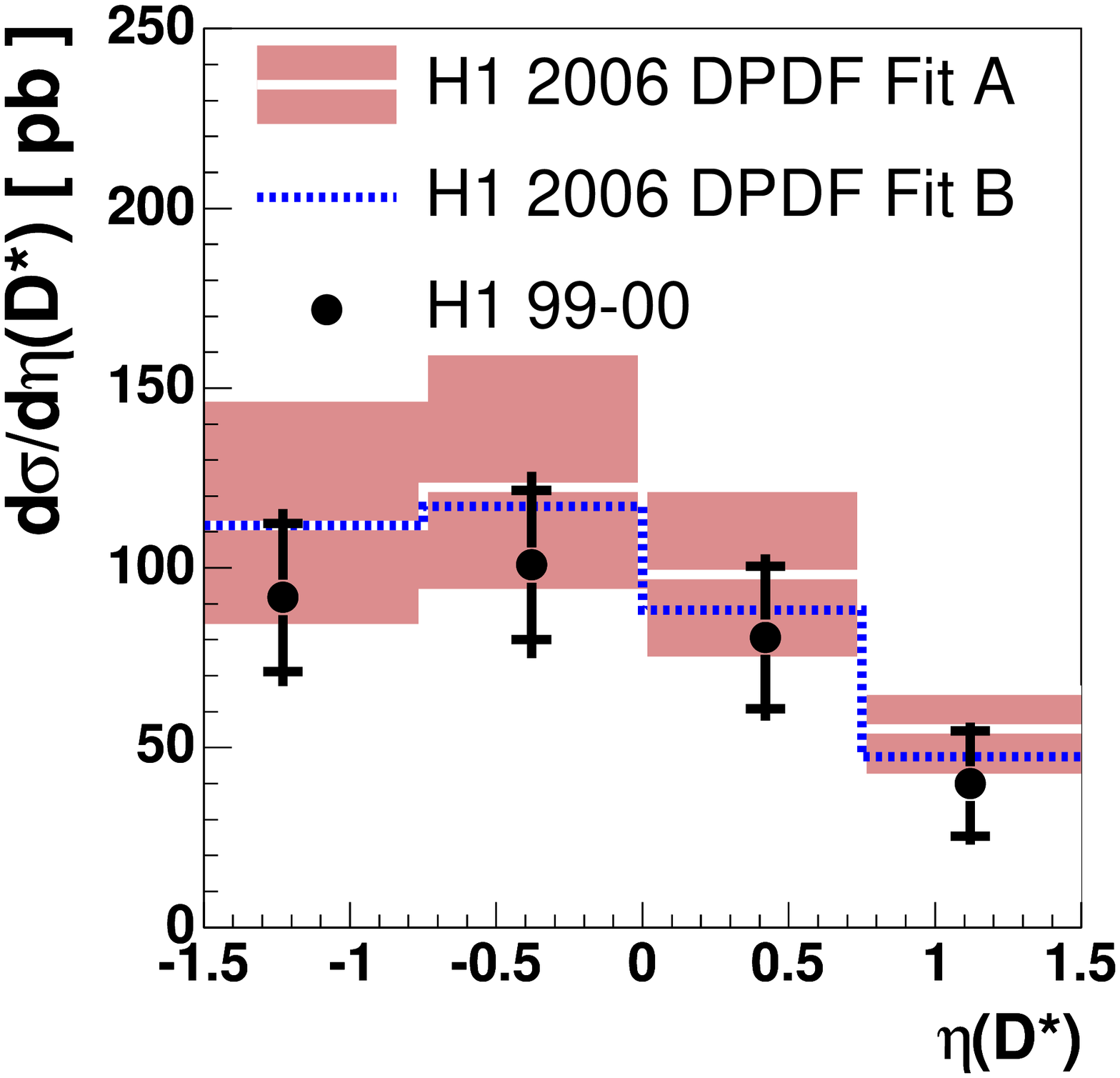}
    \includegraphics[width=.48\textwidth]{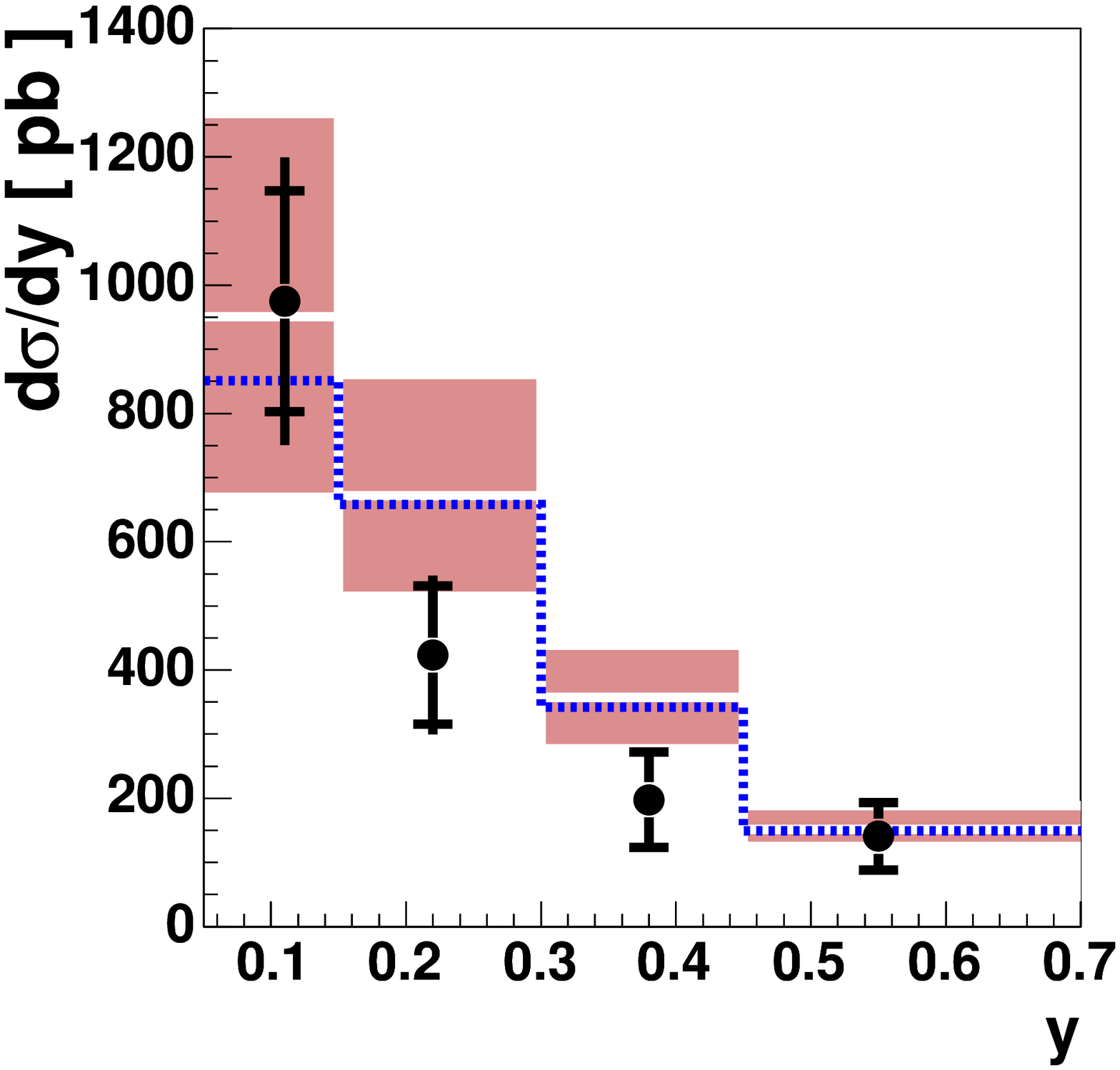}
    \includegraphics[width=.48\textwidth]{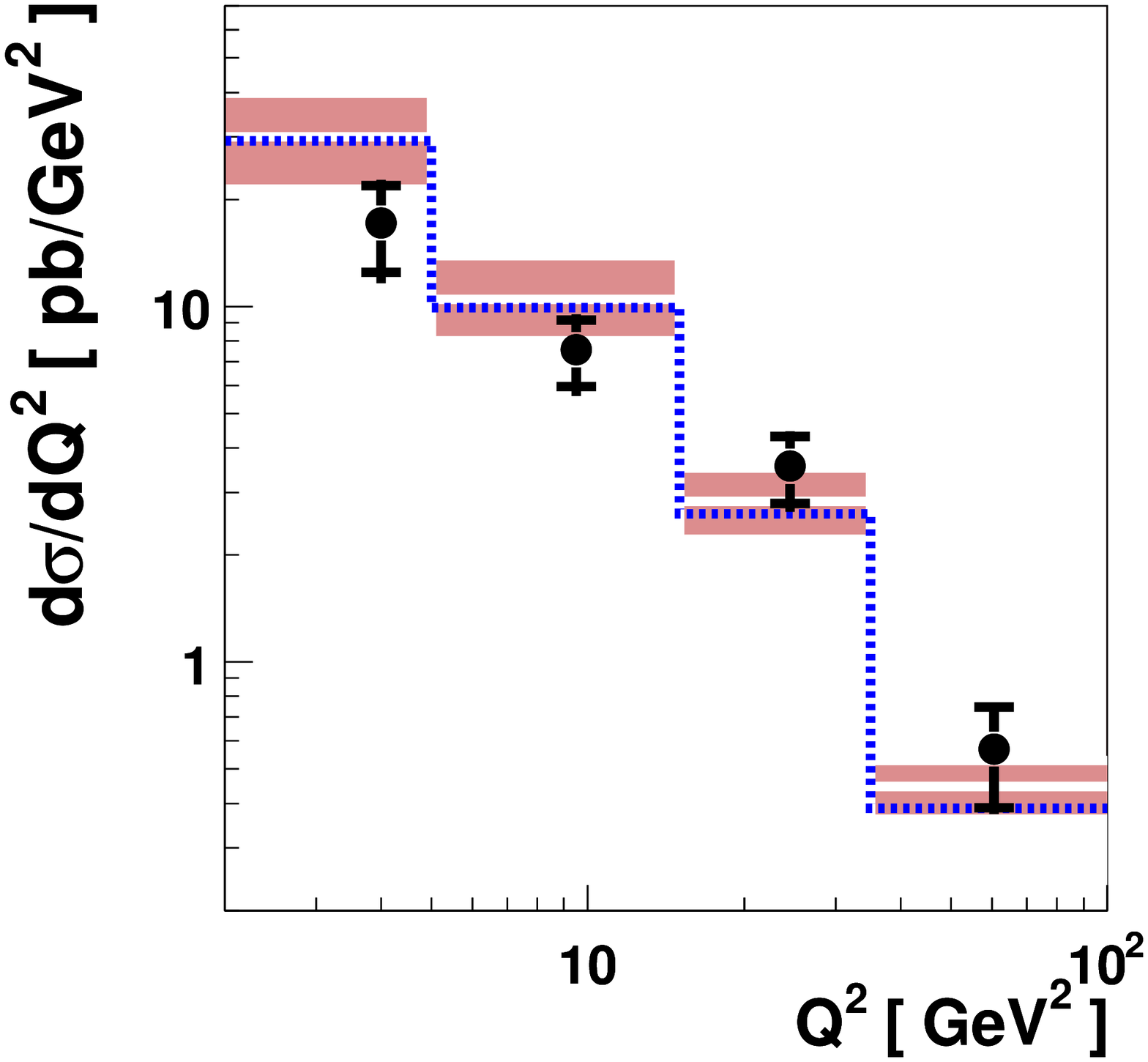}
    \setlength{\unitlength}{\textwidth}
    
    \begin{picture}(0,0)
      \begin{Large}
        \put(-0.17,0.97){\bfseries \boldmath H1 Diffractive \dstar \, in DIS}
        \put(-0.47,0.57){\bfseries a)}
        \put( 0.03,0.57){\bfseries b)}
        \put(-0.47,0.11){\bfseries c)}
        \put( 0.03,0.11){\bfseries d)}
      \end{Large}
    \end{picture}
    
    \Mycaption{
      Differential cross sections for diffractive $\dstar$ meson production in DIS as a function 
      of (a) $\ptds$\! , (b) $\etads$\! , (c) the inelasticity $y$ and (d) the photon virtuality 
      $Q^{2}$\@. The inner error bars of the data points represent the statistical uncertainties 
      of the measurement only, while the outer error bars show the statistical and systematic 
      uncertainties added in quadrature. The data are compared with a pQCD calculation in NLO using 
      two alternative sets of diffractive parton density functions (Fit A and Fit B) extracted by 
      H1~\cite{h1_diff_incl_2005}. 
      }
    \label{fig:nlo_dis_kinematics}
  \end{center}
\end{figure}

\begin{figure}[htbp]
  \begin{center}
    \includegraphics[width=.48\textwidth]{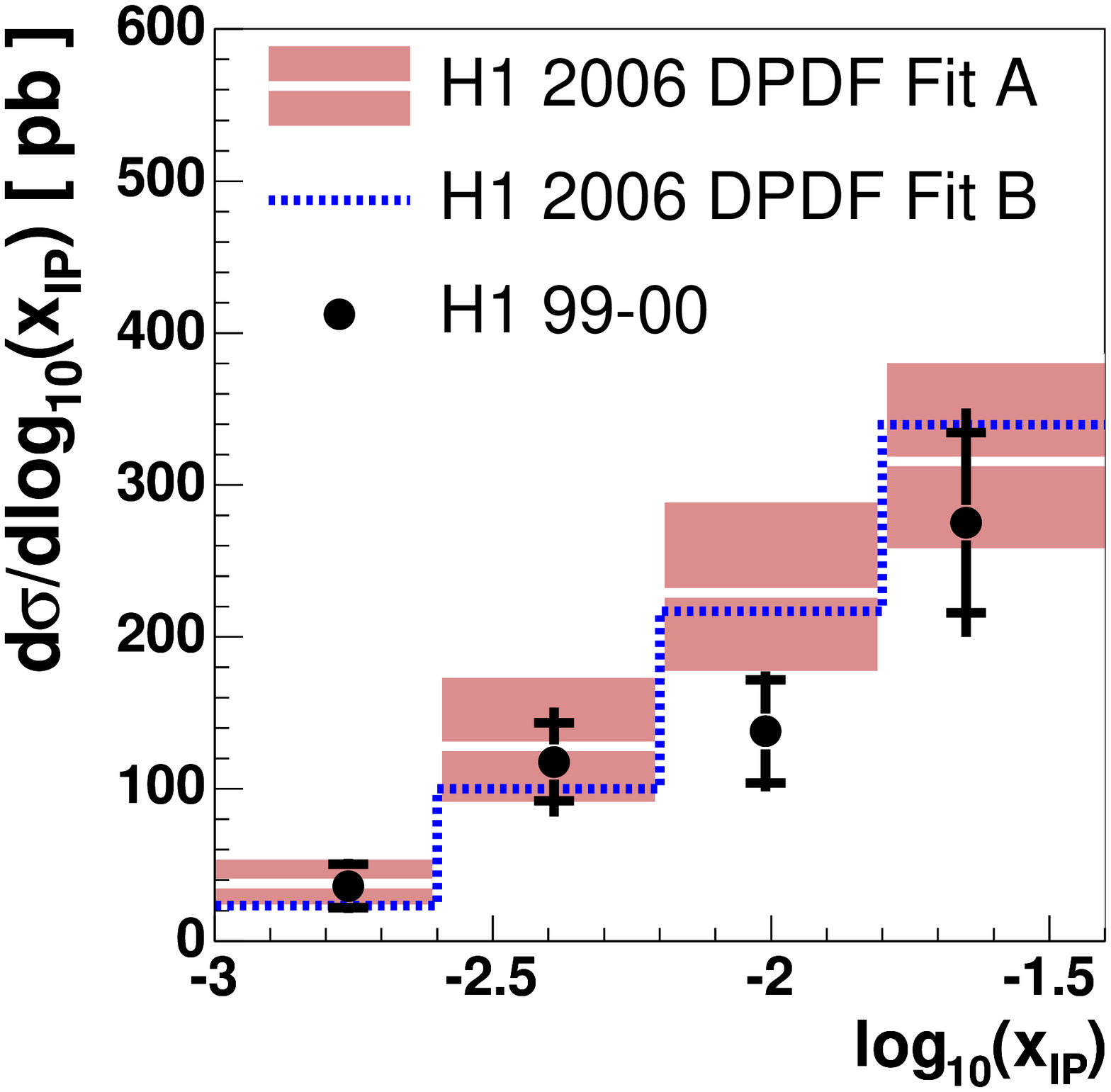}
    \includegraphics[width=.48\textwidth]{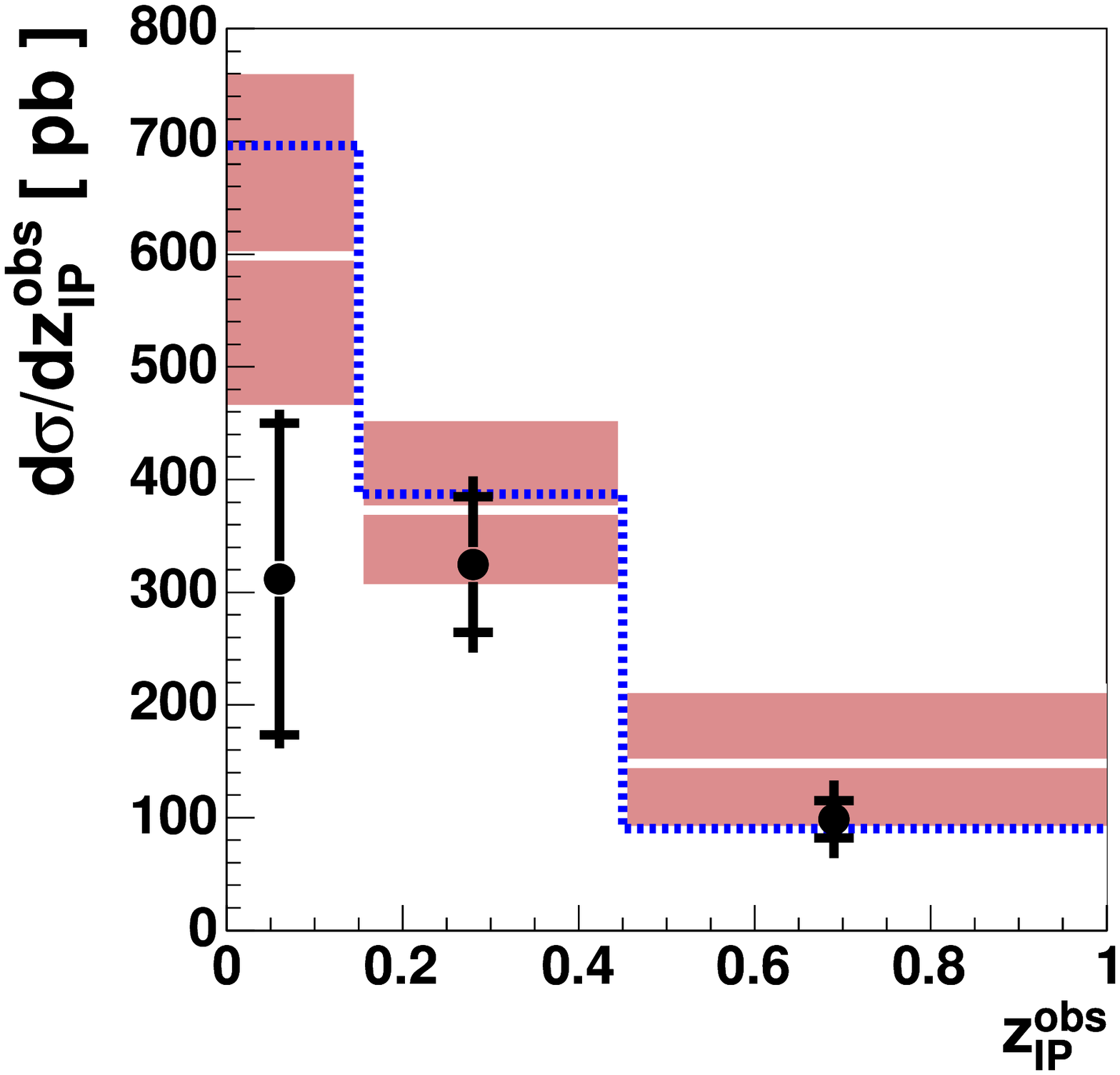}
    \includegraphics[width=.48\textwidth]{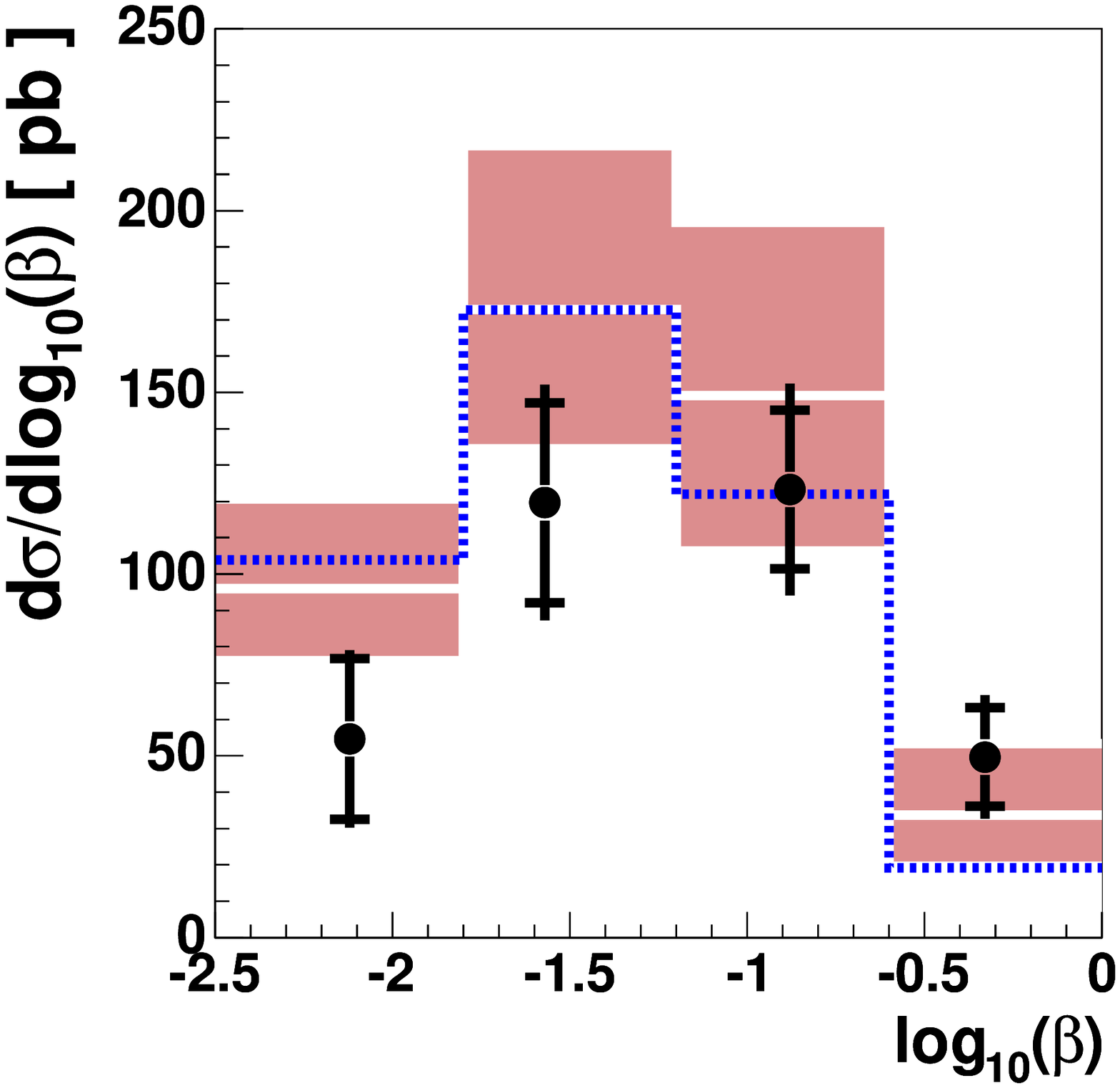}
    \includegraphics[width=.48\textwidth]{empty.eps}
    \setlength{\unitlength}{\textwidth}
    
    \begin{picture}(0,0)
      \begin{Large}
        \put(-0.17,0.97){\bfseries \boldmath H1 Diffractive \dstar \, in DIS}
        \put(-0.47,0.57){\bfseries a)}
        \put( 0.03,0.57){\bfseries b)}
        \put(-0.47,0.11){\bfseries c)}
      \end{Large}
    \end{picture}
    
    \Mycaption{
      Differential cross sections for diffractive $\dstar$ meson production in DIS as a function of 
      (a) $\xpom$\! , (b) $\zpomobs$ and (c) $\beta$\@. The inner error bars of the data points represent 
      the statistical uncertainties of the measurement only, while the outer error bars show the 
      statistical and systematic uncertainties added in quadrature. The data are compared with a pQCD 
      calculation in NLO using two alternative sets of diffractive parton density functions (Fit A 
      and Fit B) extracted by H1~\cite{h1_diff_incl_2005}.   
}
    \label{fig:nlo_dis_diffractive}
  \end{center}
\end{figure}

The $\dstar$ meson cross section in DIS is also measured differentially as a function 
of the $\dstar$ kinematic variables $\ptds$ \,and $\etads$, the DIS kinematic 
variables $y$ and $Q^{2}$, and the diffractive variables $\xpom$, $\beta$ 
and $\zpomobs$. They are listed in table~\ref{tab:xsec_dis_0} 
and shown in Figures~\ref{fig:nlo_dis_kinematics} and \ref{fig:nlo_dis_diffractive}. 
The data are compared in the figures with the predictions of the NLO QCD 
calculations. For the 
cross sections as a function of the $\dstar$ and DIS kinematic variables 
the predictions for the two sets of DPDFS are similar with both providing 
a good description of the data. 
For the comparison with the diffractive kinematic quantities the differences 
in the predictions for the two DPDFs are larger, with $\zpomobs$ showing 
the largest sensitivity, where the steeper fall-off of the gluon density 
in Fit B is reproduced. However, within the present experimental errors 
and theoretical uncertainties these differences cannot be resolved. The 
good description of the NLO QCD calculations for all of the $\dstar$, DIS 
and diffractive kinematic distributions supports the 
assumption of QCD factorization, in particular,
the compatibility of the gluon density obtained 
from scaling violations in the inclusive diffractive cross section 
with that required to describe the $\dstar$ data.

\begin{figure}[htbp]
  \begin{center}
    \includegraphics[width=.48\textwidth]{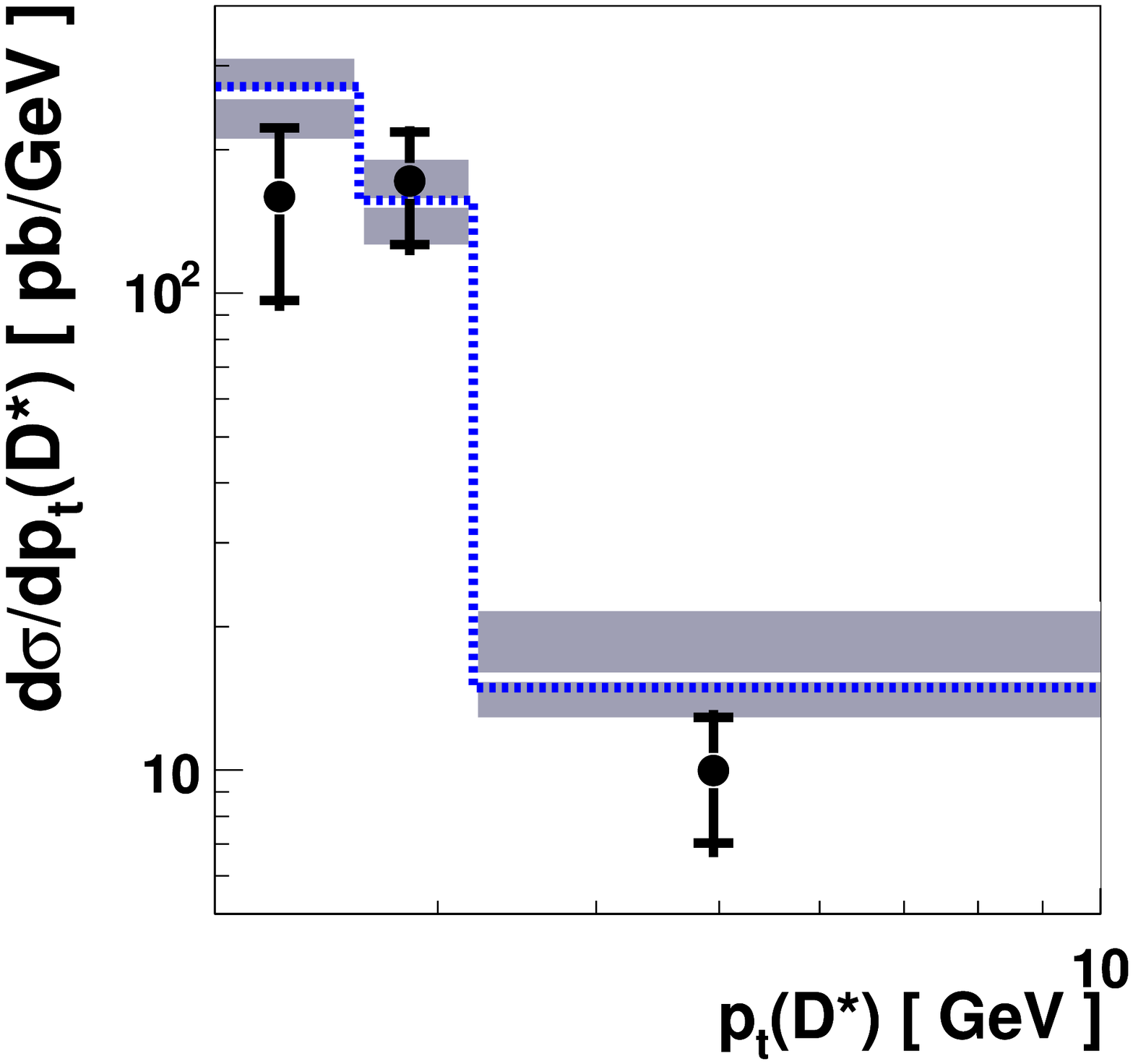}
    \includegraphics[width=.48\textwidth]{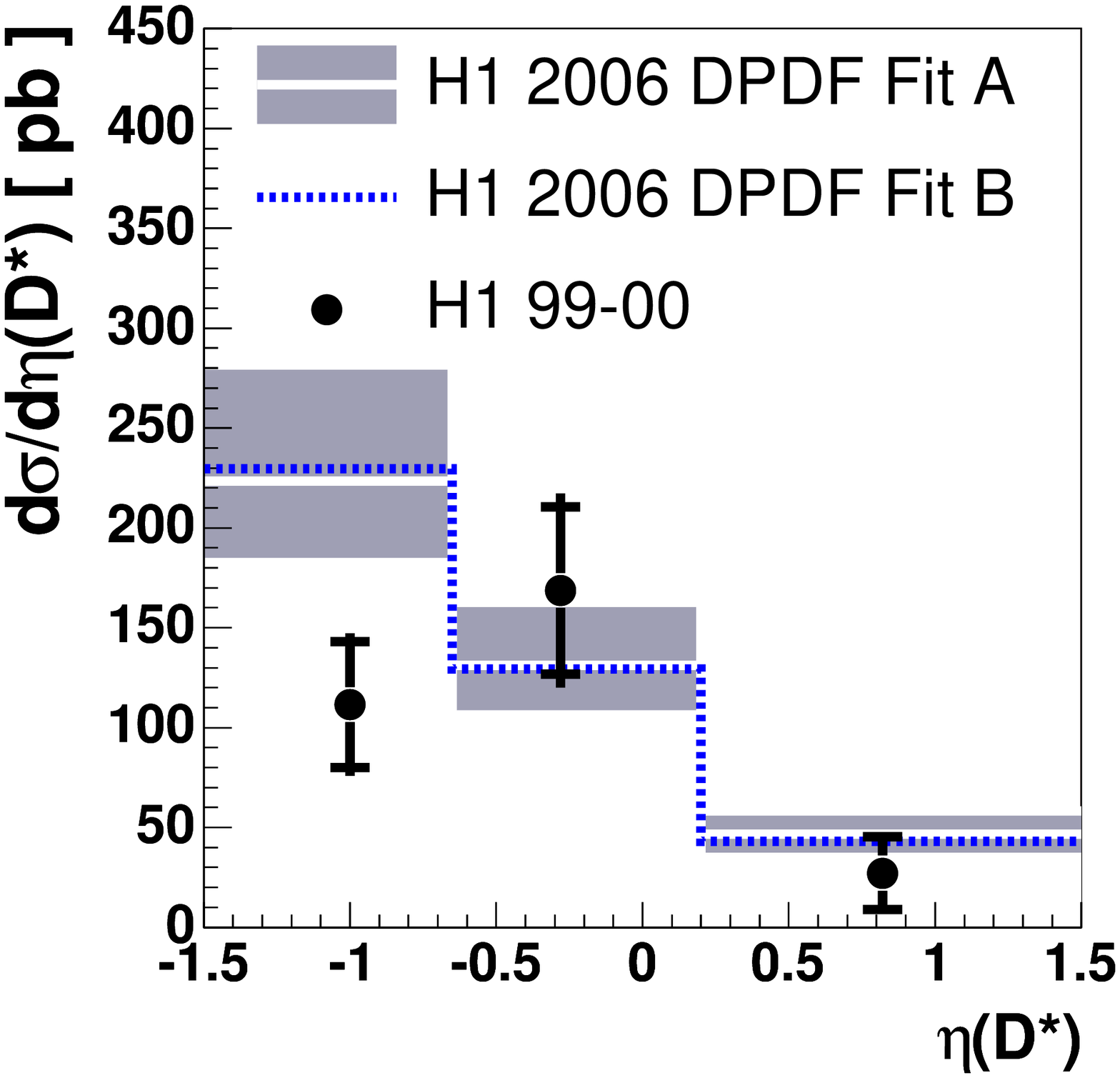}
    \includegraphics[width=.48\textwidth]{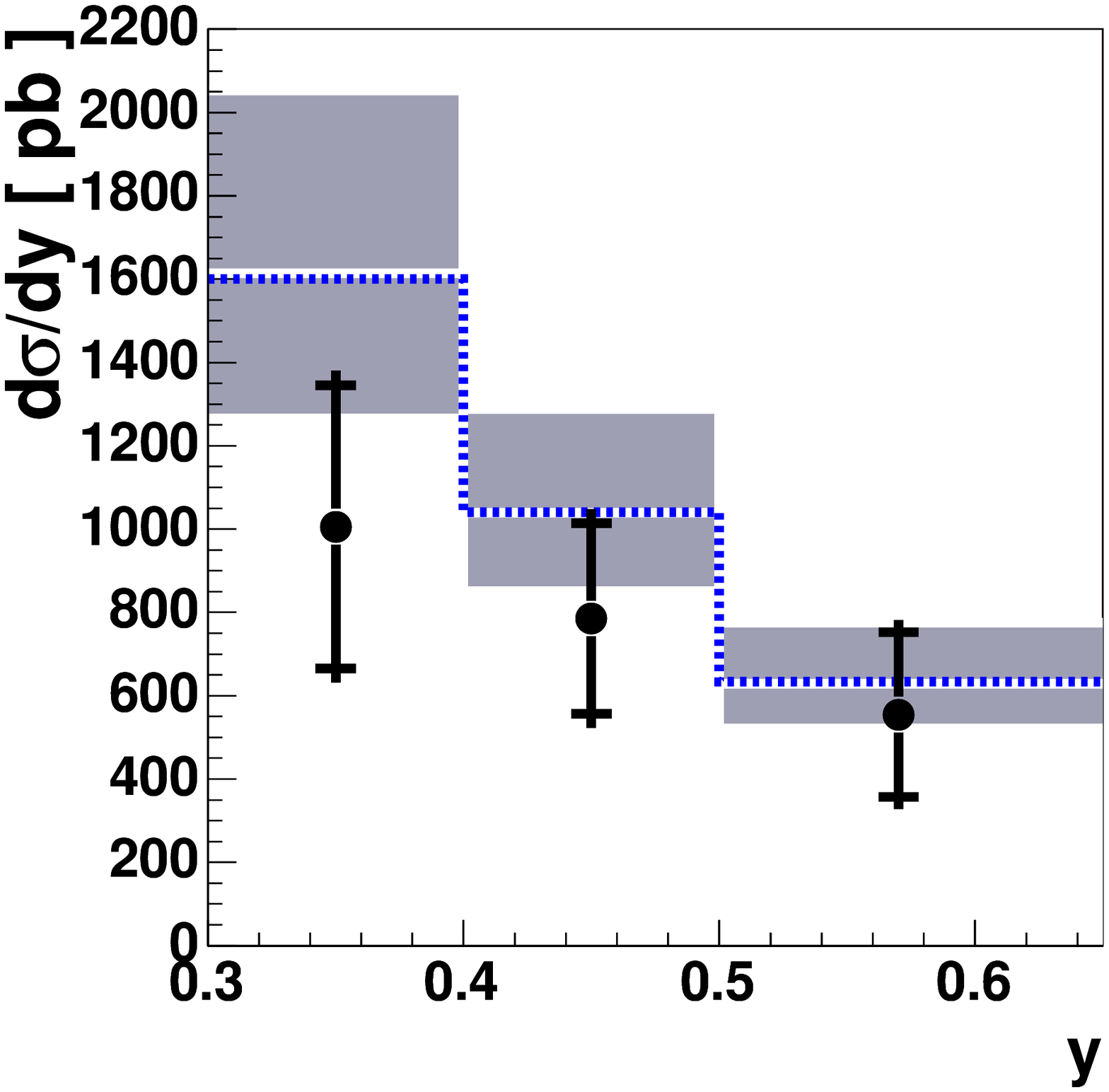}
    \includegraphics[width=.48\textwidth]{empty.eps}
    \setlength{\unitlength}{\textwidth}

    \begin{picture}(0,0)
      \begin{Large}
        \put(-0.17,0.97){\bfseries \boldmath H1 Diffractive \dstar \, in $\gamma p$}
        \put(-0.47,0.57){\bfseries a)}
        \put( 0.03,0.57){\bfseries b)}
        \put(-0.47,0.11){\bfseries c)}
      \end{Large}
    \end{picture}
    
    \Mycaption{
      Differential cross sections for diffractive $\dstar$ meson production in photoproduction as a 
      function of (a) $\ptds$\! , (b) $\etads$ and (c) the inelasticity $y$\@. The inner error bars 
      of the data points represent the statistical uncertainties of the measurement only, while the 
      outer error bars show the statistical and systematic uncertainties added in quadrature. The 
      data are compared with a pQCD calculation in NLO using two alternative sets of diffractive parton 
      density functions (Fit A and Fit B) extracted by H1~\cite{h1_diff_incl_2005}.  
      }
    \label{fig:nlo_php_kinematics}
  \end{center}
\end{figure}

\begin{figure}[htbp]
  \begin{center}
    \includegraphics[width=.48\textwidth]{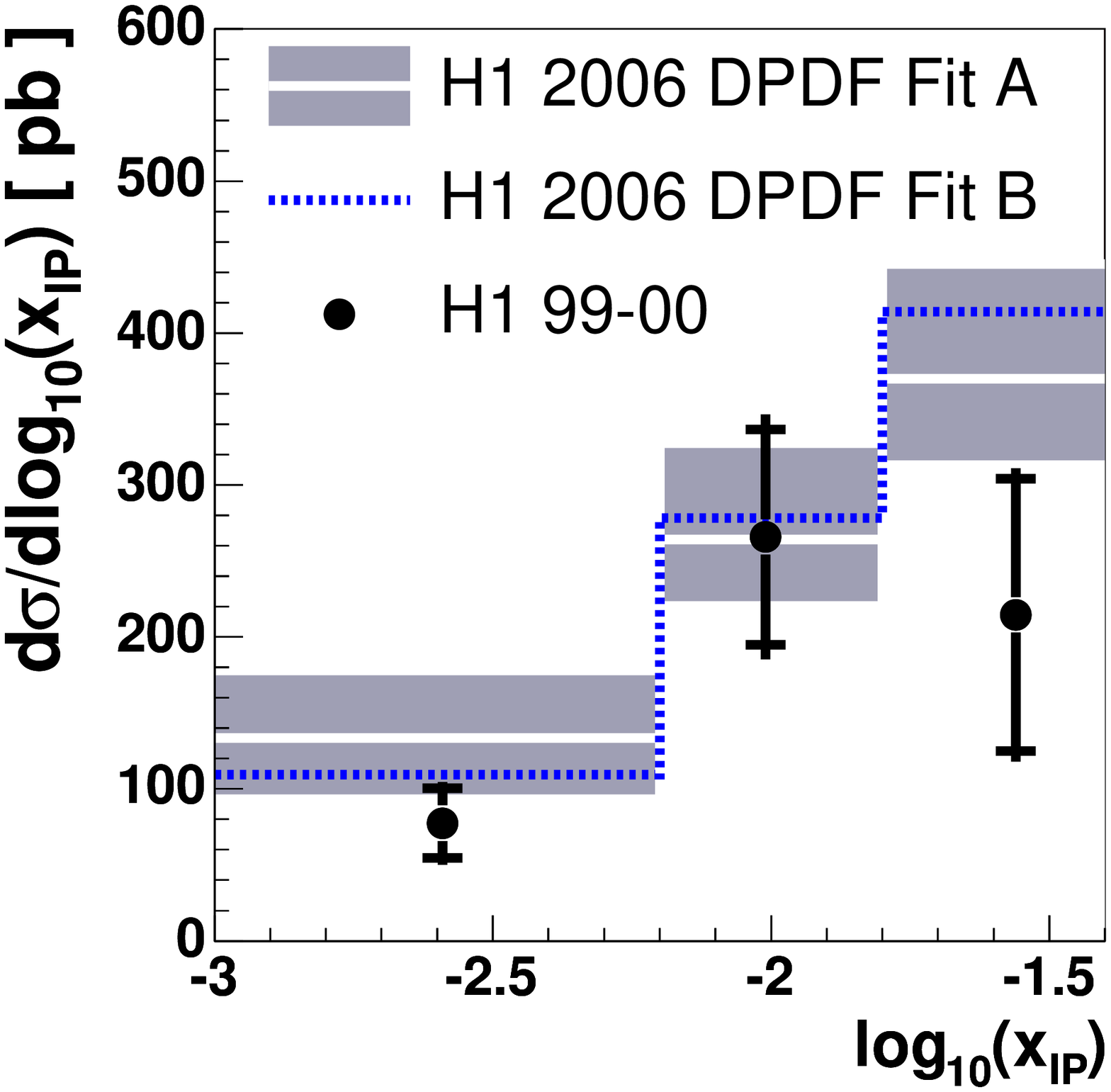}
    \includegraphics[width=.48\textwidth]{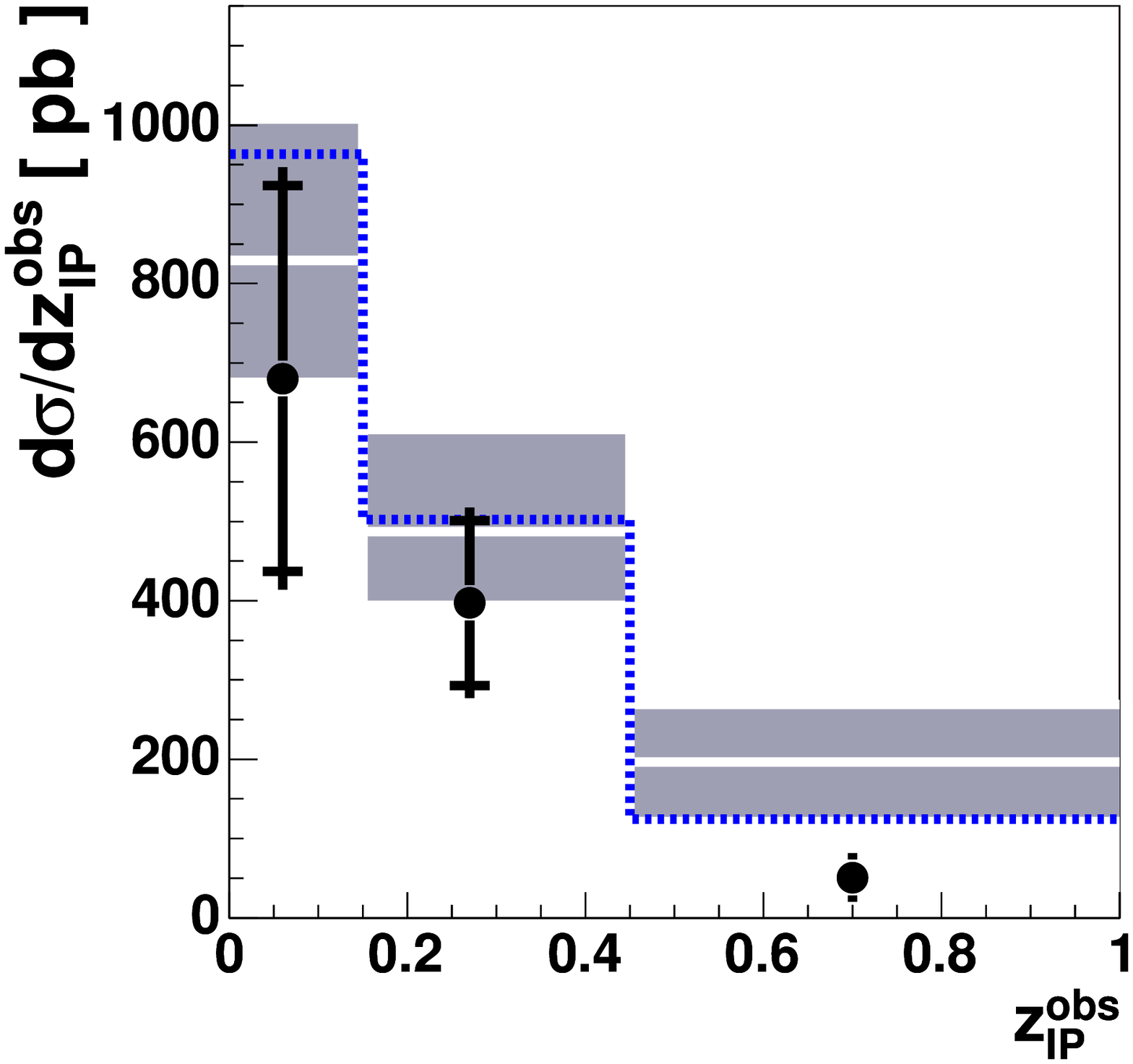}
    \setlength{\unitlength}{\textwidth}
    
    \begin{picture}(0,0)
      \begin{Large}
        \put(-0.17,0.51){\bfseries \boldmath H1 Diffractive \dstar \, in $\gamma p$}
        \put(-0.47,0.11){\bfseries a)}
        \put( 0.03,0.11){\bfseries b)}
      \end{Large}
    \end{picture}

    \Mycaption{
      Differential cross sections for diffractive $\dstar$ meson production in photoproduction 
      as a function of (a) $\xpom$ and (b) $\zpomobs$\@. The inner error bars of the data points 
      represent the statistical uncertainties of the measurement only, while the outer error 
      bars show the statistical and systematic uncertainties added in quadrature. The data are 
      compared with a pQCD calculation in NLO using two alternative sets of diffractive parton 
      density functions (Fit A and Fit B) extracted by H1~\cite{h1_diff_incl_2005}. 
      }    
    \label{fig:nlo_php_diffractive}
  \end{center}
\end{figure}

In photoproduction the $\dstar$ cross section is shown differentially as 
a function of the $\dstar$ kinematic variables $\ptds$ \,and $\etads$ and 
the kinematic variable $y$ in \fig~\ref{fig:nlo_php_kinematics} and as a 
function of the diffractive kinematic variables $\xpom$ and $\zpomobs$ in 
\fig~\ref{fig:nlo_php_diffractive}\@. The values are given in 
table~\ref{tab:xsec_php_0}\@.  The data are well described by the 
theoretical predictions within the larger experimental errors for 
photoproduction. As in DIS the largest sensitivity to the different 
parameterizations of the gluon is evident in the $\zpomobs$ distribution. 
The shapes of the $\zpomobs$ distribution for the predictions in DIS and 
$\gamma p$ are compatible which is due to the fact that both kinematic 
regimes probe the diffractive gluon density at a similar scale.

The good agreement of the NLO QCD predictions with the measured cross 
sections observed in DIS and photoproduction, both in shape and 
normalization, supports the assumption
that QCD factorization is applicable in both 
kinematic regimes. A quantity, which is less sensitive to the input of 
diffractive parton density functions and theoretical uncertainties is 
defined by

\begin{equation}
  R^{ \gamma p}_{\rm DIS} = \frac { \left( {\sigma^{\rm meas} / \sigma^{\rm theo}}  \right)_{\gamma p} }
   { \left( {\sigma^{\rm meas} / \sigma^{\rm theo} } \right)_{\rm DIS}}
\end{equation}
where $\sigma^{{\rm meas}}$ and $\sigma^{{\rm theo}}$ denote the measured 
and the predicted integrated cross section for $\dstar$ production. To 
reduce theoretical uncertainties due to extrapolations from different 
regions in $y$ the cross section in DIS is further restricted to the range 
of $0.3 < y < 0.65$ as for the photoproduction measurement. 
The DIS cross section in this range is shown in table~\ref{tab:int_cross_section}\@.
The ratio $R^{{ \gamma p}}_{{\rm DIS}}$ is found to be 
$1.15 \pm 0.40 ({\rm stat.}) \pm 0.09 ({\rm syst.})$, 
with the systematic uncertainty originating from the model 
uncertainty on the $\beta$ distribution in DIS, the fragmentation 
uncertainties and the uncertainties on the Reggeon contribution. The 
theoretical uncertainty on $R^{{ \gamma p}}_{{\rm DIS}}$ 
is $\pm 7\%$\@. The measurement of $R^{{ \gamma p}}_{{\rm DIS}}$ 
shows no evidence for a suppression of the photoproduction component 
although the statistical error of the measurement is large.

\begin{figure}[htbp]
  \begin{center}
    \includegraphics[width=.48\textwidth]{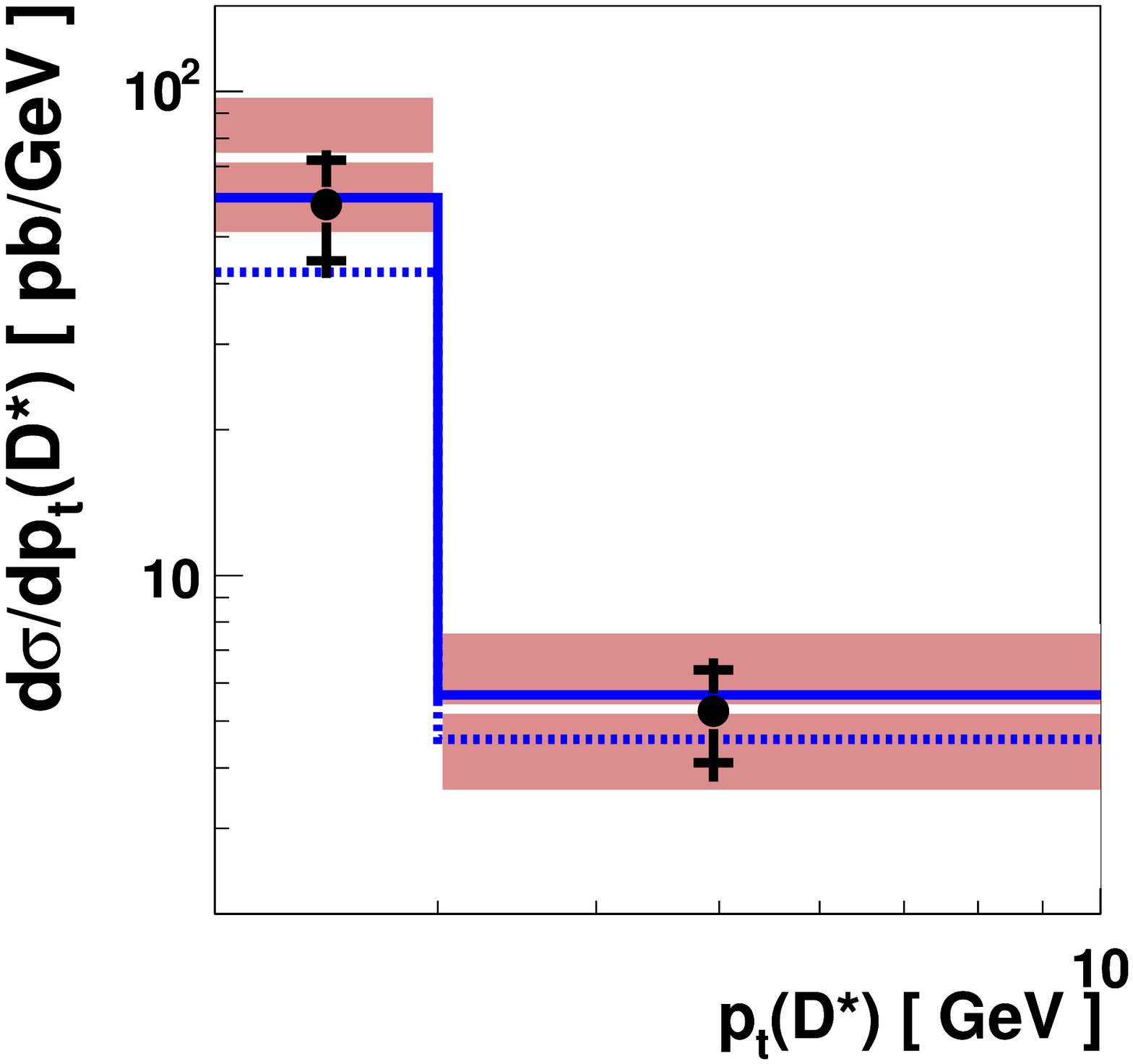}
    \includegraphics[width=.48\textwidth]{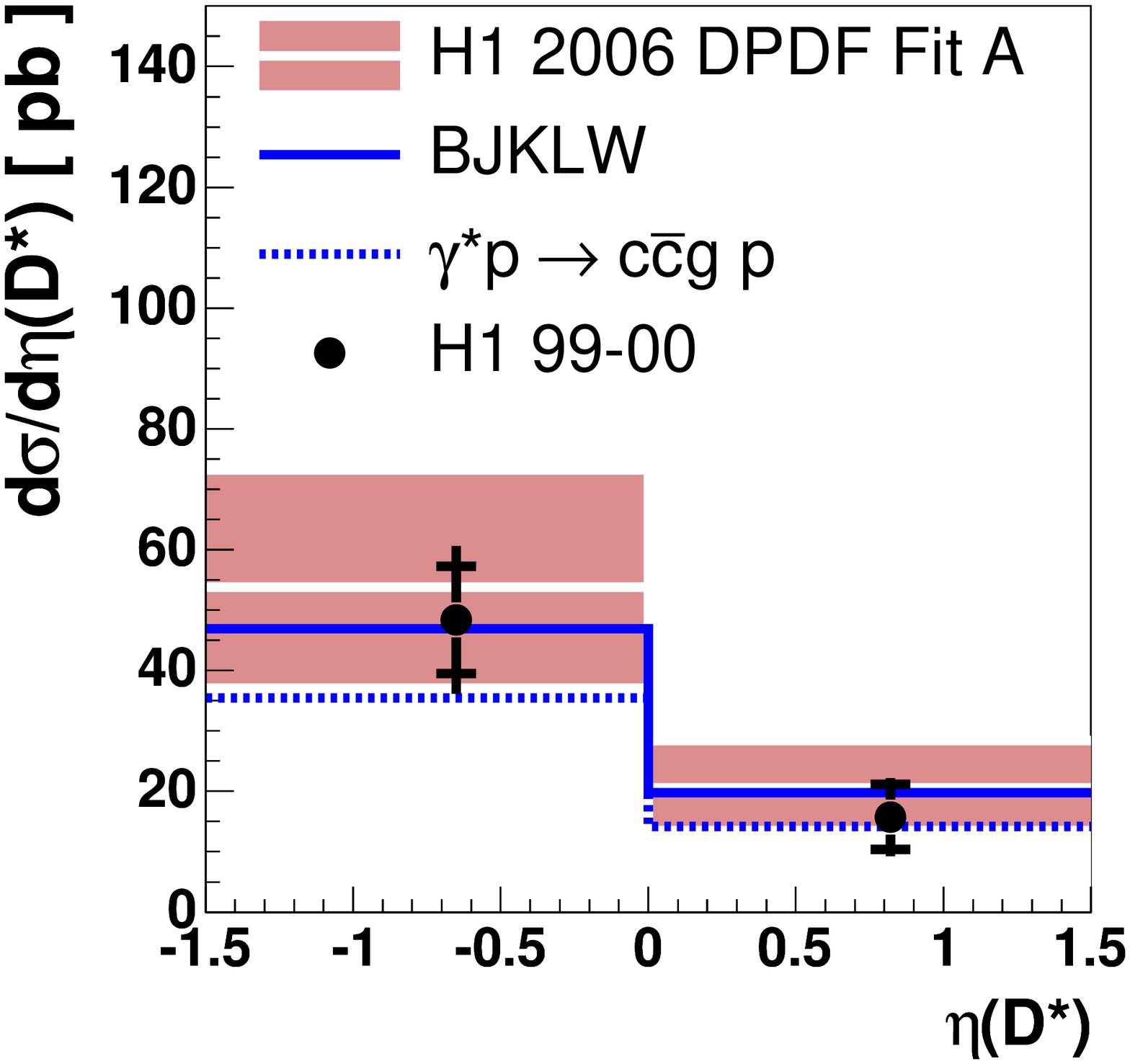}
    \includegraphics[width=.48\textwidth]{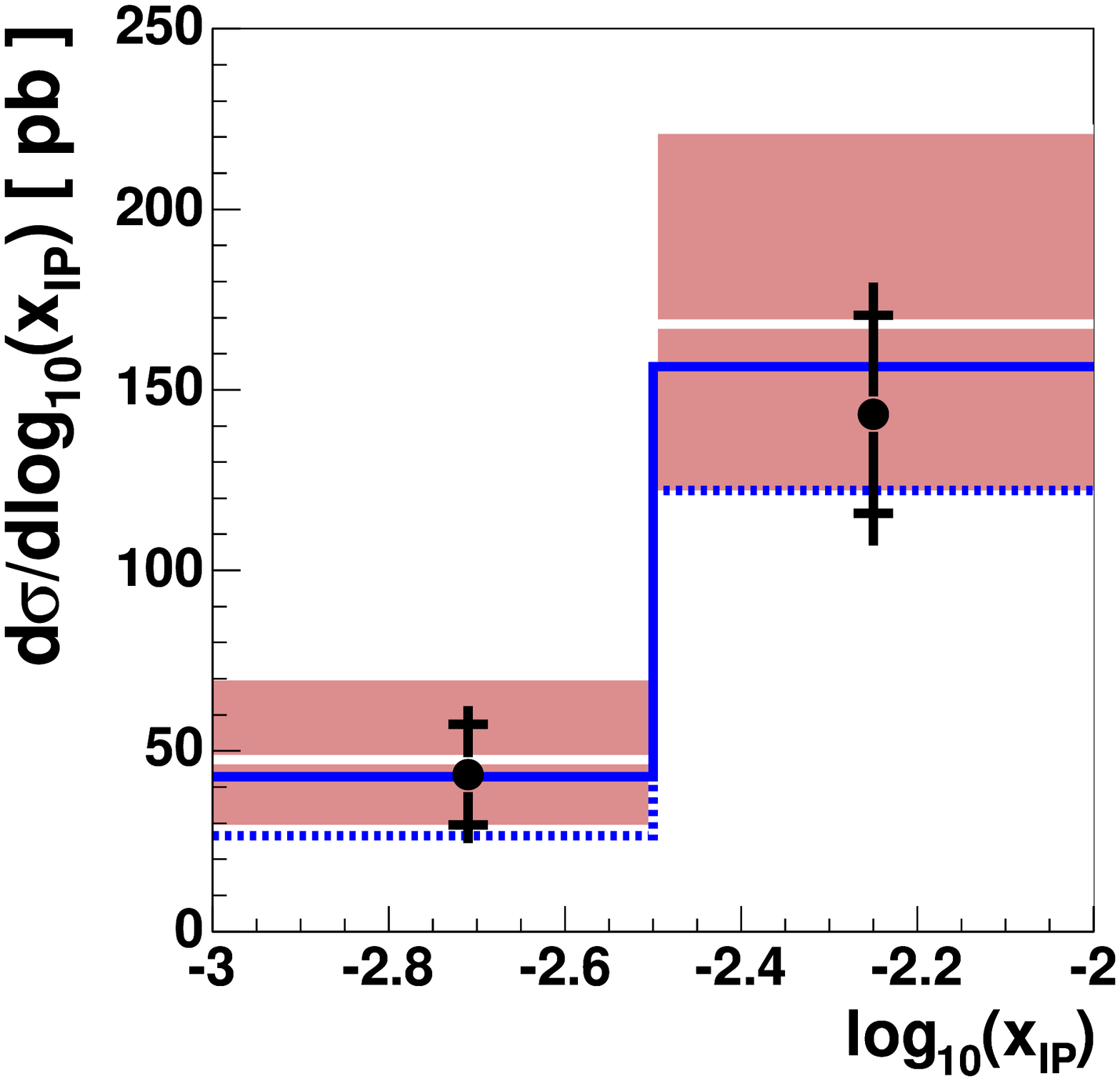}
    \includegraphics[width=.48\textwidth]{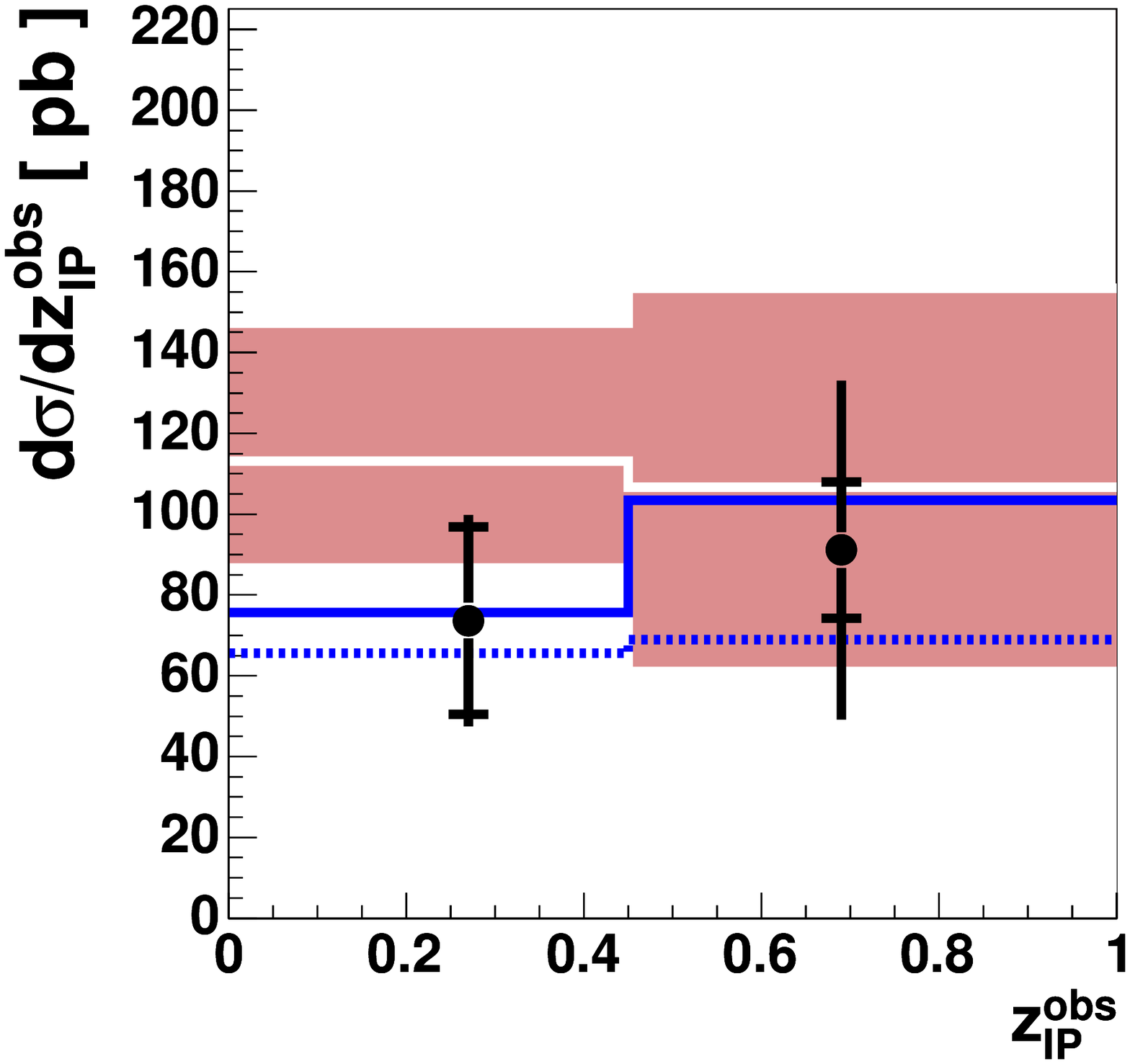}
    \setlength{\unitlength}{\textwidth}
    
    \begin{picture}(0,0)
      \begin{Large}
        \put(-0.27,0.97){\bfseries \boldmath H1 Diffractive \dstar \, in DIS ($\xpom<0.01$)}
        \put(-0.47,0.57){\bfseries a)}
        \put( 0.03,0.57){\bfseries b)}
        \put(-0.47,0.11){\bfseries c)}
        \put( 0.03,0.11){\bfseries d)}
      \end{Large}
    \end{picture}
    
    \Mycaption{
      Differential cross sections for diffractive $\dstar$ meson production in DIS, in the restricted 
      kinematic region of $\xpom<0.01$, shown as a function of (a) $\ptds$\! , (b) $\etads$\! , (c) $\xpom$
      and (d) $\zpomobs$\@. The inner error bars of the data points represent the statistical uncertainties 
      of the measurement only, while the outer error bars show the statistical and systematic uncertainties 
      added in quadrature. The data are compared with a pQCD calculation in NLO and to a prediction from the 
      perturbative two gluon approach of BJKLW~\cite{bjklw} with a cut for the 
      gluon momentum in the $\gamma^{*}p\rightarrow c \bar{c} g p$ process of $p_{t}>2.0 \, {\GeV}$\@. 
The dashed line indicates the resolved 
$\gamma^{*}p\rightarrow c \bar{c} g p$ contribution only while the solid line shows the sum of the 
$\gamma^{*}p\rightarrow c \bar{c} g p$ and the 
$\gamma^{*}p\rightarrow c \bar{c} p$ contributions.
      }
    \label{fig:xpomlt0.01}
  \end{center}
\end{figure}

In \fig~\ref{fig:xpomlt0.01} an additional comparison of both the NLO QCD 
calculations and of the prediction from the perturbative two gluon calculation 
of BJKLW~\cite{bjklw} with differential cross sections in the range of validity 
of the two gluon model (\mbox{$x_{\pom}<0.01$}) are shown. The cross sections 
are given in table~\ref{tab:xsec_dis_lowxpom}\@.
Within the uncertainties a good agreement between the data and both the NLO 
QCD calculation and the model of BJLKW is observed. For the two gluon 
calculation in this kinematic range the 
$\gamma^{*}p\rightarrow c \bar{c} g p$ contribution is 
seen to dominate with the $\gamma^{*}p\rightarrow c \bar{c} p$ 
process contributing only at high values 
of $\zpomobs$\@.
Varying the $\pt$ cut-off for the gluon in the 
$\gamma^{*}p\rightarrow c \bar{c} g p$ process by
$\pm 0.5\, {\GeV}$ leads to a variation of the cross section of
$\sim 25 \%$ and is also compatible with the data.

The measurements of the diffractive charm DIS cross sections in $Q^{2}$, $y$ 
and $\mx^{2}$ obtained from the displaced track method 
are converted to measurements in $\xpom$, $\beta$ and $Q^{2}$ using

\begin{equation}
\frac{{\rm d}^{3}\sigma^{c\bar{c}}_{D}}{{\rm d} 
\xpom\, {\rm d}\beta\,{\rm d} Q^{2}}  = 
\frac{{\rm d}^{3}\sigma^{c\bar{c}}_{D}}{{\rm d} 
Q^{2}\, {\rm d} y\,{\rm d} M_{X}^2} \frac{s y^2} {\beta}. 
\end{equation}
The diffractive charm reduced cross section is defined as

\begin{equation}
\tilde{\sigma}^{c\bar{c}}_{D} (\xpom, \beta, Q^{2}) = \frac{{\rm d}^{3}\sigma^{c\bar{c}}_{D}}{{\rm d} 
\xpom\, {\rm d}\beta\,{\rm d} Q^{2}} \frac{\beta Q^{4} } {2 \pi \alpha^{2} (1+ (1-y)^{2})},
\end{equation}
where $\alpha$ is the fine structure constant. The reduced 
cross section is approximately equal to the charm contribution 
$F_{2}^{D(3) c \bar{c}}$ to the diffractive structure function 
$F_{2}^{D(3)}$\@. The difference is due to the contribution 
from the longitudinal diffractive charm cross section, which 
is expected to be small for the data points presented in this paper. 

\begin{figure}[htbp]
  \begin{center}
    \includegraphics[width=.48\textwidth]{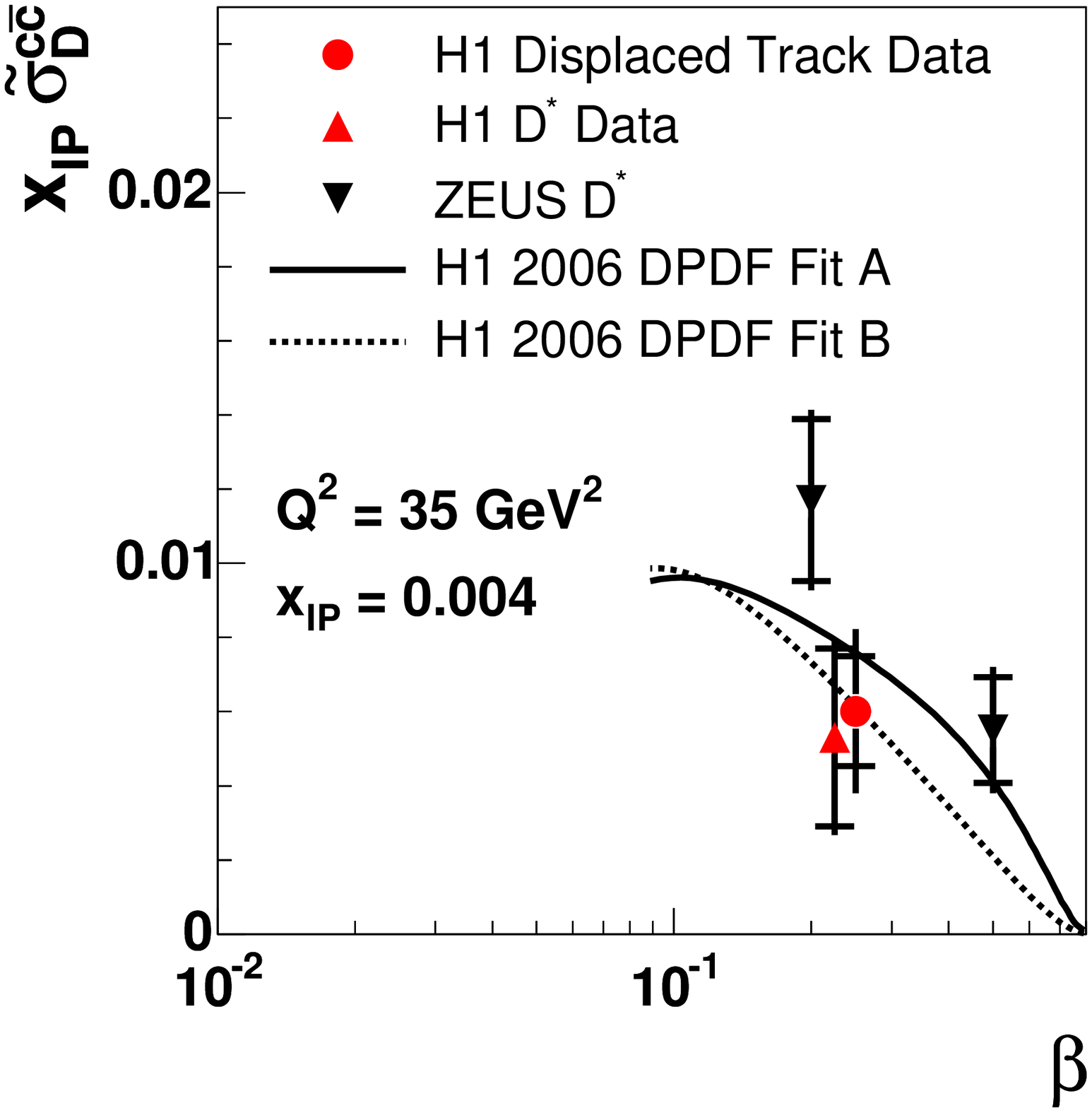}
    \includegraphics[width=.48\textwidth]{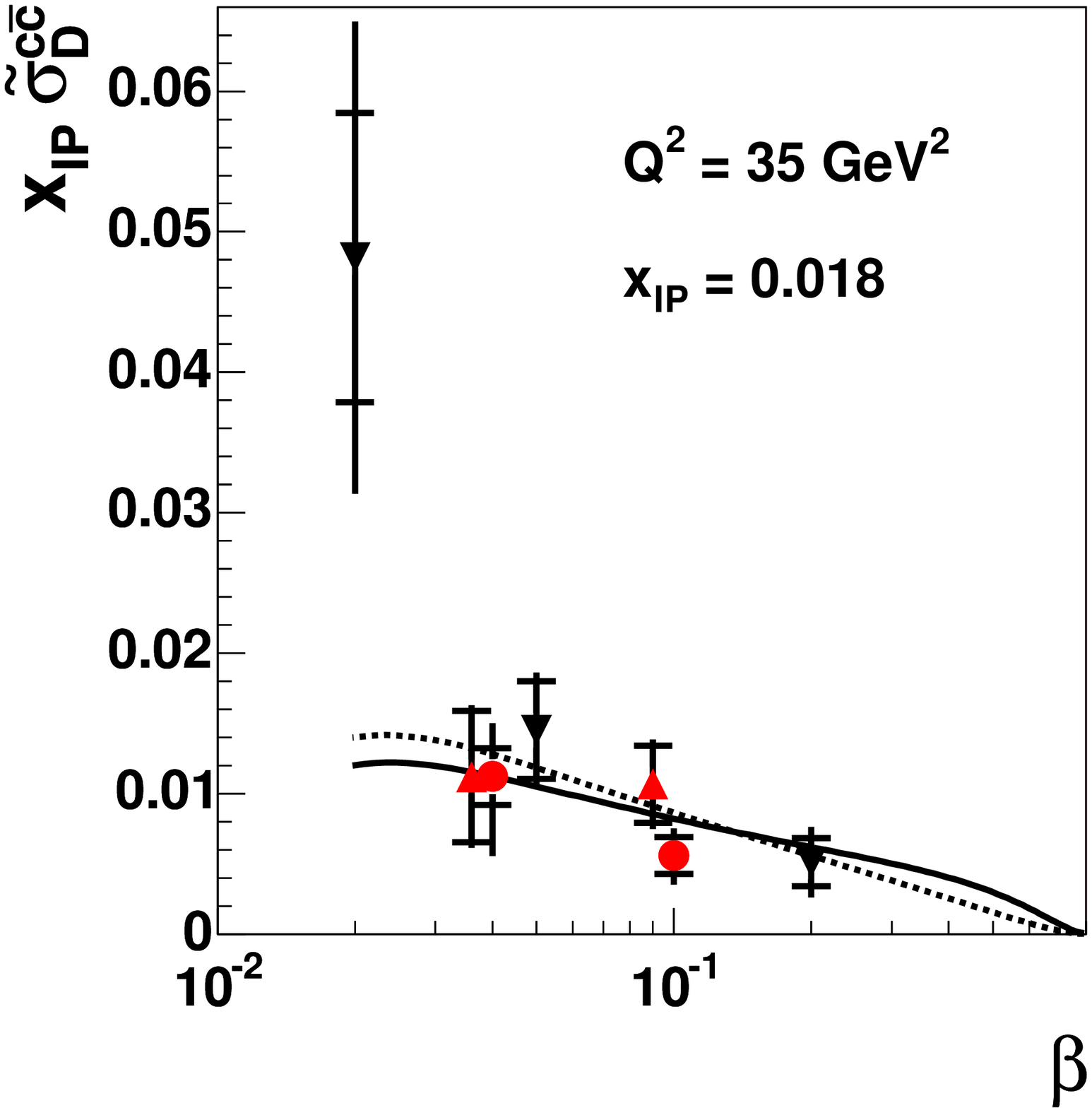}
    \setlength{\unitlength}{\textwidth}
    
    \begin{picture}(0,0)
      \begin{Large}
      \end{Large}
    \end{picture}

    \Mycaption{
      The measured reduced cross section $\xpom \, \tilde{\sigma}^{c\bar{c}}_{D}$ shown as a function 
      of $\beta$ for two different values of $\xpom$\@.  The inner error bars 
      of the data points represent the statistical 
      error, while the outer error bars represent the statistical and systematic uncertainties added in 
      quadrature. The measurements obtained from $\dstar$ mesons from H1 in this paper and from 
      ZEUS~\cite{zeus_diff_dstar_2004} are also shown. Measurements at the same values
      of $\beta$ are displaced for visibility.
      The measurements are compared with NLO
      predictions based on two alternative sets of diffractive parton density functions (Fit A and 
      Fit B) extracted by H1~\cite{h1_diff_incl_2005}. 
      }
    \label{fig:sigmarcc}
  \end{center}
\end{figure}

The measurements of $\xpom \tilde{\sigma}^{c\bar{c}}_{D}$ obtained 
from the displaced track method are listed in table~\ref{tab:sigmarcc} 
and shown in~\fig~\ref{fig:sigmarcc} as a function of $\beta$ for 
fixed values of $Q^{2}$ and $\xpom$\@. In the figure, the displaced 
track method data point measured at $\xpom = 0.01$ is interpolated 
to $\xpom = 0.018$ using a parameterization of $\tilde{\sigma}^{c\bar{c}}_{D}$ 
from the NLO QCD fit. The measured points of $\xpom \tilde{\sigma}^{c\bar{c}}_{D}$ 
are compared with the results extracted from the $\dstar$ meson 
analysis. For this purpose the $\dstar$ cross section is measured in the 
same $Q^{2}$, $y$ and $\mx^{2}$ ranges as for the displaced track 
method. The results are given in table~\ref{tab:sigmamx}\@. 
These measurements in the visible $\dstar$ kinematic range 
$\ptds > 2~\GeV$ \, and $|\etads| < 1.5$ are
extrapolated with the NLO calculation program \hvqdis \ to the full
$\dstar$ kinematic phasespace in order to extract the diffractive
open charm cross section. The extrapolation factors are found to be 
$\approx 2.5$\@. 
The NLO calculation program is also used to evaluate the bin center 
corrections, which are made to the same central values as in the 
displaced track analysis. The H1 data are also compared with $\dstar$ 
measurements from the ZEUS collaboration~\cite{zeus_diff_dstar_2004} 
which are interpolated to the same 
kinematic range as the H1 measurement using the NLO QCD fit and 
corrected with a factor of $1.23$ to account for the difference in the measured range
from $\my=m_{p}$ to $\my<1.6 ~{\GeV}$~\cite{h1_diff_elasp_2005}\@. 
The measurements for $\xpom \tilde{\sigma}^{c\bar{c}}_{D}$ from the 
displaced track analysis and the $\dstar$ extraction methods from both 
H1 and ZEUS are in good agreement. A comparison with the predictions 
of the NLO DPDFs shows a good description of the data.

\begin{figure}[htbp]
  \begin{center}
    \includegraphics[width=.48\textwidth]{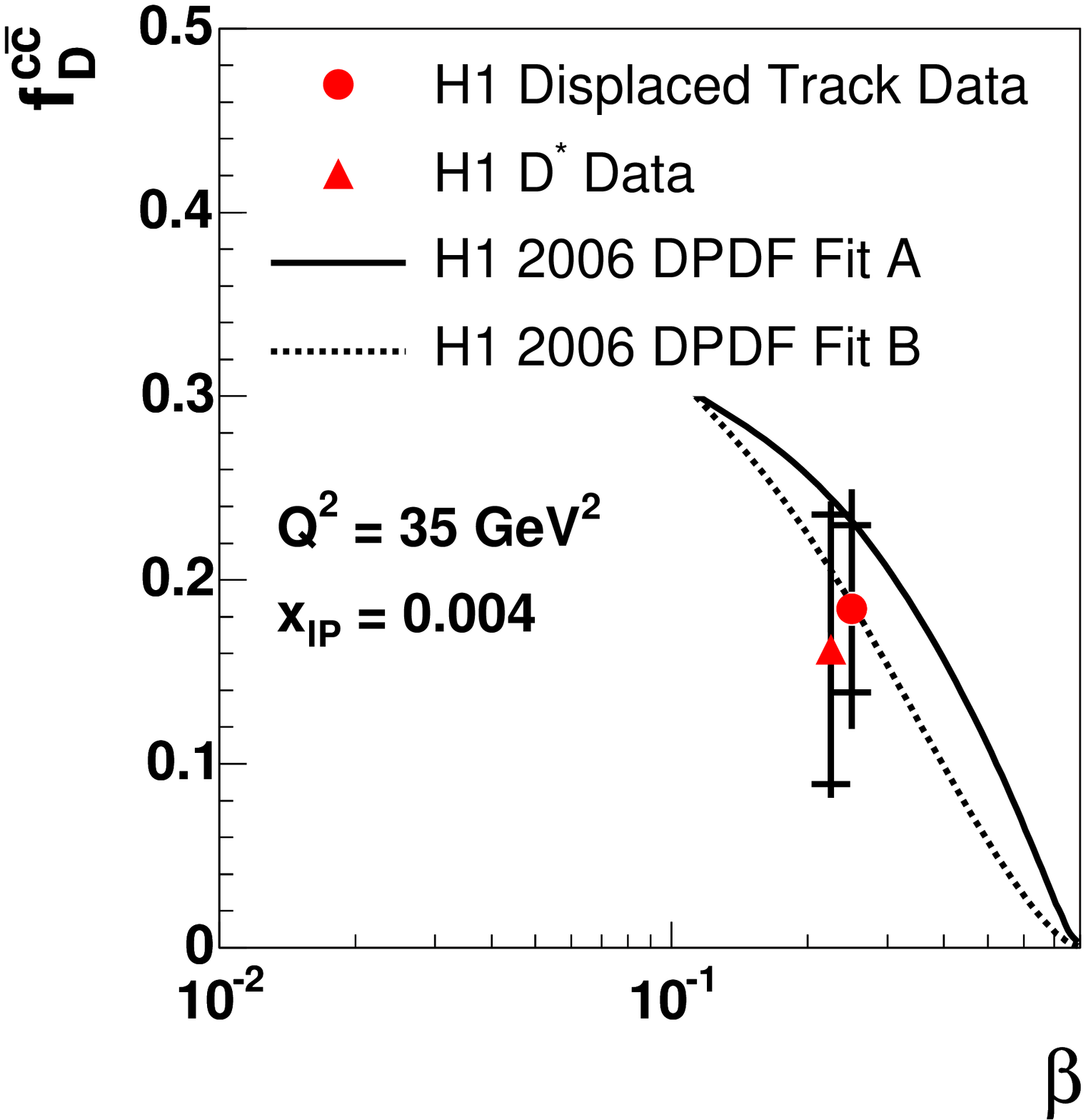}
    \includegraphics[width=.48\textwidth]{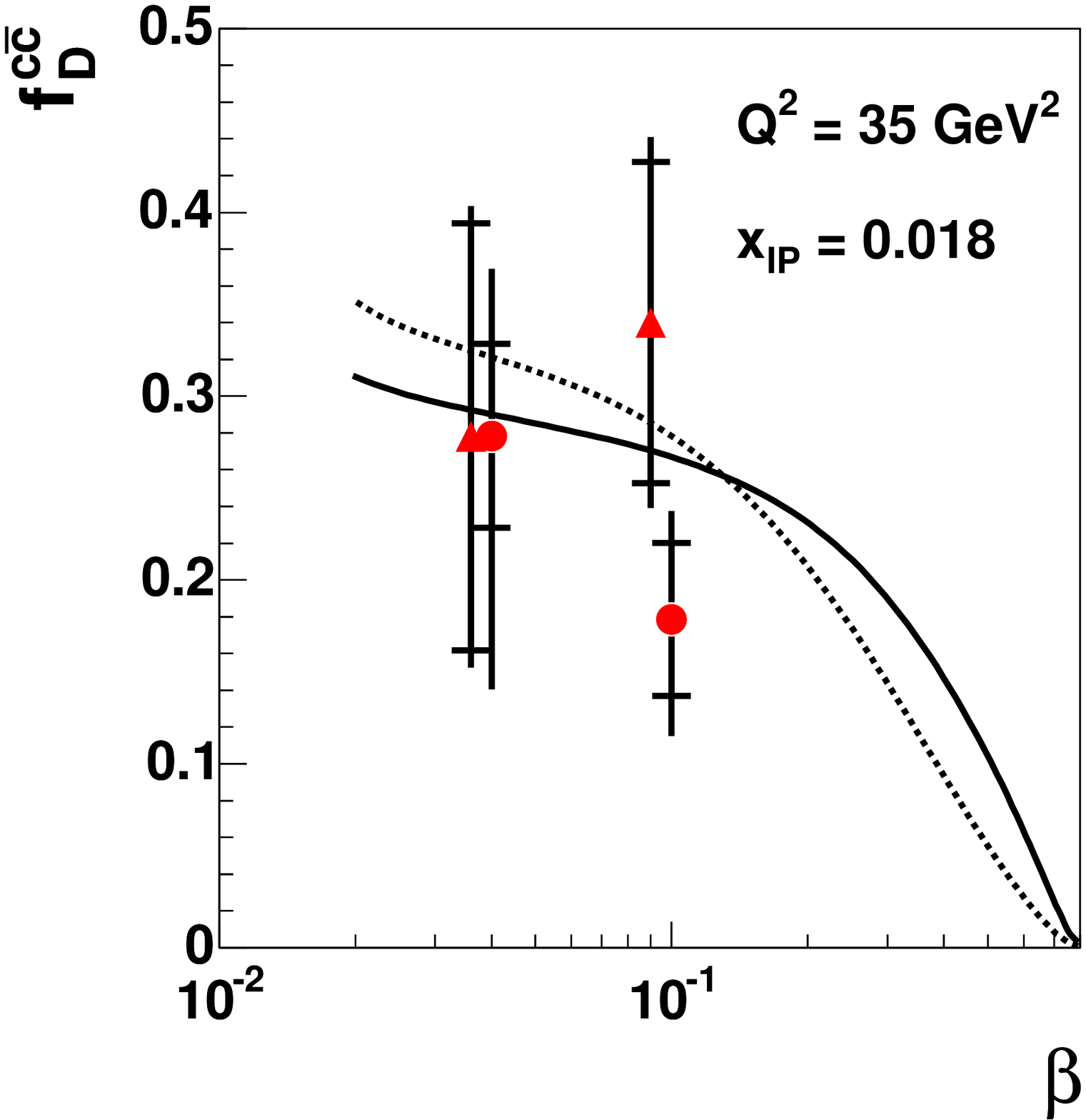}
    \setlength{\unitlength}{\textwidth}
    
    \begin{picture}(0,0)
      \begin{Large}
      \end{Large}
    \end{picture}

    \Mycaption{
      The contribution of charm quarks to the total diffractive cross section $f_D^{c\bar{c}}$ shown 
      as a function of $\beta$ for two different values of $\xpom$\@. The inner error bars 
      of the data points represent the 
      statistical uncertainties, while the outer error bars represent the 
      statistical and systematic uncertainties 
      added in quadrature. The measurements are compared with NLO predictions based 
      on two alternative 
      sets of diffractive parton density functions (Fit A and Fit B) extracted by H1~\cite{h1_diff_incl_2005}. 
      }        
    \label{fig:frac}
  \end{center}
\end{figure}

In table~\ref{tab:frac} and~\fig~\ref{fig:frac} the measurements are 
also presented in the form of the fractional contribution of charm to 
the total diffractive $ep$ cross section $f^{c\bar{c}}_D$\@. In the 
given kinematic range the value of $f^{c\bar{c}}_{D}$ is $\approx 20\%$ 
on average, which is comparable to the charm fraction in the inclusive 
cross section at low values of Bjorken $x$
for similar values of $Q^2$~\cite{h1_incl_charm_lowq2_2005}\@. 
The NLO QCD predictions shown in~\fig~\ref{fig:frac} are found to describe 
the data well.

\begin{figure}[htbp]
  \begin{center}
    \includegraphics[width=.48\textwidth]{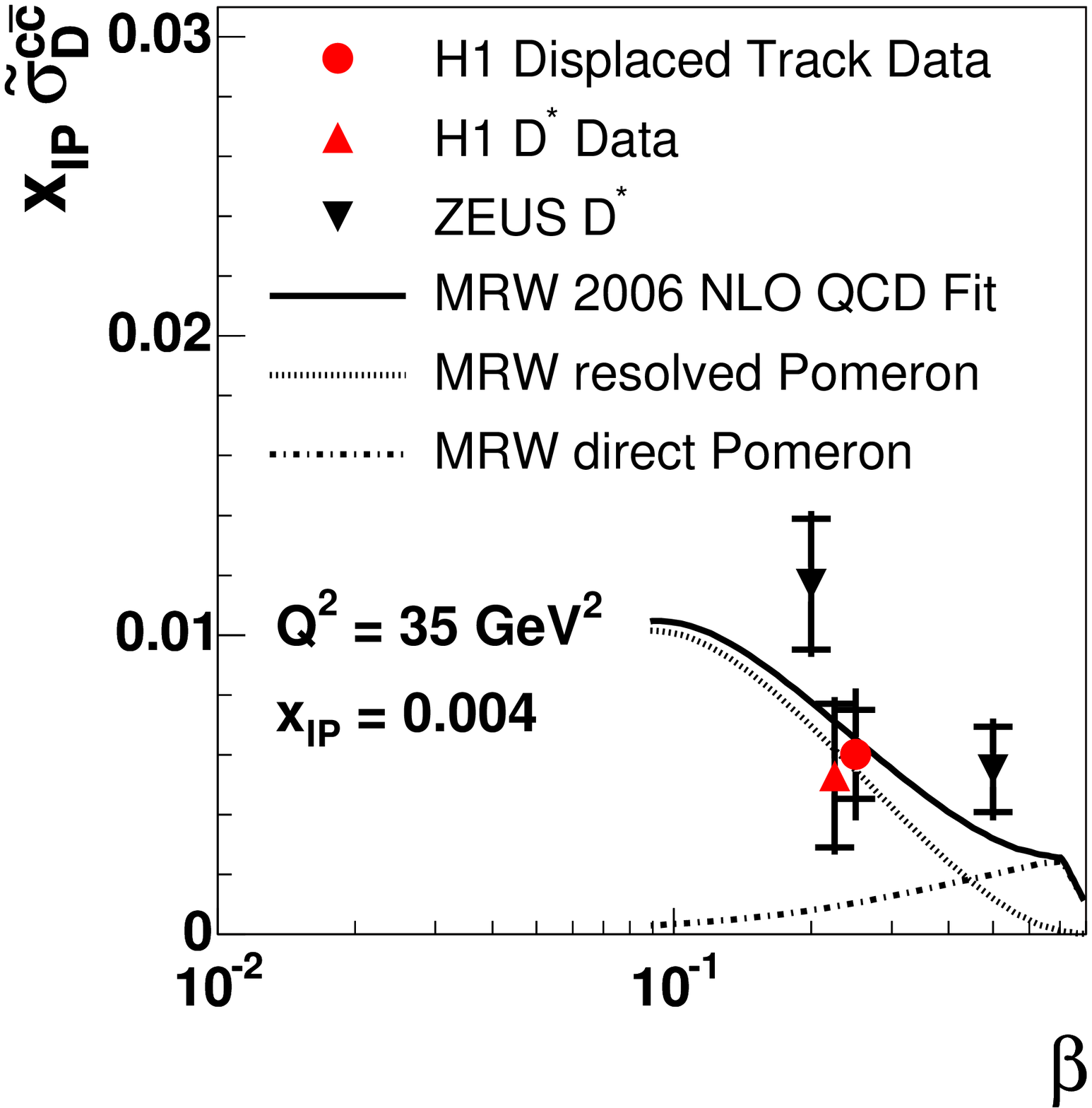}
    \includegraphics[width=.48\textwidth]{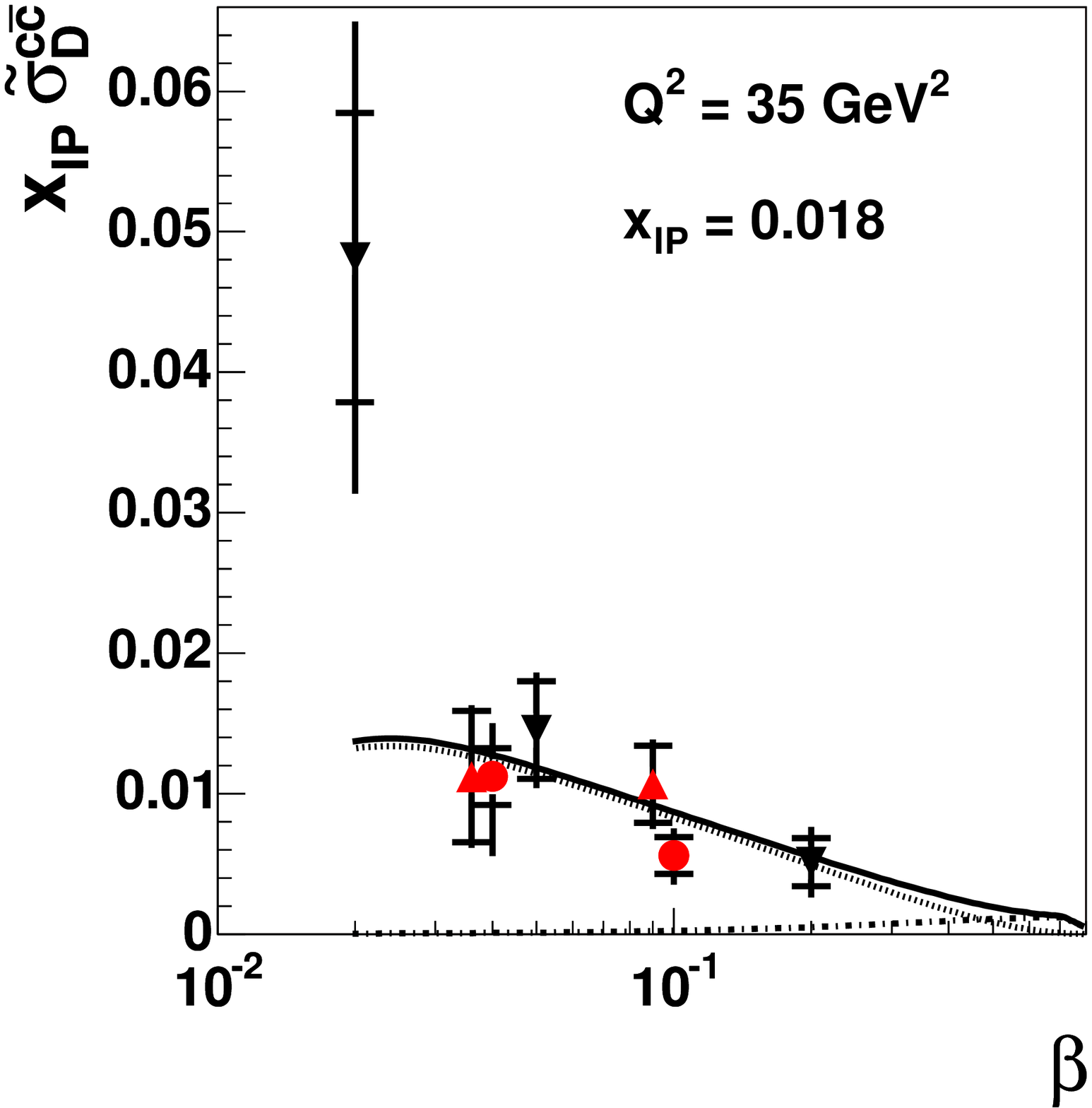}
    \setlength{\unitlength}{\textwidth}
    
    \begin{picture}(0,0)
      \begin{Large}
      \end{Large}
    \end{picture}

    \Mycaption{
      The measured reduced cross section $\xpom \, \tilde{\sigma}^{c\bar{c}}_{D}$ shown as a function 
      of $\beta$ for two different values of $\xpom$\@.  The inner error bars 
      of the data points represent the statistical 
      error, while the outer error bars represent the statistical and systematic uncertainties added in 
      quadrature. The measurements obtained from $\dstar$ mesons from H1 in this paper and from 
      ZEUS~\cite{zeus_diff_dstar_2004} are also shown. 
      The measurements are compared with the model of MRW~\cite{mrw} based on 
      perturbative two gluon exchange and DPDFs. 
      }
    \label{fig:sigmarccMRW}
  \end{center}
\end{figure}

\begin{figure}[htbp]
  \begin{center}
    \includegraphics[width=.48\textwidth]{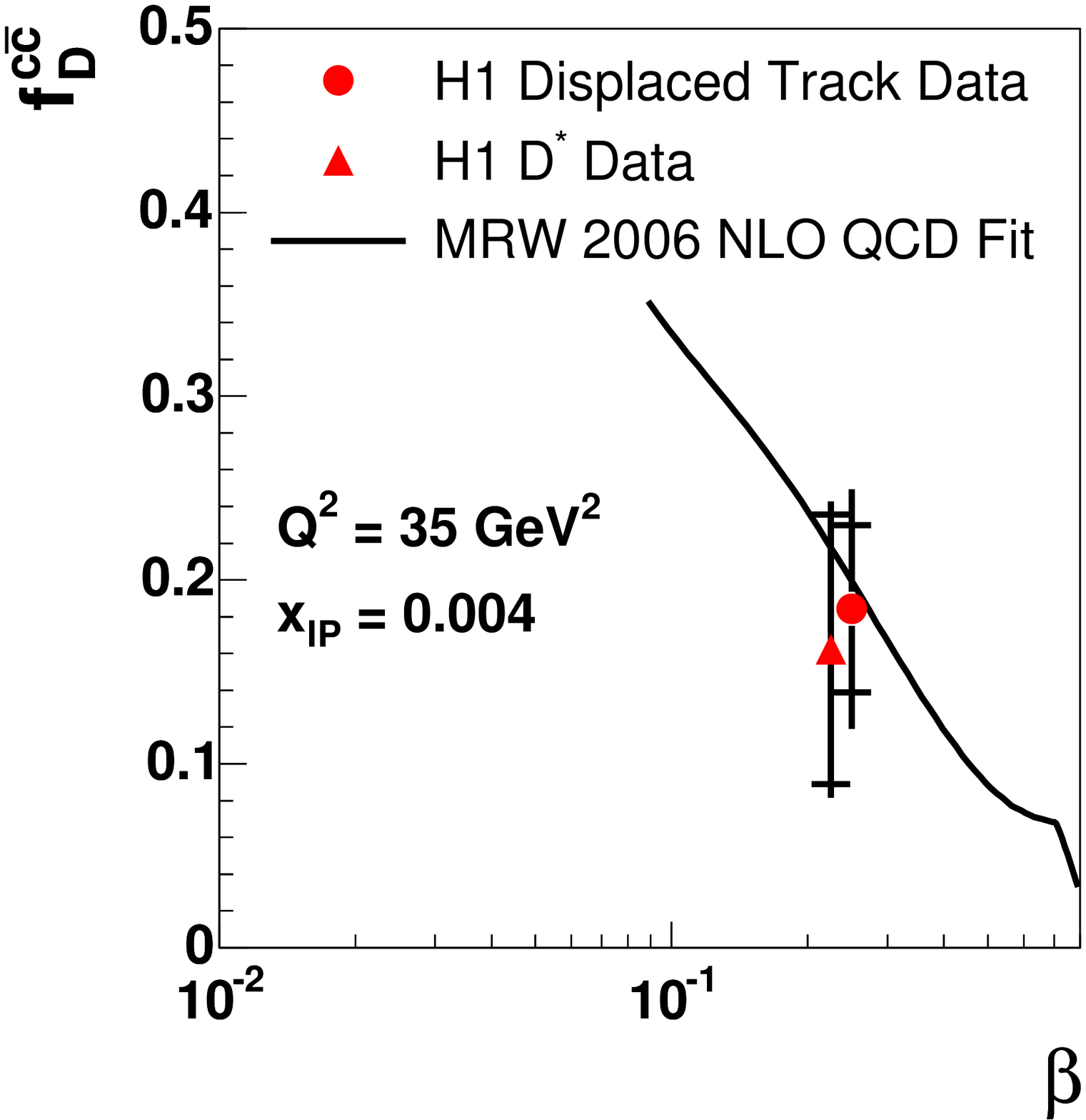}
    \includegraphics[width=.48\textwidth]{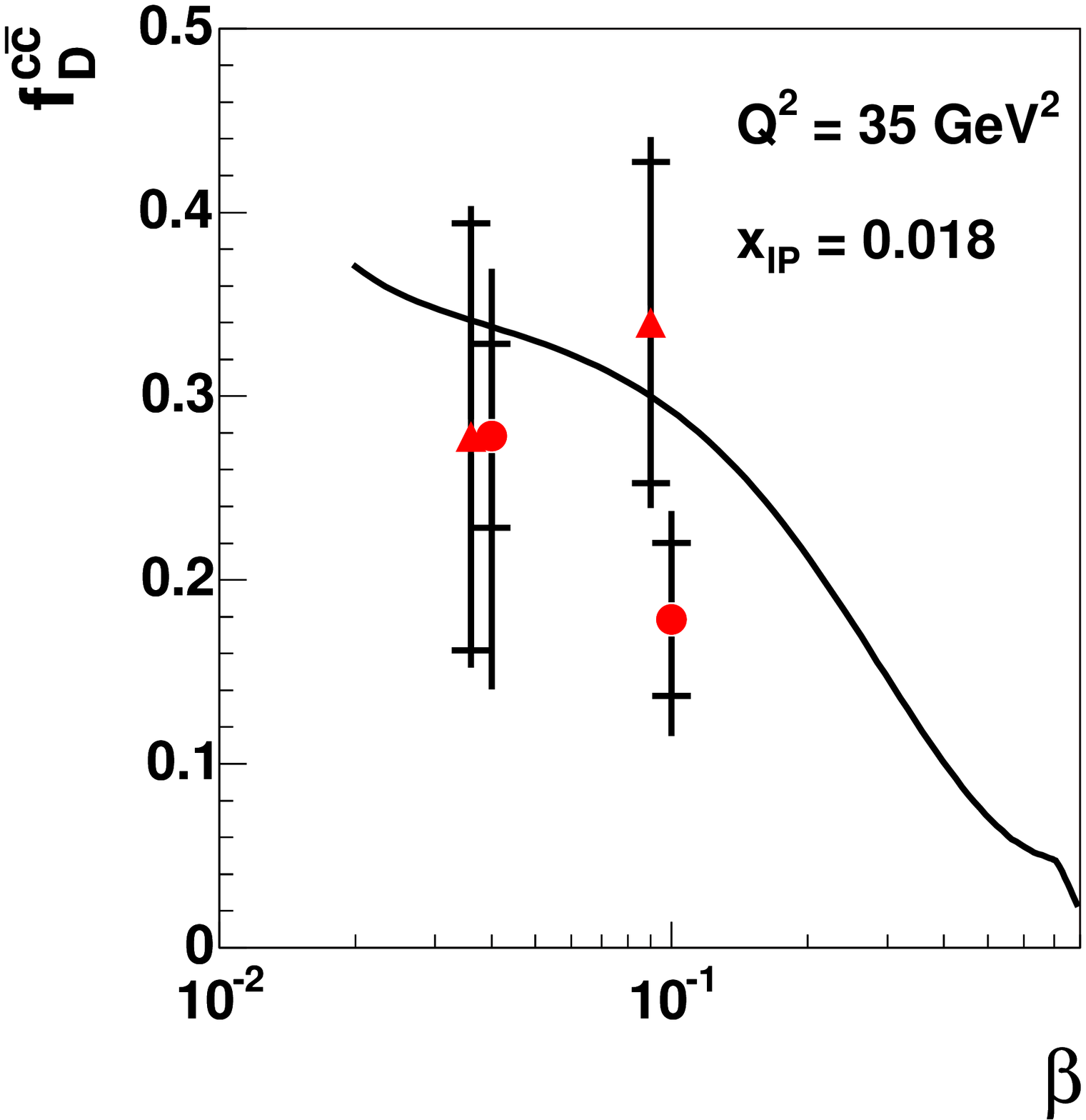}
    \setlength{\unitlength}{\textwidth}
    
    \begin{picture}(0,0)
      \begin{Large}
      \end{Large}
    \end{picture}

    \Mycaption{
      The contribution of charm quarks to the total diffractive cross section $f_D^{c\bar{c}}$ shown 
      as a function of $\beta$ for two different values of $\xpom$\@. The inner error bars 
      of the data points represent the 
      statistical uncertainties, while the outer error bars represent the 
      statistical and systematic uncertainties 
      added in quadrature. 
      The measurements are compared with 
      the model of MRW~\cite{mrw} based on 
      perturbative two gluon exchange and DPDFs. 
      }        
    \label{fig:fracMRW}
  \end{center}
\end{figure}

In Figures~\ref{fig:sigmarccMRW} and \ref{fig:fracMRW} 
the $\xpom \tilde{\sigma}^{c\bar{c}}_{D}$ and 
$f^{c\bar{c}}_{D}$ data are compared with 
the predictions of the MRW model~\cite{mrw}. 
In the kinematic range of the measurements  
the `resolved Pomeron' contribution, where charm is generated via BGF,  
is seen to dominate in the model at low $\beta$, while the `direct Pomeron' process, 
where charm is generated via `photon--Pomeron' fusion is significant at high values of 
$\beta$\@. A good description of the data is observed supporting the validity 
of the DPDFs extracted in this model.

\section{Conclusions}
\label{sec:conclusions}

Measurements are presented of the diffractive charm cross section using two 
independent methods of charm reconstruction. In the first method charm quarks 
are tagged using $\dstar$ mesons. In the second method tracks, with a significant 
displacement from the primary vertex, are reconstructed using the CST of H1. 
These displaced tracks arise due to the long lifetime of charmed hadrons.

The  diffractive $\dstar$  cross section is measured in DIS and photoproduction. 
The integrated cross section in DIS is in good agreement with a former measurement 
of H1, which was obtained from an independent dataset with less than half the 
luminosity of the present measurement. 
This is the first cross section
measurement of diffractive open charm photoproduction at \hera\@.
A comparison with QCD calculations in 
NLO based on DPDFs obtained from inclusive diffractive scattering at H1 is in 
good agreement with the measurement in both kinematic regimes. No evidence is 
observed for a suppression 
in photoproduction. In the region of $\xpom < 0.01$ the DIS $\dstar$ data are 
found to be also well described by a model based on perturbative two gluon 
exchange and $\kt$-factorization. 

The displaced track measurements are made at $Q^2=35~\GeV^{2}$ for $3$ different 
values of $\xpom$ and $\beta$\@. In this kinematic range the charm contribution 
to the inclusive diffractive cross section is found to be $\approx 20\%$ on 
average which is compatible with the charm fraction in inclusive DIS found at 
low values of Bjorken $x$ for similar values of $Q^2$\@. 
The cross sections are found to be in good agreement 
with the measurements extrapolated from the $\dstar$ cross section results and 
to be well described by the predictions of NLO QCD. At low $\xpom$, the data 
are found to be also well described by a hybrid
model based on two gluon exchange and 
diffractive parton densities.

\section*{Acknowledgements}
\label{sec:acknowladgements}
We are grateful to the HERA machine group whose outstanding efforts have 
made this experiment possible. We thank the engineers and technicians for 
their work in constructing and maintaining the H1 detector, our funding 
agencies for financial support, the DESY technical staff for continual 
assistance and the DESY directorate for support and for the hospitality 
which they extend to the non DESY members of the collaboration.

\clearpage
\bibliographystyle{mysty}
\raggedright
\bibliography{bibliography}

\clearpage

%
\begin{table}[ht]
  \begin{center}
    \begin{tabular}{|c|r@{}c@{}l|c|c|c|cc|}
      \hline
  \multicolumn{4}{|c|} {H1 99-00} & \multicolumn{2}{|c|} {N$(D^{*})$} & 
\multicolumn{3}{|c|} {Cross Section [ $\pb$ ] } \\ \cline{7-9} 
  \multicolumn{4}{|c|} {} & \multicolumn{2}{|c|} {} & Data & 
\multicolumn{2}{|c|} {H1 2006 DPDF} \\
  \multicolumn{4}{|c|} { }        & \multicolumn{2}{|c|} { }          &                         & Fit A & Fit B                    \\
      \hline \bigstrut
      DIS &0.05&$\,<y<\,$& 0.7& \multicolumn{2}{|c|}{$   122\pm 15$} & $   234\pm 29\,({\rm stat.})\pm 34\,({\rm syst.})$ & $ 287\pm^{81}_{70}$ & $ 272\pm^{78}_{71}$\\
          &0.3 &$\,<y<\,$&0.65& \multicolumn{2}{|c|}{$\,  34\pm  8$} & $\;\, 55\pm 16\,({\rm stat.})\pm \;\, 9\,({\rm syst.})$ & $ \;\,86\pm^{20}_{18}$ & $ \;\,84\pm^{20}_{18}$\\
                      &\multicolumn{3}{|c|}{ }&\multicolumn{2}{|c|}{ }& & & \\   
      ${ \gamma p}$&0.3 &$\,<y<\,$&0.65& \multicolumn{2}{|c|}{$\;\;70\pm 13$} & $\; 265\pm 50\,({\rm stat.})\pm 41\,({\rm syst.})$ & $\;360\pm^{90}_{70}$ & $\;359\pm^{93}_{75}$ \\
      \hline
    \end{tabular}
  \end{center}
  \Mycaption{
    Measured cross sections and NLO QCD predictions
    for diffractive $\dstar$ meson production in 
    the visible ranges of DIS and photoproduction (${\gamma p}$). The uncertainty on the NLO QCD 
    predictions is given by the variation of the mass, the scale and the fragmentation parameters 
    as described in the text. 
    }
  \label{tab:int_cross_section}
\end{table}
\begin{table}[ht]
  \begin{center}
    \begin{tabular}{|l|c|c|c|}
      \hline
                             & \multicolumn{3}{|c|}{ Uncertainty (\%)} \\
      Source of Uncertainty  & $\dstar $ &  $\dstar $ & Displaced  \\
                             & $(\gamma p)$ &  $({\rm DIS})$ & track $({\rm DIS})$ \\
      \hline
      \hline
      Trigger efficiency                                                                       & $5$ & $3$   & $1$  \\
      Scat. $e^{+}$ energy/angle ($1\% \oplus 1 \, {\rm mrad}$ (DIS), $1.5\%$ ($\gamma p$))    & $2$ & $2$   & $2$  \\
      Track reconstruction efficiency                                                          & $6$ & $6$   & $2$  \\
      Signal extraction method ($\dstar$)                                                      & $6$ & $6$   & $-$  \\
      Reflections ($\dstar$)                                                                   &$1.5$&$1.5$  & $-$  \\
      $\delta$ resolution ($25 \, \mu{\rm m} \, \oplus 200 \, \mu{\rm m}$)                     & $-$ & $-$   & $2$  \\
      Hadronic energy scale ($4\%$)                                                            & $1$ & $1$   & $3$  \\
      QCD model (reweights in $\xpom$, $\beta$, $Q^{2}$, $y$)                                  & $1$ & $5$   & $12-18$ \\
      Proton diss. model (reweights in $|t|$, $\my$, fraction)                                 & $4$ & $5$   & $5$  \\
      Noise in FMD and PLUG                                                                    &$1.5$&$1.5$  & $1.5$  \\
      Tagging efficiency of FMD ($10 \%$)                                                      & $1$ & $1$   & $1$  \\
      Tagging efficiency of PRT ($^{+25}_{-50} \%$)                                            & $7$ & $9$   & $9$  \\
      Non-diffractive background ($100 \%$)                                                    & $3$ & $1$   & $1$  \\
      Reggeon contribution ($100 \%$)                                                          & $4$ & $1$   & $1-9$\\
      Fragmentation of $c$ quarks                                                              & $4$ & $1$   & $7$ \\
      Branching fractions / lifetimes / track multiplicities                                   & $2.5$ &$2.5$& $3$  \\
      Asymmetry of $\delta$ for light quarks ($\pm 50\%$)                                          & $-$ & $-$   &$4-16$\\
      Beauty fraction ($^{+400}_{-100}\%$)                                                    & $-$ & $-$   &$0-40$\\
      Quark axis ($2~\grad / 5~\grad$)                                                         & $-$ & $-$   & $3$  \\
      Luminosity                                                                               &$1.5$& $1.5$ & $1.5$\\
      QED correction                                                                           &$-$& $2$ & $2$\\
      Bin center correction                                                                    &$-$& $-$ & $8-10$\\
      \hline
      \hline
      Total                                                                                    &$15$ & $15$ &$26-47$\\
      \hline
    \end{tabular}
  \end{center}
  \Mycaption{
    Systematic uncertainties for the measurement of diffractive open charm production for the 
    inclusive cross section in the visible range for the reconstruction of $\dstarpm$ mesons 
    in DIS and photoproduction and in the differential bins for the displaced track method in 
    DIS.
    }
  \label{tab:sys_uncertainties}
\end{table}
\begin{table}[ht]
  \begin{center}
    \begin{tabular}{|ccc|c|c|c|c|}

      \hline
      \multicolumn{7}{|c|}{DIS $\dstarpm$ meson cross section as a function of $\ptds$} \\
      \hline
       \multicolumn{3}{|c|}{Range (\GeV)} & Bin Center ($\GeV$) & $d\sigma / d\ptds$ ($\pb/\GeV$) & $\delta_{stat}$ (\%) & $\delta_{syst}$ (\%) \\
      \hline 
      $2.0$ & $-$ & $ 2.5$ & $2.20$ & $169$ & $25$ & $16$ \\ 
      $2.5$ & $-$ & $ 3.0$ & $2.75$ & $114$ & $25$ & $15$ \\ 
      $3.0$ & $-$ & $ 3.6$ & $3.35$ & $ 76$ & $24$ & $15$ \\ 
      $3.6$ & $-$ & $10.0$ & $5.45$ & $  8$ & $23$ & $15$ \\ 
      \hline

      \hline
      \multicolumn{7}{|c|}{DIS $\dstarpm$ meson cross section as a function of $\etads$} \\
      \hline
      \multicolumn{3}{|c|}{Range} & Bin Center & $d\sigma / d\etads$ ($\pb$) & $\delta_{stat}$ (\%) & $\delta_{syst}$ (\%) \\
      \hline 
      $ -1.5$ & $-$ & $-0.75$ & $-1.17$ & $ 92$ & $22$ & $15$ \\ 
      $-0.75$ & $-$ & $ 0   $ & $-0.33$ & $101$ & $21$ & $15$ \\ 
      $ 0   $ & $-$ & $ 0.75$ & $ 0.42$ & $ 81$ & $25$ & $15$ \\ 
      $ 0.75$ & $-$ & $ 1.5 $ & $ 1.12$ & $ 40$ & $37$ & $21$ \\ 
      \hline 

      \hline
      \multicolumn{7}{|c|}{DIS $\dstarpm$ meson cross section as a function of $y$} \\
      \hline
      \multicolumn{3}{|c|}{Range} & Bin Center & $d\sigma / dy$ ($\pb$) & $\delta_{stat}$ (\%) & $\delta_{syst}$ (\%) \\
      \hline 
      $0.05$ & $-$ & $0.15$ & $0.09$ & $ 975$ & $18$ & $15$ \\ 
      $0.15$ & $-$ & $0.30$ & $0.22$ & $ 423$ & $25$ & $15$ \\ 
      $0.30$ & $-$ & $0.45$ & $0.38$ & $ 198$ & $37$ & $16$ \\ 
      $0.45$ & $-$ & $0.70$ & $0.55$ & $ 141$ & $37$ & $18$ \\ 
      \hline 

      \hline
      \multicolumn{7}{|c|}{DIS $\dstarpm$ meson cross section as a function of $Q^{2}$} \\
      \hline
      \multicolumn{3}{|c|}{Range ($\GeV^{2}$)} & Bin Center (GeV$^{2}$) & $d\sigma / dQ^2$ ($\pb/\GeV^2$) & $\delta_{stat}$ (\%) & $\delta_{syst}$ (\%) \\
      \hline 
      $ 2.0$ & $-$ & $  5.0$ & $ 4.0$ & $ 17$ & $27$ & $17$ \\ 
      $ 5.0$ & $-$ & $ 15.0$ & $ 9.5$ & $7.6$ & $21$ & $15$ \\ 
      $15.0$ & $-$ & $ 35.0$ & $23.5$ & $3.6$ & $21$ & $14$ \\ 
      $35.0$ & $-$ & $100.0$ & $60.5$ & $0.6$ & $31$ & $14$ \\ 
      \hline

      \hline
      \multicolumn{7}{|c|}{DIS $\dstarpm$ meson cross section as a function of $\log(\xpom)$} \\
      \hline
      \multicolumn{3}{|c|}{Range} & Bin Center & $d\sigma / d\log(\xpom)$ ($\pb$) & $\delta_{stat}$ (\%) & $\delta_{syst}$ (\%) \\
      \hline 
      $-3.0$ & $-$ & $-2.6$ & $-2.79$ & $ 36$ & $39$ & $30$ \\ 
      $-2.6$ & $-$ & $-2.2$ & $-2.39$ & $118$ & $22$ & $21$ \\ 
      $-2.2$ & $-$ & $-1.8$ & $-2.01$ & $138$ & $25$ & $15$ \\ 
      $-1.8$ & $-$ & $-1.4$ & $-1.55$ & $275$ & $21$ & $17$ \\ 
      \hline 

      \hline
      \multicolumn{7}{|c|}{DIS $\dstarpm$ meson cross section as a function of $\zpomobs$} \\
      \hline
      \multicolumn{3}{|c|}{Range} & Bin Center & $d\sigma / d\zpomobs$ ($\pb$) & $\delta_{stat}$ (\%) & $\delta_{syst}$ (\%) \\
      \hline 
      $ 0  $ & $-$ & $0.15$ & $0.07$ & $312$ & $44$ & $19$ \\ 
      $0.15$ & $-$ & $0.45$ & $0.29$ & $325$ & $19$ & $15$ \\ 
      $0.45$ & $-$ & $1   $ & $0.69$ & $ 99$ & $17$ & $30$ \\ 
      \hline 

      \hline
      \multicolumn{7}{|c|}{DIS $\dstarpm$ meson cross section as a function of $\log(\beta)$} \\
      \hline
      \multicolumn{3}{|c|}{Range} & Bin Center & $d\sigma / d\log(\beta)$ ($\pb$) & $\delta_{stat}$ (\%) & $\delta_{syst}$ (\%) \\
      \hline 
      $-2.5$ & $-$ & $-1.8$ & $-2.12$ & $ 55$ & $40$ & $19$ \\ 
      $-1.8$ & $-$ & $-1.2$ & $-1.57$ & $120$ & $23$ & $14$ \\ 
      $-1.2$ & $-$ & $-0.6$ & $-0.88$ & $123$ & $18$ & $16$ \\ 
      $-0.6$ & $-$ & $ 0  $ & $-0.28$ & $ 50$ & $27$ & $21$ \\ 
      \hline 

    \end{tabular}
  \end{center}
  \Mycaption{
    Differential cross sections for diffractive $\dstarpm$ meson production in DIS, as a function of $\ptds$, 
    $\etads$, $y$, $Q^{2}$, $\xpom$, $\zpomobs$ and $\beta$, given in the range of
    $2 < Q^{2} < 100~\GeV^{2}$, $0.05 < y < 0.7$, $\xpom < 0.04$, $\my < 1.6~\GeV$, 
    $|t| < 1~\GeV^{2}$, $\ptds > 2~\GeV$ \ and $|\etads| < 1.5$\@. 
    }
  \label{tab:xsec_dis_0}
\end{table}
\begin{table}[ht]
  \begin{center}
    \begin{tabular}{|ccc|c|c|c|c|}

      \hline
      \multicolumn{7}{|c|}{$\gamma p$ $\dstarpm$ meson cross section as a function of $\ptds$} \\
      \hline
      \multicolumn{3}{|c|}{Range ($\GeV$)} & Bin Center ($\GeV$) & $d\sigma / d\ptds$ ($\pb/\GeV$) & $\delta_{stat}$ (\%) & $\delta_{syst}$ (\%) \\
      \hline 
      $2.0$ & $-$ & $ 2.6$ & $2.25$ & $160$  & $39$ & $16$ \\ 
      $2.6$ & $-$ & $ 3.2$ & $2.95$ & $172$  & $26$ & $14$ \\ 
      $3.2$ & $-$ & $10.0$ & $4.95$ & $ 10$ & $29$ & $17$ \\ 
      \hline

      \hline
      \multicolumn{7}{|c|}{$\gamma p$ $\dstarpm$ meson cross section as a function of $\etads$} \\
      \hline
      \multicolumn{3}{|c|}{Range} & Bin Center & $d\sigma / d\etads$ ($\pb$) & $\delta_{stat}$ (\%) & $\delta_{syst}$ (\%) \\
      \hline 
      $ -1.5$ & $-$ & $-0.65$ & $-1.05$ & $112$ & $28$ & $15$ \\ 
      $-0.65$ & $-$ & $ 0.20$ & $-0.28$ & $169$ & $25$ & $15$ \\ 
      $ 0.20$ & $-$ & $ 1.50$ & $ 0.82$ & $ 27$ & $67$ & $23$ \\ 
      \hline 

      \hline
      \multicolumn{7}{|c|}{$\gamma p$ $\dstarpm$ meson cross section as a function of $y$} \\
      \hline
      \multicolumn{3}{|c|}{Range} & Bin Center & $d\sigma / dy$ ($\pb$) & $\delta_{stat}$ (\%) & $\delta_{syst}$ (\%) \\
      \hline 
      $0.30$ & $-$ & $0.40$ & $0.35$ & $1010$ & $34$ & $16$ \\ 
      $0.40$ & $-$ & $0.50$ & $0.45$ & $ 785$ & $29$ & $16$ \\ 
      $0.50$ & $-$ & $0.65$ & $0.57$ & $ 555$ & $36$ & $20$ \\ 
      \hline 

      \hline
      \multicolumn{7}{|c|}{$\gamma p$ $\dstarpm$ meson cross section as a function of $\log(\xpom)$} \\
      \hline
      \multicolumn{3}{|c|}{Range} & Bin Center & $d\sigma / d\log(\xpom)$ ($\pb$) & $\delta_{stat}$ (\%) & $\delta_{syst}$ (\%) \\
      \hline 
      $-3.0$ & $-$ & $-2.2$ & $-2.59$ & $ 77$ & $30$ & $19$ \\ 
      $-2.2$ & $-$ & $-1.8$ & $-2.01$ & $266$ & $27$ & $15$ \\ 
      $-1.8$ & $-$ & $-1.4$ & $-1.61$ & $214$ & $42$ & $17$ \\ 
      \hline 

      \hline
      \multicolumn{7}{|c|}{$\gamma p$ $\dstarpm$ meson cross section as a function of $\zpomobs$} \\
      \hline
      \multicolumn{3}{|c|}{Range} & Bin Center & $d\sigma / d\zpomobs$ ($\pb$) & $\delta_{stat}$ (\%) & $\delta_{syst}$ (\%) \\
      \hline 
      $ 0  $ & $-$ & $0.15$ & $0.06$ & $680$ & $36$ & $16$ \\ 
      $0.15$ & $-$ & $0.45$ & $0.28$ & $400$ & $26$ & $15$ \\ 
      $0.45$ & $-$ & $1   $ & $0.70$ & $ 51$ & $37$ & $48$ \\ 
      \hline 

    \end{tabular}
  \end{center}
  \Mycaption{
    Differential cross sections for diffractive $\dstarpm$ meson production in $\gamma p$, as a
    function of $\ptds$, $\etads$, $y$, $\xpom$ and $\zpomobs$, given in the range of 
    $Q^{2} < 0.01~\GeV^{2}$, $0.3 < y < 0.65$, $\xpom < 0.04$, $\my < 1.6~\GeV$, $|t| < 1~\GeV^{2}$, 
    $\ptds > 2~\GeV$ \ and $|\etads| < 1.5$\@.
    }
  \label{tab:xsec_php_0}
\end{table}
\begin{table}[ht]
  \begin{center}
    \begin{tabular}{|ccc|c|c|c|c|}

      \hline
      \multicolumn{7}{|c|}{DIS $\dstarpm$ meson cross section as a function of $\ptds$} \\
      \hline
      \multicolumn{3}{|c|}{Range (\GeV)} & Bin Center ($\GeV$) & $d\sigma / d\ptds$ ($\pb/\GeV$) & $\delta_{stat}$ (\%) & $\delta_{syst}$ (\%) \\
      \hline 
      $2.0$ & $-$ & $ 3.0$ & $2.45$ & $ 58$ & $23$ & $18$ \\ 
      $3.0$ & $-$ & $10.0$ & $4.95$ & $  5$ & $22$ & $18$ \\ 
      \hline

      \hline
      \multicolumn{7}{|c|}{DIS $\dstarpm$ meson cross section as a function of $\etads$} \\
      \hline
      \multicolumn{3}{|c|}{Range} & Bin Center & $d\sigma / d\etads$ ($\pb$) & $\delta_{stat}$ (\%) & $\delta_{syst}$ (\%) \\
      \hline 
      $ -1.5$ & $-$ & $ 0.$ & $-0.65$ & $48$ & $18$ & $17$ \\ 
      $   0.$ & $-$ & $1.5$ & $-0.33$ & $16$ & $34$ & $23$ \\ 
      \hline 

      \hline
      \multicolumn{7}{|c|}{DIS $\dstarpm$ meson cross section as a function of $\log(\xpom)$} \\
      \hline
      \multicolumn{3}{|c|}{Range} & Bin Center & $d\sigma / d\log(\xpom)$ ($\pb$) & $\delta_{stat}$ (\%) & $\delta_{syst}$ (\%) \\
      \hline 
      $-3.0$ & $-$ & $-2.5$ & $-2.71$ & $ 43$ & $32$ & $30$ \\ 
      $-2.5$ & $-$ & $-2.0$ & $-2.25$ & $143$ & $19$ & $17$ \\ 
      \hline 

      \hline
      \multicolumn{7}{|c|}{DIS $\dstarpm$ meson cross section as a function of $\zpomobs$} \\
      \hline
      \multicolumn{3}{|c|}{Range} & Bin Center & $d\sigma / d\zpomobs$ ($\pb$) & $\delta_{stat}$ (\%) & $\delta_{syst}$ (\%) \\
      \hline 
      $ 0 $ & $-$ & $0.5$ & $0.27$ & $ 74$ & $31$ & $16$ \\ 
      $0.5$ & $-$ & $ 1.$ & $0.69$ & $ 91$ & $19$ & $42$ \\ 
      \hline 

    \end{tabular}
  \end{center}
  \Mycaption{
    Differential cross sections for diffractive $\dstarpm$ meson production in DIS, in the same 
    kinematic region as that given 
    in table~\ref{tab:xsec_dis_0} but further restricted to $\xpom<0.01$\@. 
    }
  \label{tab:xsec_dis_lowxpom}
\end{table}
\begin{table}[ht]
  \begin{center}

    \begin{tabular}{|c|c|c|c|c|c|c|}

      \hline
      \multicolumn{7}{|c|}{Reduced Cross Section
$\tilde{\sigma}^{c\bar{c}}_{D}(\xpom,\beta,Q^2)$ } \\
      \hline
      \multicolumn{6}{|c|}{Displaced track } & $\dstar$ \\
      \hline \bigstrut
      $Q^2 (\GeV^2)$ & $\xpom$ & $\beta$ & $\tilde{\sigma}^{c\bar{c}}_{D}$ & $\delta_{stat}$ (\%)  & $\delta_{sys}$
 (\%) & $\tilde{\sigma}^{c\bar{c}}_{D}$ \\
      \hline  \bigstrut 
      \hspace*{5.75mm} $35$ & $0.004$ & $0.25$ & $1.50$ & $25$ & $^{+27}_{-27}$ & $1.33$ \\
      $35$ & $0.010$ & $0.10$ & $0.63$ & $23$ & $^{+26}_{-29}$ & $1.20$  \\
      $35$ & $0.018$ & $0.04$ & $0.62$ & $18$ & $^{+29}_{-47}$ & $0.62$ \\
      \hline
     \end{tabular}
  \end{center}
  \Mycaption{
    The reduced cross section $\tilde{\sigma}^{c\bar{c}}_{D}(\xpom,\beta,Q^2)$ obtained from the 
    displaced track method. The last column shows the results obtained by extrapolating the $\dstar$ 
    cross sections in table~\ref{tab:sigmamx} using the H1 NLO QCD fit.
    }
  \label{tab:sigmarcc}
\end{table}

\begin{table}[ht]
  \begin{center}
    \begin{tabular}{|ccc|c|c|c|}

      \hline
      \multicolumn{6}{|c|}{DIS $\dstarpm$ meson cross section as a function
of $\mx$} \\
      \hline
      \multicolumn{3}{|c|}{Range ($\GeV$)} & $d\sigma / d\mx$ ($\pb/\GeV$) & $\delta_{stat}$ (\%) &  $\delta_{sys}$ (\%) \\
      \hline
      $ 6$ & $-$ & $12$ & $ 2.5$ & $45$ & $20$  \\
      $12$ & $-$ & $20$ & $ 5.0$ & $26$ & $15$  \\
      $20$ & $-$ & $99$ & $0.39$ & $42$ & $17$  \\
      \hline
     \end{tabular}
  \end{center}
  \Mycaption{
    The differential cross section for diffractive  $D^*$ production in DIS as a function of $\mx$ measured 
    in the range $15 < Q^2 < 100 \ {\GeV^2}$, $0.07 < y < 0.7$, $\xpom < 0.04$, $\my < 1.6~\GeV$, 
    $|t| < 1~\GeV^{2}$, $\ptds > 2.0 \ {\GeV}$ and $|\etads| < 1.5$\@.
    }
  \label{tab:sigmamx}
\end{table}
\begin{table}[ht]
  \begin{center}
    \begin{tabular}{|c|c|c|c|c|c|c|}

      \hline
      \multicolumn{7}{|c|}{Fractional charm contribution $f^{c\bar{c}}_{D}$ } \\
      \hline
      \multicolumn{6}{|c|}{Displaced track } & $\dstar$ \\
      \hline \bigstrut
      $Q^2 (\GeV^2)$ & $\xpom$ & $\beta$ & $f^{c\bar{c}}_{D}$ & $\delta_{stat}$ (\%) & $\delta_{syst}$ (\%) & $f^{c\bar{c}}_{D}$ \\
      \hline \bigstrut
    \hspace*{5.75mm}  $35$ & $0.004$ & $0.25$ & $0.184$ & $25$ & $^{+25}_{-25}$ & $0.162$ \\
      $35$ & $0.010$ & $0.10$ & $0.193$ & $23$ & $^{+23}_{-27}$ & $0.367$ \\
      $35$ & $0.018$ & $0.04$ & $0.278$ & $18$ & $^{+27}_{-46}$ & $0.278$ \\
      \hline

     \end{tabular}
  \end{center}
  \Mycaption{
The fractional charm contribution to the diffractive cross section 
$f^{c\bar{c}}_{D}$ 
obtained from the displaced track method.
The last column shows the results obtained by extrapolating the $\dstar$ 
cross sections in table~\ref{tab:sigmamx} using the H1 NLO QCD fit
and dividing by the measured total diffractive cross section.
    }
  \label{tab:frac}
\end{table}

\clearpage

%
%


%
%


%
%


%
%


%
%


%
%




%
%


%
%


%
%


%

%
%


%
%


\end{document}

%% file: h1auts.tex

A.~Aktas$^{10}$,               
V.~Andreev$^{25}$,             
T.~Anthonis$^{4}$,             
B.~Antunovic$^{26}$,           
S.~Aplin$^{10}$,               
A.~Asmone$^{33}$,              
A.~Astvatsatourov$^{4}$,       
A.~Babaev$^{24, \dagger}$,     
S.~Backovic$^{30}$,            
A.~Baghdasaryan$^{38}$
% ...

\bigskip{\it
% ...
$ ^{4}$ Inter-University Institute for High Energies ULB-VUB, Brussels;
          Universiteit Antwerpen, Antwerpen; Belgium$^{ c}$ \\
% ...
$ ^{10}$ DESY, Hamburg, Germany \\
% ...
$ ^{24}$ Institute for Theoretical and Experimental Physics,
          Moscow, Russia$^{ k}$ \\
 $ ^{25}$ Lebedev Physical Institute, Moscow, Russia$^{ e}$ \\
 $ ^{26}$ Max-Planck-Institut f\"ur Physik, M\"unchen, Germany \\
% ...
$ ^{30}$ Faculty of Science, University of Montenegro,
          Podgorica, Montenegro$^{ e}$ \\
% ...
$ ^{33}$ Dipartimento di Fisica Universit\`a di Roma Tre
          and INFN Roma~3, Roma, Italy \\
% ...
$ ^{38}$ Yerevan Physics Institute, Yerevan, Armenia \\
% ...
$ ^{\dagger}$ Deceased \\

\bigskip
% ...
$ ^c$ Supported by FNRS-FWO-Vlaanderen, IISN-IIKW and IWT
      and  by Interuniversity
Attraction Poles Programme,
      Belgian Science Policy \\
% ...
$ ^e$ Supported by the Deutsche Forschungsgemeinschaft \\
% ...
$ ^k$ Partially Supported by Russian Foundation
      for Basic Research,  grants  03-02-17291
      and  04-02-16445 \\
% ...
}

%% file: desy-06-164.bbl
\begin{thebibliography}{10}

\bibitem{collins_1998}
J.~C. Collins, {\em Phys. Rev. {\bf D}} {\bf 57}\nolinebreak
  [2]\,(1998)\nolinebreak [2]\,3051 [\mbox{Erratum}-ibid. {\bf D 61} (2000)
  019902] [hep-ph/9709499].

\bibitem{h1_diff_incl_1997}
C.~Adloff et~al., [\mbox{H1} Collaboration], {\em Z. Phys. {\bf C}} {\bf
  76}\nolinebreak [2]\,(1997)\nolinebreak [2]\,613.

\bibitem{h1_diff_incl_2005}
\mbox{H1} Collaboration, {\it ``Measurement and QCD analysis of the diffractive
  deep-inelastic scattering cross section at HERA''}, \mbox{DESY} 06-049,
  accepted by {\em Eur. Phys. J. {\bf C}} [hep-ex/0606004].

\bibitem{resolved_pomeron}
G.~Ingelman, and P.~Schlein, {\em Phys. Lett. {\bf B}} {\bf
  152}\nolinebreak [2]\,(1985)\nolinebreak [2]\,256;
\\
A.~Donnachie and P.~Landshoff, {\em Phys. Lett. {\bf B}} {\bf
  191}\nolinebreak [2]\,(1987)\nolinebreak [2]\,309
[Erratum-ibid.\ B {\bf 198} (1987) 590].


\bibitem{dglap}
V.~Gribov and L.~Lipatov, {\em Sov. J. Nucl. Phys.} {\bf 15}\nolinebreak
  [2]\,(1972)\nolinebreak [2]\,438 and 675;
\\
Y.~Dokshitzer, {\em Sov. Phys. JETP} {\bf 45}\nolinebreak
  [2]\,(1977)\nolinebreak [2]\,641;
\\
G.~Altarelli, G.~Parisi, {\em Nucl. Phys. {\bf B}} {\bf 126}\nolinebreak
  [2]\,(1977)\nolinebreak [2]\,298.

\bibitem{h1_diff_dijet}
C.~Adloff et~al., [\mbox{H1} Collaboration], {\em Eur. Phys. J. {\bf C}} {\bf
  6}\nolinebreak [2]\,(1999)\nolinebreak [2]\,421 [hep-ex/9808013];
\\
C.~Adloff et~al., [\mbox{H1} Collaboration], {\em Eur. Phys. J. {\bf C}} {\bf
  20}\nolinebreak [2]\,(2001)\nolinebreak [2]\,29 [hep-ex/0012051];
\\
S.~Sch\"atzel, \mbox{PhD thesis, Univ. Heidelberg (2004)}, available
  from http://www-h1.desy.de/publications/theses\_list.html.

\bibitem{cdf_diff_dijets_2000}
T. Affolder~et al., [\mbox{CDF} Collaboration], {\em Phys. Rev. Lett.} {\bf
  84}\nolinebreak [2]\,(2000)\nolinebreak [2]\,5043.

\bibitem{bjklw}
J.~Bartels, H.~Lotter, M.~W\"usthoff, {\em Phys. Lett. {\bf B}} {\bf
  379}\nolinebreak [2]\,(1996)\nolinebreak [2]\,239 [Erratum-ibid.\ B {\bf
  382} (1996) 449] [hep-ph/9602363];
\\
J.~Bartels, C.~Ewerz, H.~Lotter, M.~W\"usthoff, {\em Phys. Lett. {\bf B}} {\bf
  386}\nolinebreak [2]\,(1996)\nolinebreak [2]\,389 [hep-ph/9605356];
\\
J.~Bartels, H.~Jung, M.~W\"usthoff, {\em Eur. Phys. J. {\bf C}} {\bf 11}\nolinebreak
  [2]\,(1999)\nolinebreak [2]\,111 [hep-ph/9903265];
\\
J.~Bartels, H.~Jung, A.~Kyrieleis, {\em Eur. Phys. J. {\bf C}} {\bf
  24}\nolinebreak [2]\,(2002)\nolinebreak [2]\,555 [hep-ph/0204269].


\bibitem{jung_dis03}
M.~Hansson, H.~Jung, {\em ``Status of CCFM: Un-integrated gluon densities''},
  \mbox{P}roceedings of 11$^{th}$ International Workshop on Deep Inelastic
  Scattering (DIS 2003), St. Petersburg, Russia, April 2003, p 488. Edited by
  V.T. Kim and L.N. Lipatov [hep-ph/0309009].

\bibitem{mrw}
A.~D. Martin, M.~G. Ryskin, G.~Watt, {\em Eur. Phys. J. {\bf C}} {\bf
  44}\nolinebreak [2]\,(2005)\nolinebreak [2]\,69 [hep-ph/0504132];
\\
G.~Watt, Proc. of the Workshop on New Trends in HERA Physics,
ed. G.~Grindhammer et al., Ringberg Castle, 
Germany (2005)\nolinebreak [2]\, 303 [hep-ph/0511333];
\\
A.~D.~Martin, M.~G.~Ryskin,  G.~Watt,
{\it ``Diffractive parton distributions from H1 data''},
  hep-ph/0609273.


\bibitem{h1_diff_dstar_2001}
C.~Adloff et~al., [\mbox{H1} Collaboration], {\em Phys. Lett. {\bf B}} {\bf
  520}\nolinebreak [2]\,(2001)\nolinebreak [2]\,191 [hep-ex/0108047].

\bibitem{zeus_diff_dstar_2004}
S.~Chekanov et~al., [\mbox{ZEUS} Collaboration], {\em Nucl. Phys. {\bf B}} {\bf
  672}\nolinebreak [2]\,(2003)\nolinebreak [2]\,3 [hep-ex/0307068].

\bibitem{h1_incl_charm_highq2_2004}
A.~Aktas et~al., [\mbox{H1} Collaboration], {\em Eur. Phys. J. {\bf C}} {\bf
  40}\nolinebreak [2]\,(2005)\nolinebreak [2]\,349 [hep-ex/0411046].

\bibitem{h1_incl_charm_lowq2_2005}
A.~Aktas et~al., [\mbox{H1} Collaboration], {\em Eur. Phys. J. {\bf C}} {\bf
  45}\nolinebreak [2]\,(2006)\nolinebreak [2]\,23 [hep-ex/0507081].


\bibitem{h1_detector_1}
I.~Abt et~al., [\mbox{H1} Collaboration], {\em Nucl. Instrum. Meth.} {\bf
  A386}\nolinebreak [2]\,(1997)\nolinebreak [2]\,310 and 348.

\bibitem{H1_cst}
D.~Pitzl et~al., {\em Nucl. Instrum. Meth.} {\bf A454}\nolinebreak
  [2]\,(2000)\nolinebreak [2]\,334 [hep-ex/0002044].

\bibitem{h1_spacal}
R.~D. Apphuhn et~al., [\mbox{H1} Collaboration], {\em Nucl. Instrum. Meth.}
  {\bf A386}\nolinebreak [2]\,(1997)\nolinebreak [2]\,397.

\bibitem{geant_manual}
R.~Brun et~al., {\em \mbox{GEANT 3} User's Guide}, 1987,
  \verb+CERN-DD/EE/84-1+.

\bibitem{rapgap}
H.~Jung, {\em Comp. Phys. Comm.} {\bf 86}\nolinebreak [2]\,(1995)\nolinebreak
  [2]\,147, (see also http://www-h1.desy.de/~jung/rapgap.html).

\bibitem{frag_lund_bowler}
B.~Andersson, G.~Gustafson, G.~Ingelman, T.~Sj\"ostrand, {\em Phys. Rept.} {\bf
  97}\nolinebreak [2]\,(1983)\nolinebreak [2]\,31.

\bibitem{heracles}
A.~Kwiatkowski, H.~Spiesberger, H.~J. M\"ohring, {\em Comp. Phys. Comm.} {\bf
  69}\nolinebreak [2]\,(1992)\nolinebreak [2]\,155.

\bibitem{res_phot_grv}
M.~Gl\"uck, E.~Reya, A.~Vogt, {\em Phys. Rev. {\bf D}} {\bf 45}\nolinebreak
  [2]\,(1992)\nolinebreak [2]\,3986;
\\
M.~Gl\"uck, E.~Reya, A.~Vogt, {\em Phys. Rev. {\bf D}} {\bf 46}\nolinebreak
  [2]\,(1992)\nolinebreak [2]\,1973.

\bibitem{diffvm_1}
B.~List, \mbox{Diploma thesis, Techn. Univ. Berlin (1993)}, available from
  http://www-h1.desy.de/publications/theses\_list.html.

\bibitem{h1_diff_elasp_2005}
\mbox{H1} Collaboration, {\it "Measurement of Diffractive Deep-Inelastic
  Scattering with a Leading Proton at HERA"}, \mbox{DESY} 06-048, accepted by
  {\em Eur. Phys. J. {\bf C}} [hep-ex/0606003].

\bibitem{pythia}
T.~Sj\"ostrand, {\em Comp. Phys. Comm.} {\bf 135}\nolinebreak
  [2]\,(2001)\nolinebreak [2]\,238 [hep-ph/0010017].

\bibitem{pdg_2004}
S.~Eidelman et~al., {\em Phys. Lett. {\bf B}} {\bf 592}\nolinebreak
  [2]\,(2004)\nolinebreak [2]\,1.

\bibitem{h1_mass_reflections_2000}
S.~Hengstmann, \mbox{PhD thesis, Univ. Z\"urich (2000)}, available from
  http://www-h1.desy.de/publications/theses\_list.html.

\bibitem{kt_algorithm}
S.~D. Ellis, D.~E. Soper, {\em Phys. Rev. {\bf D}} {\bf 48}\nolinebreak
  [2]\,(1993)\nolinebreak [2]\,3160 [hep-ph/9305266];
\\
S.~Catani, Y.~L. Dokshitzer, M.~H. Seymour, B.~R. Webber, {\em Nucl. Phys. {\bf
  B}} {\bf 406}\nolinebreak [2]\,(1993)\nolinebreak [2]\,187.

\bibitem{h1_incl_dstar_php_2005}
\mbox{H1} Collaboration, {\it ``Inclusive D*-Meson Cross Sections and D*-Jet
  Correlations in Photoproduction at HERA''}, \mbox{DESY} 06-110, submitted to
  {\em Eur. Phys. J. {\bf C}} [hep-ex/0608042].

\bibitem{frag_peterson}
C.~Peterson, D.~Schlatter, I.~Schmitt, P.~M. Zerwas, {\em Phys. Rev. {\bf D}}
  {\bf 27}\nolinebreak [2]\,(1983)\nolinebreak [2]\,105.

\bibitem{charm_fractions}
L.~Gladilin, {\it ``Charm hadron production fractions''}, hep-ex/9912064.

\bibitem{h1_incl_dmesons_2005}
A.~Aktas et~al., [\mbox{H1} Collaboration], {\em Eur. Phys. J. {\bf C}} {\bf
  38}\nolinebreak [2]\,(2005)\nolinebreak [2]\,447 [hep-ex/0408149].

\bibitem{track_multiplicities}
D.~Coffman et~al., [\mbox{MARK III} Collaboration], {\em Phys. Lett. {\bf B}}
  {\bf 263}\nolinebreak [2]\,(1991)\nolinebreak [2]\,135.

\bibitem{h1_nlo_fit}
E.~Laenen, S.~Riemersma, J.~Smith, W.~L. van Neerven, {\em Nucl. Phys. {\bf B}}
  {\bf 392}\nolinebreak [2]\,(1993)\nolinebreak [2]\,162;
\\
E.~Laenen, S.~Riemersma, J.~Smith, W.~L. van Neerven, {\em Nucl. Phys. {\bf B}}
  {\bf 392}\nolinebreak [2]\,(1993)\nolinebreak [2]\,229.

\bibitem{hvqdis_incl}
B.~W. Harris, J.~Smith, {\em Nucl. Phys. {\bf B}} {\bf 452}\nolinebreak
  [2]\,(1995)\nolinebreak [2]\,109 [hep-ph/9503484].

\bibitem{hvqdis_diff}
L.~Alvero, J.~C. Collins, J.~J. Whitmore, {\em PSU-TH-} {\bf 200}\nolinebreak
  [2]\,(1998)\nolinebreak [2]\,13 [hep-ph/9806340].

\bibitem{fmnr_incl_1}
S.~Frixione, M.~L. Mangano, P.~Nason, G.~Ridolfi, {\em Phys. Lett. {\bf B}}
  {\bf 348}\nolinebreak [2]\,(1995)\nolinebreak [2]\,633 [hep-ph/9412348].

\bibitem{fmnr_incl_2}
S.~Frixione, P.~Nason, G.~Ridolfi, {\em Nucl. Phys. {\bf B}} {\bf
  454}\nolinebreak [2]\,(1995)\nolinebreak [2]\,3 [hep-ph/9506226].

\bibitem{frag_nason_oleari}
P.~Nason, C.~Oleari, {\em Nucl. Phys. {\bf B}} {\bf 565}\nolinebreak
  [2]\,(2000)\nolinebreak [2]\,245.

\bibitem{ccfm}
M.~Ciafaloni, {\em Nucl. Phys. {\bf B}} {\bf 296}\nolinebreak
  [2]\,(1988)\nolinebreak [2]\,49;
\\
S.~Catani, F.~Fiorani, G.~Marchesini, {\em Phys. Lett. {\bf B}} {\bf
  234}\nolinebreak [2]\,(1990)\nolinebreak [2]\,339;
\\
S.~Catani, F.~Fiorani, G.~Marchesini, {\em Nucl. Phys. {\bf B}} {\bf
  336}\nolinebreak [2]\,(1990)\nolinebreak [2]\,18.



\end{thebibliography}
